\documentclass[a4paper,11pt]{article}
\pdfoutput=1 

\usepackage{jheppub} 

\usepackage[T1]{fontenc} 
\usepackage[toc]{appendix}
\usepackage{caption}
\usepackage{stackengine}

\usepackage{color,amsmath,amssymb,multirow,hyperref,booktabs,graphicx,mathtools} 
\usepackage{color,xcolor}
\definecolor{darkblue}{rgb}{0,0,0.5}
\definecolor{darkred}{rgb}{0.5,0,0}
\definecolor{darkgreen}{rgb}{0,0.5,0}
\newcommand{\cw}[1]{\textbf{\textcolor{darkblue}{(#1 --cw)}}}

\counterwithin{equation}{section} 

\title{The $z_g$ distribution  for heavy flavour jets in dense quark-gluon plasma }

\author[a]{Boris Blok}
\author[a]{Chang Wu}

\affiliation[a]{Department of Physics, Technion – Israel Institute of Technology,\\Haifa, Israel}

\emailAdd{blok@physics.technion.ac.il}
\emailAdd{chang.wu@campus.technion.ac.il}

\abstract{
We study the $z_g$ distribution for heavy flavour, i.e. bottom and charm quark, jets propagating through the dense QCD medium. We extend the late emission approximation for Armesto-Salgado-Wiedemann (ASW) formula to heavy flavours. %
We consider both the normalised and $N_{\rm jet}$ normalised $z_g$ distributions,  and the ratio of the latter distributions to the vacuum ones, called the R ratio. We demonstrate that since there is no collinear singularity in the medium, there is no principal need to use the Sudakov safety technique for medium-induced emission (MIE). We see that contrary to the vacuum case, the normalised $z_g$ distribution is sensitive to the dead cone angle value. In particular, for the case of $N_{jet}$ normalised distribution with $\theta_g\le \theta_{cut}$, it is possible to directly probe the dead cone gluons in the medium.

Our results can be useful to guide the experimental measurements of the heavy flavour jet substructure in dense QCD medium.
}

\def\beq{\begin{equation}}
\def\eeq{\end{equation}}
\def\beeq{\begin{eqnarray}}
\def\eeeq{\end{eqnarray}}

\def\2GPD{$_2\mbox{GPD}$}

\def\12{$1\otimes 2$}
\def\22{$2 \otimes 2$}

\def\Qsep{Q_{\mbox{\rm\scriptsize sep}}}
\def\Qsep2{Q^2_{\mbox{\rm\scriptsize sep}}}
\begin{document} 
\maketitle
\flushbottom

\section{Introduction}
\label{sec:intro}
\par The study of heavy flavour, i.e. b- and c-jets, is an important 
subject in the study of quark-gluon plasma (QGP). In particular, a lot of work was done to study these jets inGyulassy-Levai-Vitev (GLV) ~\cite{Gyulassy:2000fs,Gyulassy:1999zd,Gyulassy:2000er} (dilute medium) and Baier-Dokshitzer-Mueller-Peigne-Schiff -Zakharov (BDMPS-Z) ~\cite{Baier:1996kr,Baier:1996sk,Baier:1998yf,
Baier:1998kq,Baier:2001yt,Baier:2000mf,Zakharov:1996fv,Zakharov:1997uu} (dense medium) approaches. The issue attracted a lot of attention due to the seeming tension between the experimental results, which showed a very slow decrease in energy loss with the increase of dead cone angle, up to rather small energies of order 25 GeV, and the theoretical results. For the dense medium, which is the subject of this article, the first work was done in the dead cone approximation ~\cite{Dokshitzer:2001zm} and showed a rather rapid decrease in energy loss by heavy quark with the increase of the dead cone angle. This result was found to be in contradiction with the experimental data. %
A more careful analysis was done in ~\cite{Armesto:2003jh}, where the so-called ASW formula was derived, and in ~\cite{Aurenche:2009dj}. The net result was that the energy loss remains the same as for the massless case if we increase the dead-cone angle up to 0.05~\cite{Armesto:2003jh} and starts to decrease for a larger dead-cone angle, but the rate of the energy loss decreases much slower than in ~\cite{Dokshitzer:2001zm}. The physical reason is that in the medium the dead cone effect is weakened, and the collinear gluons "fill" the dead cone.
\par The aim of this paper is to show that the study of the jet substructure, in particular of the  $z_g$ distribution, gives the possibility to check if indeed the dead cone effect is weakened in the media, relative to vacuum, both qualitatively and quantatively.
\par The ASW formula for massless quarks can be rewritten in large L limit (where L is the length of the medium) in the form of the convolution of the broadening term and BDMPS-Z-like radiation term, with the peak at transverse momenta $k_{f}$, where 
\begin{equation}
  k_{f}^2=\sqrt{\hat q\omega},  
  \label{one}
\end{equation}
$\hat{q}$ being the jet quenching coefficient and $\omega$ the radiation frequency ~\cite{Mehtar-Tani:2012mfa}.This picture is easily extended to the heavy quark case. The filling of the dead cone, i.e. nonzero low $k_t$ limit of the transverse momentum distribution of gluons radiated by the heavy quark, comes from the re-scattering of radiated soft gluons on the medium centres. 
\par One of the ways to study the properties of jets in quark-gluon plasma is to study the jet substructure observables. These observables on one hand  can be directly measured in the experiments and on the other hand, can be studied with both 
using MC simulations and calculated analytically. 
These observables are often associated with different jet grooming techniques, for example, the soft drop algorithm ~\cite{Larkoski:2014wba}. 
\par In this paper, we shall calculate the $z_g$ distribution semianalytically for the specific $\beta=0$ case that corresponds to the modified mass drop tagger~\cite{Marzani:2019hun}, which can probe the medium splitting function.
The $z_g$ distribution in the quark-gluon plasma was recently studied for massless partons ~\cite{Mehtar-Tani:2016aco,Caucal:2019uvr}.
\par We extend the calculation of \cite{Mehtar-Tani:2016aco,Caucal:2019uvr}.
to heavy flavour case, using late emission approximation, i.e. large L limit of the ASW formula, which provides the extension of BDMPS-Z formalism to heavy flavours. Our results have a strong dependence on the dead cone angle, contrary to the vacuum case, where such dependence is absent.  Thus, our results can be used for experimental measurements to check the results in Ref.~\cite{Armesto:2003jh} that the dead cone effect is strongly reduced in the medium due to re-scattering.
\par The basic physical picture that follows from the late emission approximation is explained in detail in Sec.\ref{sec:theory}, where the heavy parton radiation in the medium is naturally split into two parts. First, it is the late emission part, which includes the gluons that are radiated at times $t\ge t_f$, where $t_f$ is the  formation time for medium-induced emissions (MIE).   Note that this contribution scales with the medium length.
\par The second part of the radiation that we take into account is vacuum-like emissions (VLE) \cite{Caucal:2018dla}. This is the conventional vacuum emission of gluons by heavy quark, that occur before heavy quark scatters in the medium centres and acquires 
Landau-Pomeranchuk-Migdal (LPM) effect related phase, i.e. at times $t\le t_f$, i.e. medium formation time. These radiated gluons then propagate through the medium remaining inside the jet cone, and acting as the sources of MIE.
\par Thus, we are working on the approximation that the full shower is the sum of VLE and MIE parts. Such division naturally implies the factorisation picture since these two parts of the shower were created in different time intervals.
\par The ASW formula contains the terms that have different behaviour as a function of media scale $L$. 
There are terms that increase like L, and the terms that do not scale, i.e. have a finite limit at large L, but are numerically large for characteristic medium lengths L=4-6 fm.
\par First there are the emissions radiated for $t\le t_f$. These emissions as we shall see in the next chapter can be split into two parts: 
First, there is the hard gluon contribution, which is twice the usual vacuum contribution. This contribution is suppressed at transverse momenta $k_t\le Q_s=\sqrt{\hat q L}$, where L is the medium scale, and $Q_s$ is the characteristic momenta acquired after rescattering \cite{Blaizot:2013hx}.  %
We shall see that as in the massless limit, this contribution cancels out with the boundary term in the ASW formula. This cancellation becomes better and better numerically with the increase of the dead cone angle. This cancellation is exact if we do not take into account the broadening factor, the latter leads to small corrections in the $k_t\le Q_s$ region.

Consequently, we can neglect these two terms, including the boundary term in the ASW formula.  Due to this cancellation, we may argue that although the boundary term is of the order of the bulk term numerically for intermediate $L=4-6 $ fm, it can be neglected and 
the corrections that do not scale with L can be neglected relative to the late emission approximation part even for intermediate L.
\par  The second contribution, as we shall see in the next chapter, is due to BDMPS-Z-like gluons radiated and emitted 
at $t<t_f$. This is just the extrapolation of BDMPS-Z formalism to this kinematic region, and we shall neglect them. Numerically they give negligible contribution for all the kinematic regions we consider $\omega <<\omega_c=\hat qL^2/2$. 
\par The detailed numerical comparison between the  ASW spectrum and the BDMPS-Z spectrum and the late emission approximation for massless quark jets is given in the next section and we see a rather good qualitative and quantitative agreement.
\par Let us note that we use the late emission but not the full ASW expression for calculations for a number of reasons.
First, in the late emission approximation is in good numerical agreement with ASW in the phase space regions we consider.
Moreover the numerical accuracy of late emission approximation increases with the increase of the heavy quark mass as we shall see in the next section from direct numerical comparison.
Second the  integrand in the integral representation of energy   distribution in late emission approximation 
can be be made exponentially decreasing by contour rotation (see next section)
while the integrals we encounter in ASW formula are highly oscillating and it takes considerably more machine time to 
calculate them. Third, late emission approximation naturally represents the full emission probability 
as a sum of VLE and MIE.
\par
\begin{figure}
\centering
\includegraphics[width=0.48\textwidth]{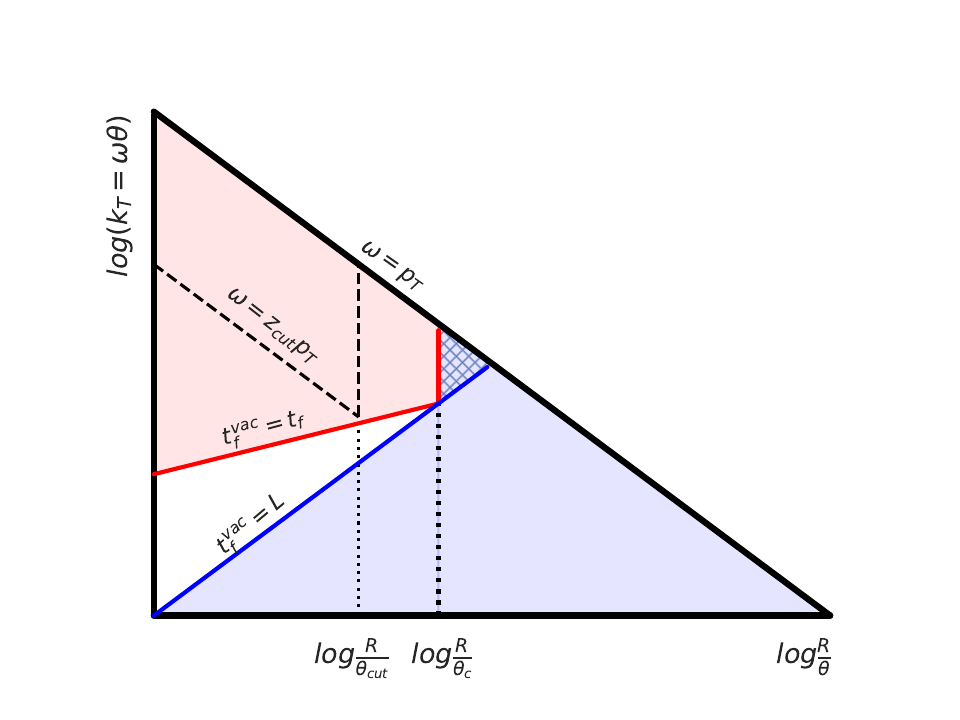}
\includegraphics[width=0.48\textwidth]{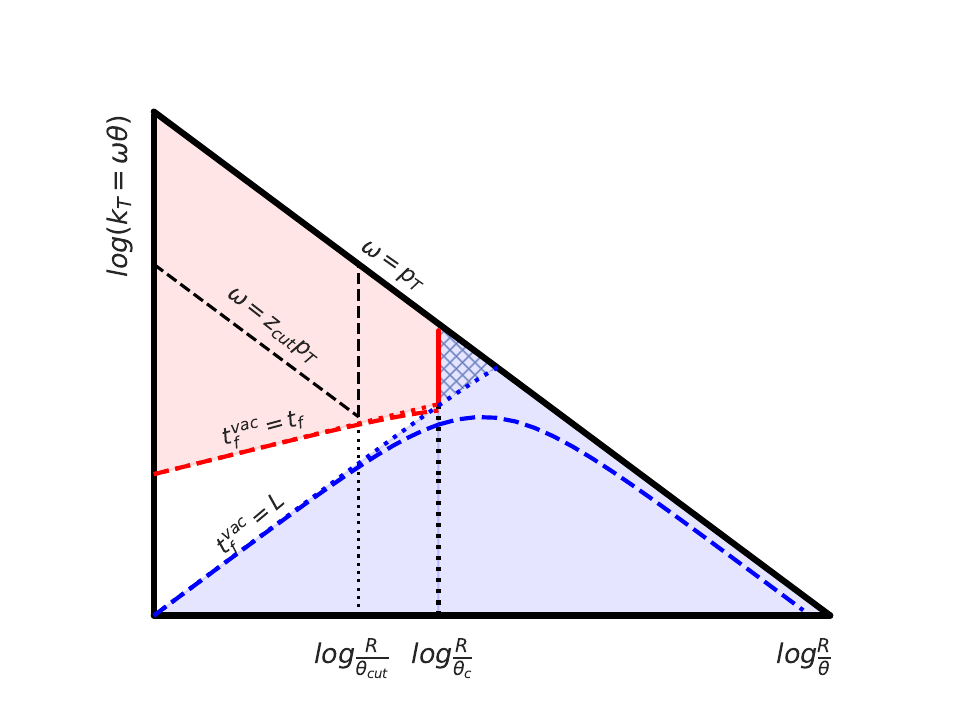}
\caption{Lund diagram representation of the phase space for the in-medium radiation for massless case (left) and heavy flavour jets (right) with c-jets (dotted line) and b-jets (dashed line).
 \label{fig:lund}}
\end{figure}

 In this work, we keep our calculations at double logarithms accuracy, so that the factorisation in time between VLE and MIE still holds for heavy flavour cases. The phase space of this factorisation picture is depicted in the Lund diagram shown in Fig.\ref{fig:lund}, where the phase space for heavy flavour jets is shown on the right side and its massless limit is shown on the left side. The phase space consists of several different regions as follows:
\begin{itemize}
    \item Blue region: This is the region corresponding to $t^{vac}_f>L$, which means radiations created outside of the medium, i.e. gluons outside of the medium, the blue crossed region\footnote{For heavy flavour Lund diagram, we only showed the blue crossed region for c-jets, to avoid overlapping between b- and c-jets cases} is between $t^{vac}_f<L$ and $\theta<\theta_c$, i.e. not resolved by the medium
\footnote{Following \cite{Caucal:2019uvr},\cite{Mehtar-Tani:2012mfa},\cite{Casalderrey-Solana:2011ule} we
 start integration over the angle $\theta$ from  $\theta_c$, such that $\theta_c=2/\sqrt{\hat q L^3}$.
In terms of this angle $t_{coh}(\theta_c)=L$,where $t_{coh}=(4/(\hat q\theta_{\bar q q}^2))^{1/3} $, so that the decoherence factor
for colour dipole is 
$$S(t)=e^{-\frac{1}{6}(\frac{t}{t_{coh}})^3}.$$
For color dipole with opening angle $\theta_{q\bar q}$ the  interference starts
to be strongly suppressed for $\theta\ge \theta_c$. The decoherence lengths  are of course determined 
up to numerical constant of order 1--2. We expect that changing of $\theta_c$ by this factor will lead to only minor numerical changes.}.  These vacuum gluons,  unlike VLE gluons inside the medium, lose energy coherently.
    \item Red region: For VLE gluons inside the medium, the corresponding kinematic condition is 
    \beq
    t_f^{\rm vac}\le t_f,
    \eeq 
    where the vacuum formation time for heavy quark is given by 
    \beq
    t_{f}^{\rm vac}=\frac{\omega}{k_{t}^2+(\theta_{0}\omega)^{2}},
    \eeq
    and the medium formation time is 
    \beq
    t_f\sim \sqrt{\omega/\hat q},
    \eeq
    here $\theta_0=m/p_T$ is the dead cone angle,
    $p_T$ is the energy of the jet and m is the mass of the heavy flavour quark.
    \item White region: For $L\gg t_{f}^{vac}>t_{f}^{med}$, the VLEs are vetoed.
\end{itemize}
\par Here $\theta_c\sim 2/\sqrt{\hat qL^3}$ is the coherence angle which is 0.04 and 0.02 for $L=4$ fm and  $L=6$ fm respectively, where we use jet quenching coefficient value $\hat{q}= 1.5\text{GeV}^2$/fm.  So for charmed quark dead cone angle is typically smaller than the coherence angle, while for b-jet the dead-cone angle is larger than the coherence angle for $p_T\le 125$ GeV for L=4 fm and  $p_T\le 70$ GeV for L=6 fm. 

\par The phase space satisfying the Soft Drop condition is shown as the region within the black dash lines at the top left corner, more about Soft Drop will be discussed in Sec.\ref{sec:z_g}.

\par Furthermore, the phase space for heavy flavour extension is shown on the right-hand side of Fig.\ref{fig:lund}, where the factorisation still holds. We plotted $\theta_0=0.2$ and $\theta_0=0.05$, and we see that if we increase the dead cone angle, the phase space for the outside medium VLE region decreases, due to the dead cone effect, as expected.

\par We use the late emission approximation for the semi-analytical calculation of $z_g$ distribution for heavy flavour jet. We consider two ways to calculate this distribution: First, $N_{\rm jet}$ normalised, without the use of Sudakov safety technique.
Second, the close analogue of Sudakov safety technique. Such an approach is needed to have the smooth limit to the massless jet case. %
We multiply only VLE contribution on the corresponding Sudakov factor since only this part of the radiation acquires collinear singularity in the massless limit. We then normalise this distribution by demanding a total integral over $z_g$ equal to 1.
This is a close analogue of the Sudakov safe distribution. Such an approach was used in \cite{Mehtar-Tani:2016aco} for the massless case.
\par We shall see that with the increase of the dead cone angle the relative role of MIE increases, since the dead cone effect is much stronger in the vacuum than in the MIE. This 
means that the study of the heavy flavour jet substructure permits  the experimental study of both MIE radiation and  the gluon dynamics in the dead cone.
\par Note that a somewhat  connected study where carried in \cite{Caletti:2023spr}, where however a different 
substructure observable $\theta_g$ distribution was calculated, using a different grooming procedure, the so called
late $k_t$ jet grooming procedure, which is different from from the soft drop procedure used in this article. Indeed the former eliminates parts of collinear radiation while soft drop eliminates part of soft radiation. Thus we expect our to studies will be complimentary to each other. 
\par The paper is organised in the following way. In Chapter~\ref{sec:theory}, we extend
 the late emission approximation to heavy flavour medium-induced radiation and explain why it is a good approximation for the energy spectrum. %
In Chapter~\ref{sec:energy-loss}, we describe the energy loss by heavy flavour jets. 
In Chapter~\ref{sec:z_g}, we start by reviewing the basic formalism needed for $z_g$ distribution calculation. We demonstrate that in the massless limit, our formula is in agreement with those used in \cite{Caucal:2019uvr} where the $z_g$ distribution of light partons was calculated using Sudakov safety technique. We show that for massive quark jets, there is no need to use Sudakov safety technique, unless we are interested in a smooth massless limit. 
In Chapter\ref{sec:pheno}, we study the heavy flavour $z_g$ distribution of both b and c jets with realistic energies and dead cone angles, we consider both normalised $z_g$ distribution ,
and $N_{\rm jet}$ normalised distribution. 
Our conclusions are summarised in Chapter~\ref{sec::conclusion}.
\par In Appendix A  we compare numerically the energy spectrum for the late emission approximation with the ASW~\cite{Armesto:2003jh}, BDMPS-Z, broadening approach ~\cite{Blaizot:2013hx,Caucal:2019uvr} and dead cone factor approach~\cite{Dokshitzer:2001zm}. 

\section{Basic formalism: gluon radiation and energy loss of a heavy quark}
\label{sec:theory}
\subsection{Late emission approximation}
\par Recall that the spectrum of gluon radiation from the heavy quark 
with the frequency  $\omega$ and transverse momentum $\vec k_t$ is described by the so-called
ASW formula, that can be written as  ~\cite{Armesto:2003jh}
 \begin{eqnarray}
 \omega \frac{dI}{d\omega d^2k_t }&=&\frac{C_F\alpha_s}{(2\pi)^2\omega^2}2Re\int d^2y \int ^\infty_0dt'\int^{t'}_0 dt e^{-i\vec k_t\vec y}\nonumber\\[10pt]
&\times& e^{-\int^\infty_{t'} dsn(s)V(\vec y(s))}\partial_{\vec x}\partial _{\vec y}(K(\vec y,t',\vec x,t)-K_0(\vec y,t';\vec x,t))\vert_{\vec x=0}.\nonumber\\[10pt]
\label{e1}
 \end{eqnarray}
 Here $K$ is the propagator of the  particle in the medium with the two-dimensional imaginary effective potential V due to the scattering centres, and 
 $K_0$ is the corresponding propagator of the free particle in the vacuum. The effective two-dimensional potential is given by 
 \beq
 V(\vec\rho )=i\int \frac{d^2q_t}{(2\pi)^2}(1-\exp(i\vec q_t\vec\rho))\frac{d^2\sigma_{el}}{d^2q_t}.
 \eeq 
 Here $d^2\sigma_{el}/d^2q_t$ is the cross-section of elastic scattering of high energy particles. 
 on the medium centre.   We assume the static medium of the brick form with the length L
 \beq
  n(s)=U(L-s)U (s)
  \eeq
  where $U=1$ if $s\ge 0$ and 0 if $s<0$ is a conventional step function.
 The medium is described by the Gyulassy-Wang model  ~\cite{Gyulassy:1993hr}.
 The effective potential in the momentum space is given by 
  \beq
\frac{ d\sigma(\vec q_t )}{d^2 q_t}=\frac{4\pi\alpha_sm^2_DT}{(q_t^2+\mu^2)^2}\equiv \frac{g^4n}{(q_t^2+\mu^2)^2},
 \eeq
 where the parameter $\mu\sim m_D$, and the Debye mass $m_D$ is given by 
 \beq
 m_D^2\sim  4\pi\alpha_sT^2(1+N_f/6)=\frac{3}{2}g^2T^2
 \eeq
 for $N_f=3$ light quarks, T is the  temperature of the created medium-QGOP. The density of the scattering centres in the Gyulassy-Wang  model is given by $n=\frac{3}{2}T^3$,
 and the strong coupling is $\alpha_s=\frac{g^2}{4\pi}$.
 The effective potential in the coordinate space is 
  \beq
 V(\rho)=\frac{\hat q}{4N_c}(1-\mu \rho K_1(\mu \rho ))\sim \frac{\hat q\rho^2}{4N_c}(\log(\frac{4}{\mu^2\rho^2})+1-2\gamma_E)\label{pot},
 \eeq 
 where $\gamma_E=0.577$ is the Euler constant, and the bare quenching coefficient is
 \beq
 \hat q=4\pi\alpha_s^2N_cn.
 \eeq
 Note that $\hat q$ is fully determined by medium properties, and does not depend on the quark mass. 
\par 
In this paper we shall work on the harmonic approximation:
\beq
V_{HO}(\rho)=\hat q\rho^2/2.
\eeq
Then the formula \ref{e1} can be represented as a sum of two contributions: the bulk one and the boundary one ~\cite{Armesto:2003jh}.
The first one is the bulk contribution and is given by 
 \begin{eqnarray}
 \omega \frac{dI^{HO\,\,\,Bulk}}{d\omega d^2k_t }&=&\frac{\alpha_sC_F}{(2\pi)^2\omega^2}2Re\int d^2y \int ^L_0dt'\int^{t'}_0 dt e^{-i\vec k_t\vec y}\nonumber\\[10pt]
&\times& e^{-1/4\hat q (L-t')y^2}\partial_{\vec x}\partial _{\vec y}(K(\vec y,t',\vec x,t)-K_0(\vec y,t';\vec x,t))\vert_{\vec x=0}.,\nonumber\\[10pt]
\label{e1a}
 \end{eqnarray}
where K is the heavy quark propagator (more rigorously, the propagator of the heavy quark--gluon system)  in harmonic oscillator approximation given by Eq. \ref{e2}. The medium length L is the length of the medium sample, i.e. the distance quark goes through the QGP. This term corresponds to the gluons that 
are created and absorbed inside the medium, at times $0\le t,t_1\le L$
\par The second contribution is a boundary term given by 
 \begin{eqnarray}
 \omega \frac{dI^{\rm HO\,\,\, boundary}}{d\omega d^2k _t}&=&\frac{\alpha_sC_F}{(2\pi)^2\omega^2}2Re\int d^2y \int ^\infty_L dt'\int^{L}_0 dt e^{-i\vec k_t\vec y}\nonumber\\[10pt]
&\times& \partial_{\vec x}\partial _{\vec y}(K(\vec y,t',\vec x,t)-K_0(\vec y,t';\vec x,t))\vert_{\vec x=0},\nonumber\\[10pt]
\label{e1b}
 \end{eqnarray} 
 This term corresponds to the gluons that were emitted in the medium and absorbed afterward in vacuum and vice versa.
\par The  heavy  quark propagator in the imaginary two-dimensional potential $iV_{HO}$  is given by  ~\cite{Zakharov:1996fv}:

\begin{eqnarray}
K_{HO}(\vec y,t';\vec x,t)&=&\frac{i\omega\Omega}{2\pi \sinh \Omega (t'-t)}\exp(\frac{i\omega 
\Omega }{2}\{\coth\Omega  (t'-t)(\vec x^2+\vec y^2)-\nonumber\\[10pt]
&-&\frac{2\vec x\vec y }{\sinh\Omega (t'-t)}\})\exp(-i\theta_0^2\omega (t'-t)/2),\nonumber\\[10pt]
\label{e2}
\end{eqnarray}
where
\beq
\Omega=\frac{(1+i)}{2}\sqrt{\frac{\hat q}{\omega}},
\eeq
and   $\theta_0=m/p_T$ is the dead cone angle, m is a quark mass and $p_T$ is a jet/heavy quark energy.
In the limit when there is no medium this propagator reduces to free heavy quark propagator 
\beq
K_0(\vec y,t';\vec x,t)=\frac{i\omega}{2\pi }\exp(i\frac{\omega (\vec x-\vec y)^2}{2(t'-t)})\exp(-i\theta_0^2\omega (t'-t)/2).\label{e3}
\eeq
 Explicitly the  bulk term is given by:
\begin{eqnarray}
 \omega \frac{dI^{\rm HO\,\,\, Bulk}}{d\omega d^2k_t }&=&-2{\rm Re}\int^L_0dt'\int^L_{t'}dt \frac{\alpha_sC_F\Omega^2}{\pi^2R^2\sinh{\Omega(t-t')}^2}(q(L-t)-\frac{2ik_t^2\omega\Omega\coth{\Omega(t'-t)}}{R})\nonumber\\[10pt]
&\times& \exp(i\theta_0^2\omega(t'-t)/2)\exp(-k_t^2/R),\nonumber\\[10pt]
\label{k1}
 \end{eqnarray}
 where 
 \beq
 R=q(L-t)-2i\omega\Omega\coth{\Omega (t'-t)},
 \eeq
 The boundary term is given by 
 \beq
  \omega \frac{dI^{\rm HO\,\,\, boundary}}{d\omega d^2k _t}=\int ^L_0dt\frac{-i\alpha_sC_Fk^2_t}{(k_t^2+\theta_0^2\omega^2)(\pi)^2\omega}\frac{\exp(\frac{-ik^2_t\tanh\Omega(L-t)}{2\omega\Omega})\exp(i\theta_0^2\omega(t-L)/2}{\cosh \Omega(L-t)^2}
  \label{g1}
  \eeq
These two equations form the so-called ASW formula.
\par The bulk part can be rewritten in a suggestive way ~\cite{Mehtar-Tani:2012mfa}
\begin{eqnarray}
 \omega \frac{dI^{\rm HO\,\,\,Bulk}}{d\omega d^2k_t }&=&\frac{2Re}{4\pi^2\omega^2}\int\frac{d^2k}{4\pi^2}\int^L_0dt' \int ^{t'}_0dt P(\vec k_t-\vec k,t',L)\nonumber\\[10pt]
 &\times&\exp{(-\frac{ik^2\tanh{\Omega \xi}}{2\omega \Omega})}\exp{(-i\theta_0^2\omega \xi/2)}
 /\cosh{(\Omega \xi/2)}^2\nonumber\\[10pt],
 \label{s1}
\end{eqnarray}
where
\beq
P(\vec{k},t,L)=\frac{4\pi}{q(L-t)}\exp\left[-\frac{\vec{k}^{2}}{q(L-t)}\right]
\label{eq:gaussian}
\eeq
and $\xi=t'-t$,
Let us  use the identity:
\begin{equation}\label{change}
\int_{0}^{L}dt'\int_{0}^{t'}dt=\int_{0}^{L}d\xi\int_{\xi}^{L}dt'
\end{equation}

Note now that integral in $\xi=\Delta t\equiv t'-t$ is exponentially 
suppressed for $\xi>t_f$ independent whether we have or not nonzero dead cone value.
We can then split, like it was done in the massless case  the integration in $dt'$ into two regions: $t'>t_f$ and $t'<t_f$. The first region corresponds to $\Delta t\ge t_f$ and the second region to $\Delta t<t'<t_f$. For the first kinematic region  the formula 
\ref{s1} acquires the form \cite{Mehtar-Tani:2011lic}
\begin{eqnarray}
 \omega \frac{dI^{HO\,\,\,Bulk}}{d\omega d^2k_t }&=&\frac{2Re}{4\pi^2\omega^2}\int\frac{d^2k}{4\pi^2}\int^L_{t_f}dt \int^L_0 d\xi P(\vec k_t-\vec k,t,L)\nonumber\\[10pt]
 &\times&\exp{(-\frac{ik^2\tanh{\Omega \xi}}{2\omega \Omega})}\exp{(-i\theta_0^2\omega \xi/2)}
 /\cosh{(\Omega \xi/2)}^2\nonumber\\[10pt],
 \label{s1a}
\end{eqnarray}
In this formula we assume as in massless case \cite{Mehtar-Tani:2011lic}, that $\Delta t\ge t_f$
The formula  \ref{s1} describes the radiation of a gluon with momenta of order $k^2\sim k_f^2=\sqrt{\omega \hat q}$,
that in turn goes through diffusion process by scattering on the medium centers, acquiring momentum of order $k_t^2\sim \hat qL$,
described by the broadening factor \ref{eq:gaussian}. Note that only contribution over this kinematic region scales like L.
This is the late emission contribution. The integration in two variables $t$ and $\xi$ can now be disentangled, 
 \par
\begin{center}
 \begin{figure}
\centering
 \includegraphics[width=0.6\textwidth]{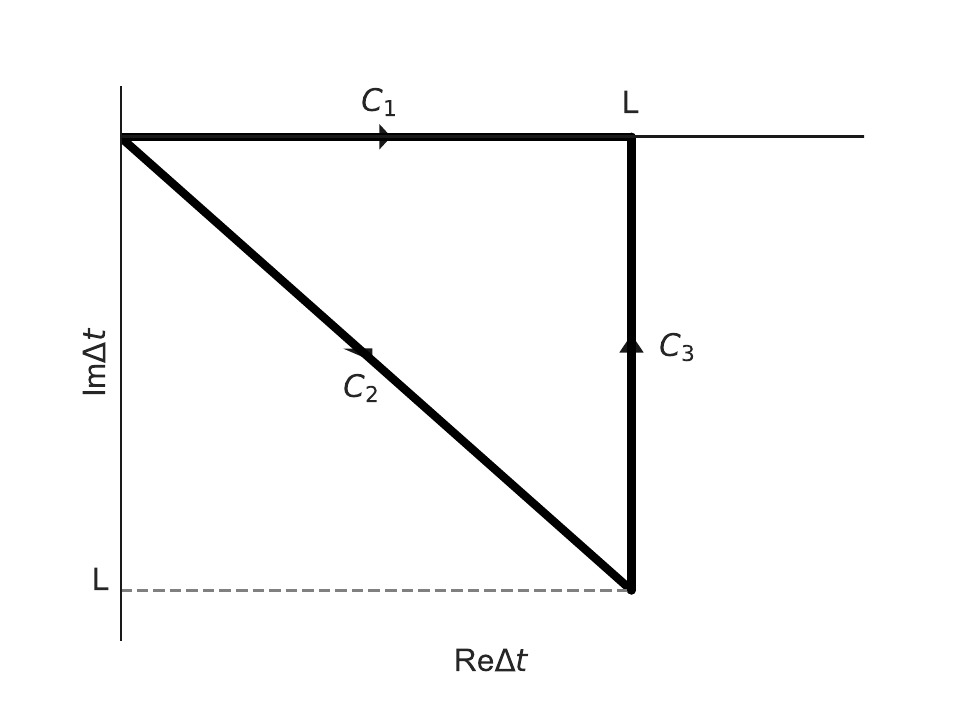}
\caption{The integration contour in Eq. \ref{lea1}.
 \label{contour}}
\end{figure}
\end{center}

 The representation \ref{s1a} can be improved by rotating the integration contour in the complex $\xi$ plane
 like it was done in the massless case \cite{Mehtar-Tani:2012mfa}.
It is  convenient to turn the integration contour round by $\pi/4$ as shown in Fig. \ref{contour},
 It was shown in ~\cite{Mehtar-Tani:2012mfa} that the correction  to the contribution of the rotated contour $C_2$ relative to
 integral over the real axis contour $C_1$) is due to a contour $C_3$ the contribution of the latter is however exponentially small. These estimates can be easily
 expanded to nonzero $\theta_0$. Thus, we shall carry the integration along the contour $C_2$ instead of contour $C_1$, as depicted in Fig. \ref{contour}.

\par The final expression for the distribution of gluon radiation in the late emission approximation is given by 
 \begin{eqnarray}
 \omega\frac{dI^{\text{LEA}}}{d\omega d^{2}k_{t}}=\frac{2}{4\pi^2\omega^2}Re(1-i)\int_{t_{f}}^{L}dt\int_{0}^{L}d\xi\int\frac{d^{2}k}{4\pi^{2}}P(\vec{k}-\vec{k}_{t},t,L)\nonumber\\[10pt]
 k^{2}\exp\left[{-(1+i)\frac{k^{2}}{2k_{f}^{2}}\tanh\frac{\xi}{t_{f}}}\right]\frac{e^{-\frac{1+i}{2}\xi\theta_0^{2}\omega}}{\cosh^{2}\frac{\xi}{t_{f}}}
 \label{lea1}
 \end{eqnarray}
 Note that such contour is optimal for numerical calculation due to a rapid decrease of the integrand with $\xi$:  not $e^{-\frac{\xi}{2t_{f}}}$ if we will not rotate the contour, but  $e^{-\frac{\xi}{t_{f}}}$. Recall here that 
\beq 
k_{f}^{2}=\sqrt{q\omega},\,t_{f}=\sqrt{\frac{\omega}{\hat{q}}}.
\eeq
 \par It is easy to check that for $\omega\le \omega_{DC}$, where $\omega_{DC}=(\frac{\hat q}{\theta_0^{4}})^{1/3}$ (and in particular for all energies of radiated gluons for massless  quark) the Eq. \ref{lea1} can be simplified: 
 \begin{eqnarray} \label{lea2}
\omega\frac{dI}{d\omega d^2k_t}&=&2Re(1-i)\int_{t_{f}}^{L}dt\int_{0}^{t_{f}}d\xi\frac{d^{2}k}{4\pi^{2}}P(\vec{k}-\vec{k}_{t},t,L)\nonumber\\[10pt]
 &\times&\frac{k^{2}\exp\left[-(1+i)\frac{(k^{2}+k_{0}^{2})}{2\omega}\xi\right]}{4\pi^{2}\omega^{2}}
 \end{eqnarray}
where $k_0=\theta_0 \omega$. Here, we continue the integration in $\xi$ only up to $t_f$, and in this region $\tanh(\xi/t_f)$ is taken into account in linear approximation. Note that the integrand in Eq. \ref{lea2} is positive.
\par Recall ~\cite{Blok:2019uny} that the condition $\omega\le \omega_{DC}$ means that the radiation goes essentially outside the dead cone. If we move to frequencies $\omega\ge \omega_{DC}$, the radiation essentially goes into the dead cone
This  corresponds to the condition that the heavy quark vacuum coherence length $t_c\sim \frac{1}{\theta_0^2\omega}\le t_f$ -medium formation length.
Mathematically, the integrand in \ref{lea2} starts to oscillate and use of \ref{lea2} leads to unphysical, in particular negative results
for differential distributions.
Then one needs to use the entire  expression  \ref{lea1}, with 
that will be done in this paper.
\par We shall see in the next subsection that this expression gives rather good approximation for the full ASW formula
for all radiated energies $\omega\le \omega_c$ already for intermediate values of L=4-6 fm.
\par For the energy distribution we obtain
\begin{eqnarray}
\omega\frac{dI}{d\omega}&=&2Re(1-i)\int_{t_{f}}^{L}dt\int_{0}^{L}d\xi\frac{1}{A+B}\left(1-e^{-\omega^{2}\frac{AB}{A+B}}\right)\nonumber\\[10pt]
&+&\frac{Ae^{-i\theta_0^{2}\omega\xi/2}}{B\cosh^{2}\frac{\xi}{t_{f}}}\left[1-e^{-\omega^{2}\frac{AB}{A+B}}\left(1+\omega^{2}\frac{AB}{A+B}\right)\right]
\label{lea3}
\end{eqnarray}
where  we include the region of integration $k_{t}\le\omega$.
\beq
A=\frac{1}{\hat{q}\xi},\,B=\frac{1+i}{2k_{f}^{2}}\tanh\frac{\xi}{t_{f}}
\eeq

\subsection{Corrections to the late emission approximation}
\label{sec:early-emm}
Let us now consider the contributions to ASW formula finite for large L limit.
Let us first comment on integration over t in \ref{s1}. There are two integration regions:$t\le t_f$, $t\ge t_f$. The late emission approximation includes only the second region. The first region describes radiation that occurred before medium formation time. This contribution is maximal for small values of $\theta$, and rapidly decreases with $\theta$. Using the calculation similar to that of ~\cite{Mehtar-Tani:2012mfa} it is easy to see that this contribution includes both hard gluons that essentially do not scatter and the relatively soft ones, with $k\le k_f$. Both 
contributions scale as $t_f$ and not like $L$:
\begin{eqnarray}
\omega\frac{dI^{early}}{d\omega d^{2}k_{t}}&\sim2Re\int\frac{d^{2}k}{4\pi^{2}}P(k_{t}-k,L)\left[\frac{k^{2}}{(k^{2}/2\omega)+(\theta_0^{2}\omega/2)^{2}}\left(1-e^{{-(1+i)\frac{k_{t}^{2}+k_{0}^{2}}{2k_{f}^{2}}}}\right)\right.
\nonumber\\[10pt]
&\left.-2\sqrt{\frac{\omega^{3}}{q}}\left(1-(1+2i)e^{{-(1+i)\frac{k_{t}^{2}+k_{0}^{2}}{2k_{f}^{2}}}}\right)\right]
\label{ad}
\end{eqnarray}
\par Recall also that pure vacuum contribution is given by ~\cite{Dokshitzer:1991fd}
\beq
\omega\frac{dI}{d\omega d^{2}k_{t}}=\frac{\text{\ensuremath{\alpha_{s}C_{F}}}}{\pi^{2}}\frac{k_{t}^{2}}{(k_{t}^{2}+\theta_{0}^{2}\omega^{2})^{2}},
\label{vac1}
\eeq

\begin{figure}
\centering
\includegraphics[width=0.48\textwidth]{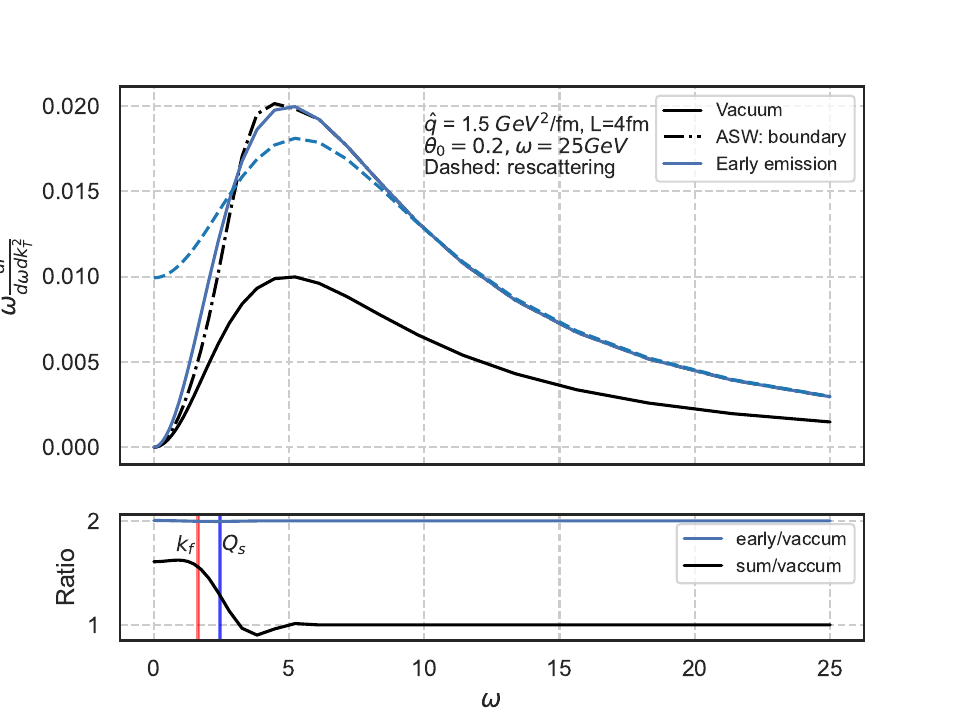}
\includegraphics[width=0.48\textwidth]{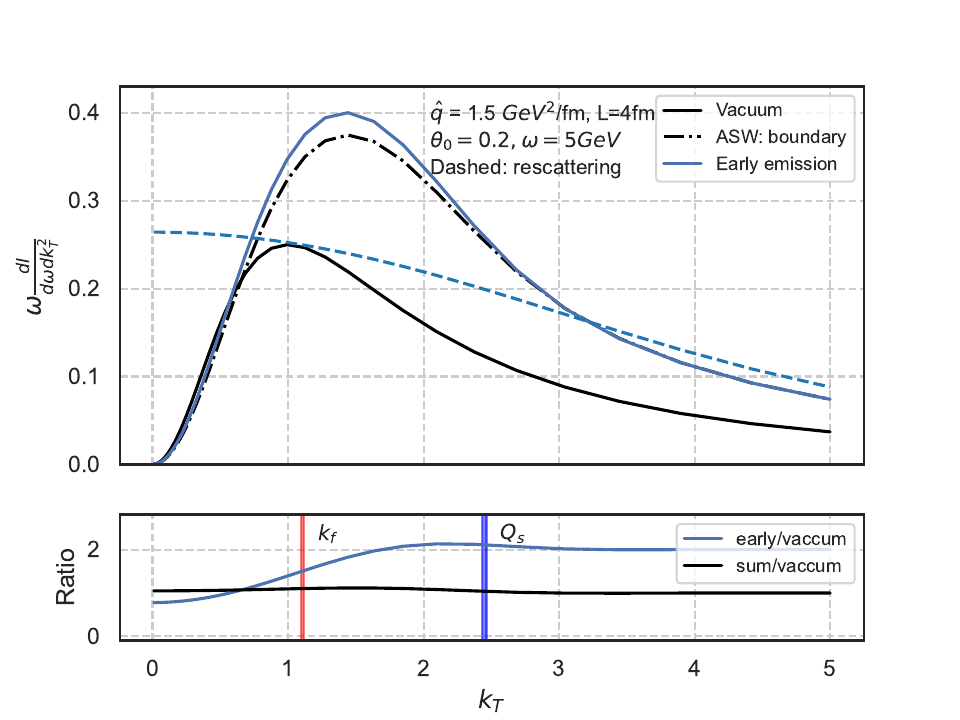}\\
\includegraphics[width=0.48\textwidth]{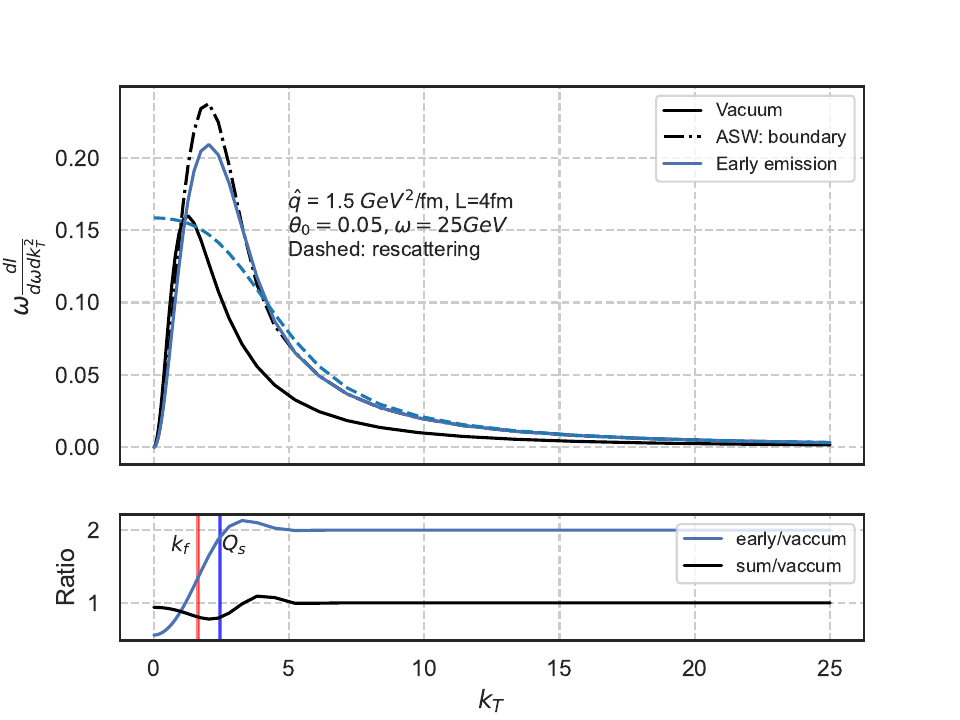}
\includegraphics[width=0.48\textwidth]{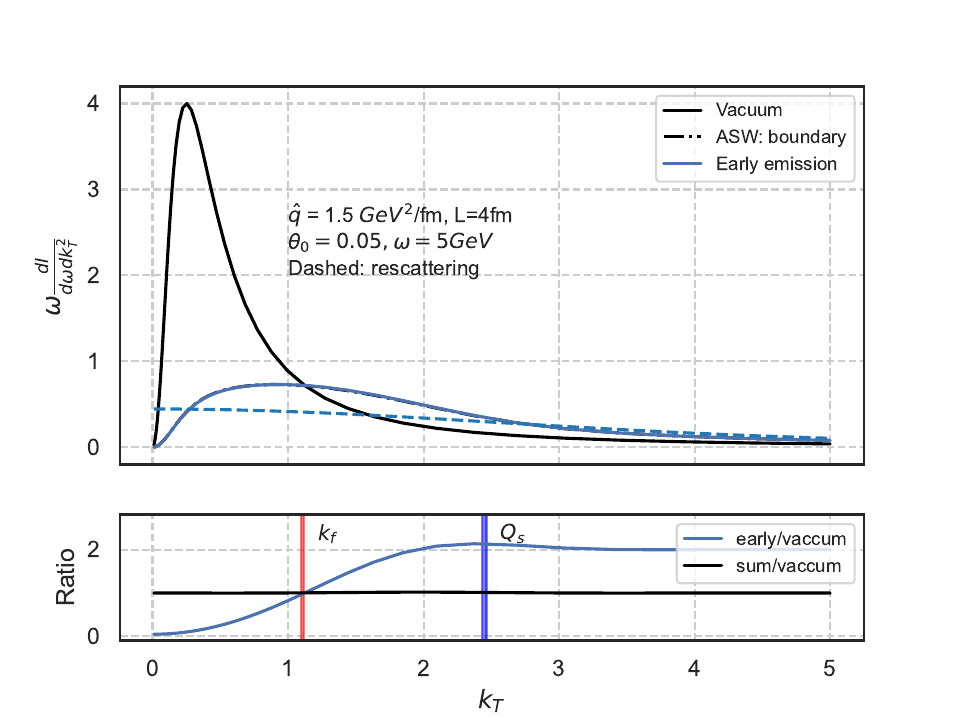}
\caption{The $k_t$ dependence of the ASW boundary contribution, early hard gluon radiation, and vacuum radiation: 
$\omega=25 $GeV (left), $\omega =5$ Gev (right). Upper figures: large $\theta_0=0.2$; Lower row of figures: relatively small $\theta_0 =0.05$. We see that cancellation of boundary and hard early radiation improves with the increase of dead cone angle.
 \label{fig:early}}
\end{figure}

 \par Note now that there is an additional contribution to gluon radiation in ASW formalism (and energy loss) that does not scale with L, and this is the boundary contribution, given by \ref{g1}.
This contribution, as can be checked numerically, is with good accuracy, in absolute value,  twice as large as vacuum contribution \ref{vac1}, ,
the agreement actually improves with the increase of dead cone angle $\theta$.
Note that this contribution is saturated for $L\sim 4-5$ fm, and does not change with the increase of L. Its contribution is negative and is concentrated near the medium endpoint L. We can compare this contribution 
with the positive contribution of semi-hard and hard gluons given by the first line in Eq. \ref{ad}. The latter equation is also 
in good agreement with twice vacuum contribution, especially for $k>k_f$, where k is the momentum that is integrated over,
and thus cancels boundary contribution, the effect that was first studied analytically in ~\cite{Mehtar-Tani:2011lic} for zero dead cone angle.
In Fig.\ref{fig:early} we depict the vacuum, ASW boundary contribution, and contribution given by the first line of Eq.\ref{ad}, note that, in order to make the comparison, we plotted the absolute value of ASW boundary term. The no re-scattering case,
meaning we approximate the broadening factor P by delta function $\sim \delta(\vec k_t-\vec k)$ is shown as the solid line. The results do not depend on L since all the contributions considered do not change for L bigger than 3-4 fm.  Without inclusion of the broadening term, we see that there is a virtually perfect agreement between twice vacuum contribution and early hard emission of ASW gluons, which is exactly cancelled by the boundary term.  We see this both for small and large $\omega$, i.e. inside and beyond the dead cone.
The cancellation is slightly less exact for small $\theta_0$, i.e. for massless cases, but is still rather good numerically. 
\par The inclusion of the broadening term P leads to a perfect cancellation for $k_t>Q_s=\sqrt{\hat q L}$,
for smaller transverse momenta the cancellation becomes less perfect. However, the accuracy of cancellation increases with the increase of dead cone angle, and for $\theta_0\sim 0.2$ is valid for $k_t>k_f$, as it is seen from numerical fit.
\par  This cancellation has immediate physical consequences.,First we can neglect these two contributions in our analysis.
Second due to smallness of this finite L contribution, the late emission approximation will work already for intermediate L.
\par Third cancellation of boundary contributions also justifies the factorisation of the full shower into vacuum-like and medium contributions.
\par Finally we have the second line in Eq.\ref{ad}, which corresponds to BDMPS-Z-like gluons.  Indeed, this distribution is cut by $k_f$, and inclusion of this contribution means that we integrate in Eq.\ref{lea2} from zero, and not from $t_f$. However, numerically, small for $\omega <<\omega_c$, relative to the late emission contribution. and we neglect this contribution relative to VLE contributions that are radiated in the same time.

\par  In summary the physical picture is that early hard gluons are cancelled out by boundary term and can be neglected. These two contributions are 
actually absent from BDMPS-Z formalism, where the boundary contribution is negligible for intermediate L (we use  in this paper $\hat q\sim 0.3\text{GeV}^3 $), and thus this cancellation is already present in the BDMPS-Z  and broadening possibility approximations for the massless case.

\par For the full shower, in addition to MIE part, we need to add the VLE,  which are emitted at $t\le t_f$ and then go through a re-scattering process due to a broadening term. This gives us an extension of the factorisation picture for a full shower consisting of VLE and MIE parts from the massless case~\cite{Caucal:2019uvr} to the massive one.
\par In other words, these results justify the use of the scheme suggested in ~\cite{Caucal:2019uvr} also for finite dead cone angle $\theta$.

\section{Energy loss and multiplicities for heavy flavour jets}
\label{sec:energy-loss}
In the previous section, we considered the single gluon energy spectrum by heavy quark. In this section, we consider the multiple radiations, which one needs to include  when $\omega<\omega_{br}$,
as it was originally done in Ref.~\cite{Baier:2001qw}.  Here $\omega_{\rm br}$ is defined by 
\begin{equation}
    \int^{\omega_c}_{\omega_{\rm br}}\frac{dI}{d\omega}d\omega \sim O(1)
    \label{obr}
\end{equation}

\subsection{Multiplicity of vacuum-like emissions}
\label{sec:multi}
We start with vacuum-like emissions. These are the emissions that occurred for $t\le t_f$. As it was argued in ~\cite{Caucal:2019uvr} these emissions occur when
heavy quark radiates inside the medium, but not yet acquired the Landau-Pomeranchuk-Migdal (LPM) phase, i.e. for $t<t_f$. The heavy quark emits the gluons that stay inside the jet and travel through the medium in the same way as a leading heavy quark, as independent gluon subjets. Multiplicity can be estimated by integrating over VLE density as
\beq\label{eq:avg-multi}
\nu^{DLA}(E,R)=\int_{\theta_{cut}}^{R}d\theta\int_{zp_{T}}^{p_{T}}d\omega.\frac{dN}{d\omega d\theta}.
\eeq
\par The multiplicity of heavy quark radiation was studied in detail in \cite{Dokshitzer:1991fd},\cite{Dokshitzer:1988bq},\cite{Khoze:2000iq} .
It is dominated by the gluon shower due to primary gluon emission..
The basic result is that in this approximation 
\beq
\frac{dN_Q(p_T)}{d\omega d\theta}=\frac{dN_q(p_T)}{d\omega d\theta}-\frac{dN_q(m_Q)}{d\omega d\theta}
\label{upt}
\eeq
In other words the multiplicity and fragmentation functions of heavy quark jet are equal to the difference of the light quark 
multiplicty (fragmentation function)  at the jet energy $p_T$ and light quark multiplicity (fragmentation function)
at the scale of heavy quark mass $m_Q$. We refer the reader to \cite{Khoze:2000iq},\cite{Khoze:1996dn} for detailed proof.
\par For light quark multiplicity in the DLA we use \cite{Dokshitzer:1988bq}
\beq
\frac{dN_q(p_T)}{dy d\log{\theta}}=\frac{c_F}{c_V}a^2I_0(2a\sqrt{(Y-y)y_\theta})
\label{eq:dla-density}
\eeq
where $a^2=2c_V\alpha_s/\pi$,
\beq
y=\log{\omega/Q_0},\,\,\,y_\theta=y-\log{1/\theta}\,\,\,Y=\log{p_T/Q_0}
\eeq
Here $Q_0$ is the scale where jet consists from only one heavy quark, and we use in this paper 
the so called limiting spectrum, $Q_0\sim \Lambda_{\rm QCD}$, which is in good agreement with LEP
phenomenological data \cite{Khoze:2000iq}. The function $I_0$ is the standard modified Bessel function.
We consider these formulae both with fixed and running coupling constant.

\begin{figure}
\centering
\includegraphics[width=0.48\textwidth]{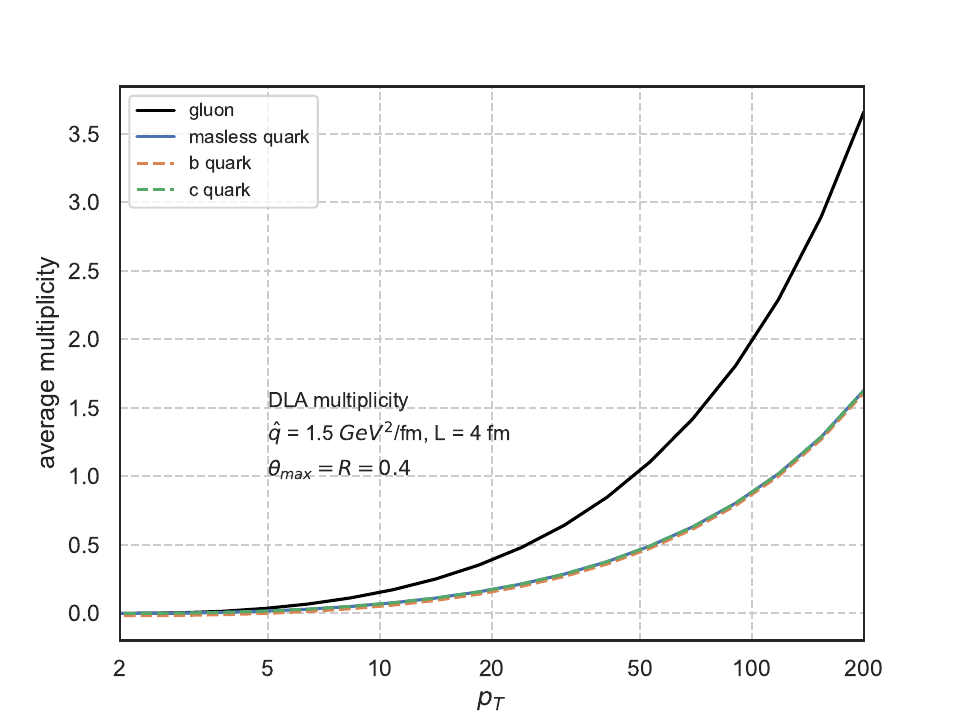}
\caption{Average jet multiplicity as a function of the jet energy $p_T$ \label{fig:multi}}
\end{figure}

 We depict the multiplicity of massless and heavy flavour jets in Fig.\ref{fig:multi}.  We included in  Eq.\ref{eq:dla-density} the 1-loop running coupling 
 \begin{equation}    
 \alpha_{s}\left(k_{t}\right)=\frac{\alpha_{s}\left(M_{z}\right)}{1-\alpha_{s}\left(M_{z}\right)\beta_{0}log\left(k_{t}/M_{z}\right)}.
 \end{equation}
 \par In this paper we shall consider both DLA with fixed coupling which we shall refer to simply as DLA and 
 DLA with running coupling.
For the average multiplicity, we integrate over the whole phase space of VLE inside the medium shown in Fig.\ref{fig:lund}, we see that there is no large rise of multiplicity for both c- and b-jets.  

 
Note that according to the Lund plane in Ref.~\cite{Cunqueiro:2022svx} obtained from MC simulation, contrary to the basic understanding of medium-induced emissions filling the dead-cone, the density of medium-induced Lund plane of c-jets vanished at $\theta<\theta_{0}$, this is due to the hierarchy of the coherence angle $\theta_c$ and the dead cone angle $\theta_0$ since c-jets have small dead-cone angle. Accounting for this is beyond the scope of this article. Note that $\theta_c$ value is mass independent \cite{Calvo:2014cba}
\subsection{Multiple emissions and  $\omega_{br}$ for heavy flavour jet.}

\par In order to calculate the energy loss for the jet, we need to evaluate $\omega_{br}$ defined by Eq. \ref{obr}. Recall that in the BDMPS-Z approach, the energy spectrum for the radiation of a quark  or gluon is 
\beq
\frac{dI^{\rm BDMPS-Z}}{d\omega}=\frac{\alpha_s C_R}{\pi}\sqrt{\frac{2\omega_c}{\omega^3}}.
\eeq
where $C_R=C_A=3$ for gluonic jet and $C_R=C_F=\frac{N^2_c-1}{2N_c}=4/3$ for quark jet.
Then for BDMPS-Z we obtain 
\beq
\omega_{br}^{R}\sim \frac{\alpha_{s}^{2}C_AC_R}{\pi^2}\omega_{c}
\label{eq:w_br}
\eeq

\begin{figure}[!htb]
\begin{center}
\includegraphics[width=0.48\linewidth]{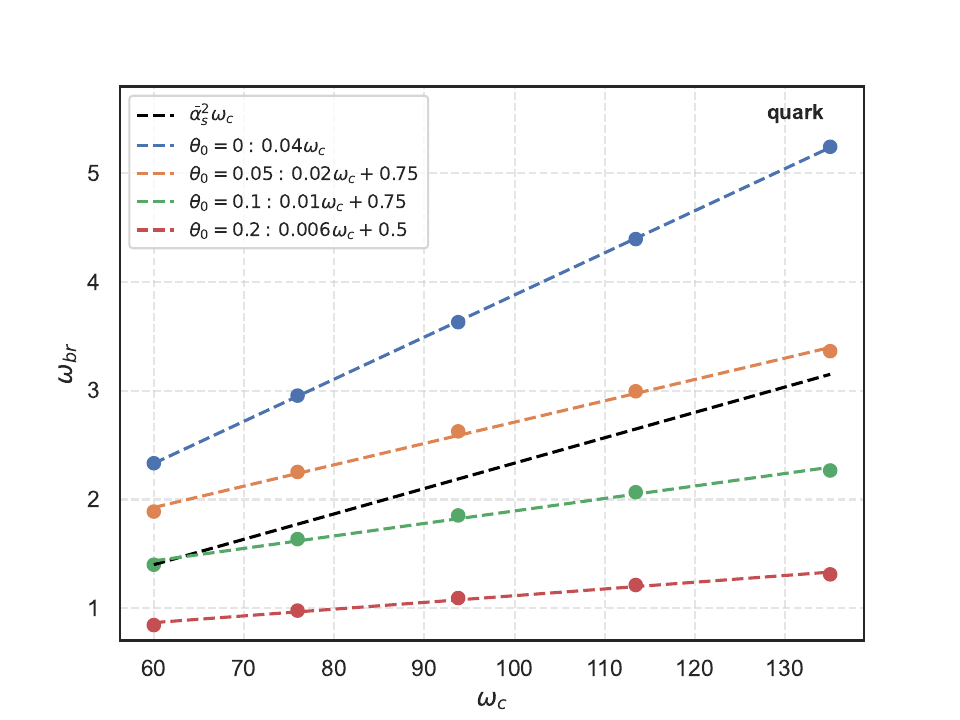}
\caption{The relation between $\omega_{br}$ and $\omega_c$ for quark jets}
\label{fig:wbr-fit}
\end{center}
\end{figure}

For our calculation, we just replace the BDMPS-Z spectrum $dN^{\rm BDMPS-Z}/d\omega$ with the late emission approximation. The numerical results are collected in Tab.\ref{tab:wbr-eloss1}. For $\omega_{br}$ at the massless limit, the late emission calculation agrees well with  Eq. \ref{eq:w_br}, i.e. we obtain 1.4 GeV and 3.2 GeV for $L=\left(4,6\right)$ fm respectively. Moreover, we found that $\omega_{br}$ keeps the linear relation with $\omega_c$ as shown in Fig.\ref{fig:wbr-fit}. 

 \begin{table}[!h]
\begin{center}
\begin{tabular}{ |p{0.8cm}||p{1.cm}|p{1.cm}|p{1.cm}|p{1.cm}|p{1.cm} |p{1.cm}|p{1.cm}|p{1.cm}|p{1.cm}|p{1.cm}|  }
 \hline
 & \multicolumn{4}{|c|}{$L=4$ fm}& \multicolumn{4}{|c|}{$L=6$ fm} \\
 \hline
 $p_T$& 200    & 100&  50& 25& 200 &100&  50 & 25\\
 \hline
$\omega^{b}_{br}$&   2.3  & 1.9   &1.4& 0.8  & 5.2  & 3.3   &2.3 & 1.3\\
 \hline
$\omega^{c}_{br}$&   2.3  & 2.2   &2.1& 1.4  & 5.2  & 4.8   &4.1 & 3.1\\

 \hline
\end{tabular}
\caption{\label{tab:wbr-eloss1}$\omega_{br}$ for heavy flavour quark jet with $\theta_0=m_q/p_T$, $\alpha_{s}=0.24$ and R=0.4}
\end{center}
\end{table}

 \par It is interesting to note that for all realistic jet energies $p_T>25$ GeV for $z_{\rm cut}=0.1$ we always probe the frequencies  $\omega>\omega_{\rm br}$, both for charm and bottom jets. 
 \par Note also that typically we can probe the dead cone radiation for b-jets, but not for charm jets. Indeed the condition for the bulk of radiation to go to the dead cone is :
 \beq
 \omega\ge \omega_{\rm DC}=(\frac{\hat q}{\theta_0^4})^{1/3}
 \eeq
This means that for $p_T\ge m^4/\hat q$, $\omega_{\rm DC}>p_T$, i.e. the characteristic BDMPS-Z angle lies outside the dead cone, and radiation of the massive quark is essentially the same as for the light quark jet. For charm quark, this is the energy of order 20 GeV, while for the bottom one is of order 400 GeV. thus we expect that charm jets at LHC will behave very similar to massless quarks, while for b quarks there will be strong dependence on dead cone.

\subsection{Energy loss by the jet.}
\label{ref:intro:energy-loss}
\par
\begin{figure}[!htb]
\includegraphics[width=0.48\linewidth]{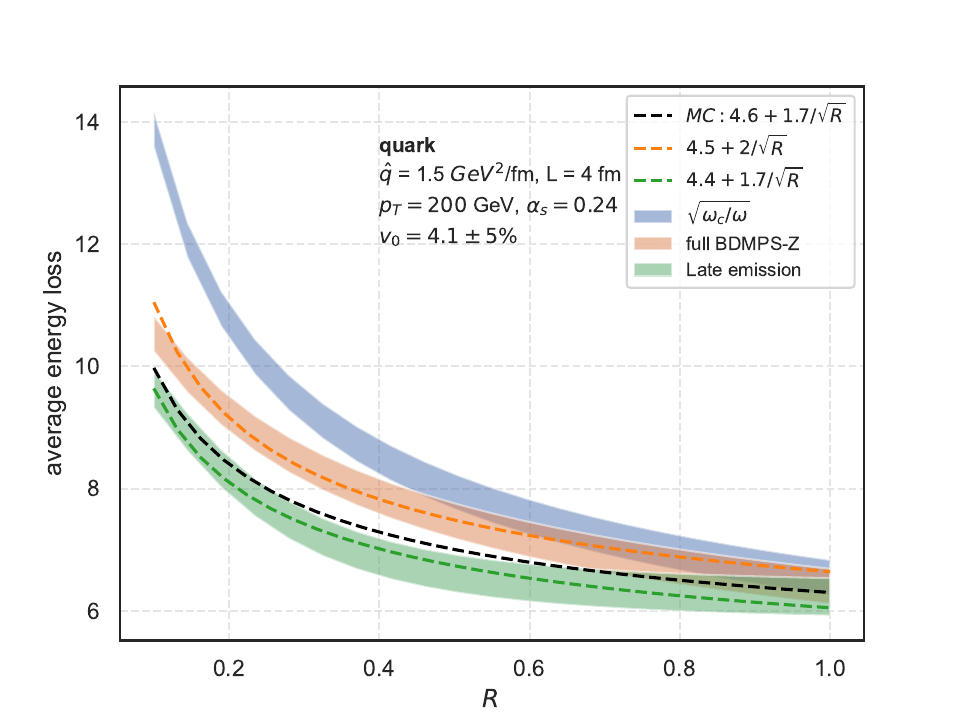}
\includegraphics[width=0.48\linewidth]{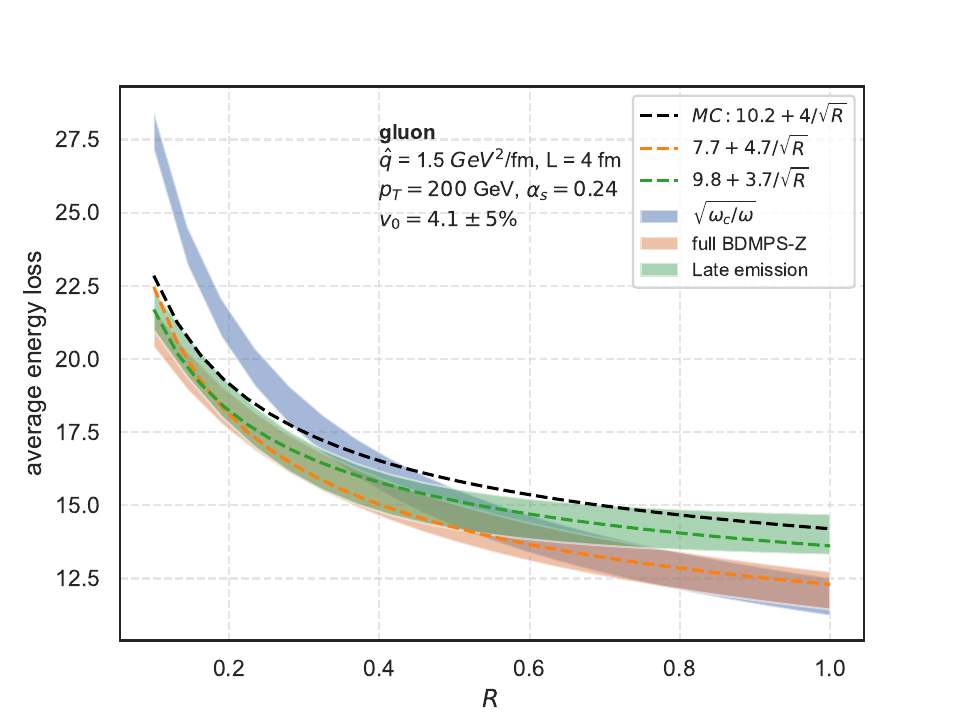}
\caption{The average energy loss by quark jet (left) and gluon jet (right), with $\alpha_s=0.24$, $p_T=200$ GeV and $L=4$ fm}
\label{fig:energy-loss}
\end{figure}

 It was argued in \cite{Blaizot:2013hx} that the jet energy loss consists of two parts: the soft turbulent cascade for multiple radiation with$\omega<\omega_{\rm br}$ and semi-hard radiation outside the dead cone angle that can be included in the single gluon approximation. The turbulent cascade gives the energy loss:
\begin{equation}\label{eq:MIE-eloss-flow}
\epsilon_{flow}\left(E\right)=E\left(1-e^{-v_{0}\frac{\omega_{br}}{E}}\right),
\end{equation}
which is independent of the jet radius. We also need to add the semi-hard emission energy loss with energy larger than $\omega_{br}$ which can be obtained by integrating the emission spectrum over $\omega$
\begin{equation}\label{eq:MIE-eloss-spec}
\epsilon_{spec}\left(R\right)=\int_{\omega_{br}}^{\bar{\omega}}d\omega\omega\frac{dI}{d\omega}e^{-\frac{v_{0}\omega_{br}}{E}},
\end{equation}
where $\bar{\omega}$ is the average frequency defined by $\bar{\omega}\equiv Q_{s}/R$, 
because $\omega= k_t/\theta$ and the emission angle is larger than the jet radius. We choose the infrared cut-off\footnote{more detailed calculation, i.e. Tab.\ref{tab:wbr-eloss} in the next section will show that, for larger dead-cone angle with L=4 fm, $\omega_{br}$ is lower and closer to $1$ GeV, and since this evaluation of $\omega_{br}$ is not very accurate, we think this is a good choice for the comparison. } at $\omega_{br}$. 
The total MIE energy loss is calculated as $\epsilon_{MIE}=\epsilon_{flow}+\epsilon_{spec}$, and the result depends on the relative energy scale of jet energy $E$, $\omega_{br}$ and $\omega_c$, which namely separate three physical regions: 1) high energy $E\gg\omega_{c}$, 2) intermediate $\omega_{c}\geq E\gg\omega_{br}$ and 3) low energy $E<\omega_{br}$. For $E\gg v_{0}\omega_{br}$, the energy loss becomes independent of energy $E$
\begin{equation}\label{eq:MIE-eloss}
\epsilon_{MIE}\left(R\right)=v_{0}\omega_{br}+\epsilon_{spec}\left(R\right),
\end{equation}
where $v_0=4.96$ for the low energy region and $v_{0}\approx3.9$ for $\alpha_s=0.24$ ~\cite{Fister:2014zxa} for high energy regions respectively. In Ref.~\cite{Caucal:2019uvr} the average energy loss is calculated via the MC fit, and the comparison between the MC fit and the numerical calculation of the above formula is shown in Fig. \ref{fig:energy-loss}. In our calculation, we have set the jet energy $E = 200$ GeV. Given that the value of $v_0$ is only known at the high and low energy limit, we chose $v_{0}=4.1\pm5\%$ so that the lower bound coincides with the high energy limit, and in order to compare with the MC fit (black dash line), we also made a line fit for the energy loss evaluated from the full BDMPS-Z spectrum with $v_{0}=4.1$ (orange dash line), we found it agrees with the MC fit very well.

For the full shower, we need to include the multiplicity shown in the previous section, and the total energy loss read as
\begin{equation}\label{eq:eloss-full}
\epsilon_{jet}(E,R)=\epsilon_{MIE}^{\rm q}(E,R)\\
+\int_{\theta_{cut}}^{R}\int_{\omega_{0}(\theta)}^{E}{d\omega}{d\theta} \frac{dN_Q}{d\omega d\theta},
\end{equation}
where $ \frac{dN_Q}{d\omega d\theta}$ is given by Eq. \ref{upt}
where, according to our physics picture, in the second term we convoluted the VLE multiplicity with the gluon jet energy $\epsilon_{MIE}^{\rm g}(\omega,R)$ loss from the early emissions. The full shower energy loss evaluated with different approaches is shown in Fig.\ref{fig:eloss-full}, our result is in fair agreement with the MC fit shown in Fig.\ref{fig:eloss-full}.

\subsection{Numerical study: energy loss for heavy flavour jets}
 
 The energy loss is calculated from Eq.\ref{eq:MIE-eloss} for the high energy jet with $p_T=200$ GeV, similar to the discussion before, the numerical result with $v_{0}=4.1\pm5\%$. From those two tables we immediately see  that $E\gg v_{0}\omega_{br}$ is true for all dead-cone configurations, because $\theta_{0}=\left(0.05,0.1,0.2\right)$ is corresponding to $p_{T}=\left(100,50,25\right)$ GeV for b quark.

\begin{table}[!h]
\begin{center}
\begin{tabular}{ |p{0.8cm}||p{1.cm}|p{1.cm}|p{1.cm}|p{1.cm}|p{1.cm} |p{1.cm}|p{1.cm}|p{1.cm}|p{1.cm}|p{1.cm}|  }
 \hline
 & \multicolumn{4}{|c|}{$L=4$ fm}& \multicolumn{4}{|c|}{$L=6$ fm} \\
 \hline
 $\theta_0$& 0    & 0.05&  0.1& 0.2& 0 &0.05&  0.1 & 0.2\\
 \hline
$\epsilon^{LE}_{q}$&   10.2  & 8.4   &6.4& 3.9 & 22.1 & 14.8& 10.5 & 6.2\\
$\epsilon^{LE}_{jet}$&  19.6  & 11.1   &7.7& 4.7 & 35.2 & 18.3& 12 & 7.2\\
 \hline
\end{tabular}
\caption{\label{tab:wbr-eloss}The average energy loss for heavy flavour quark jet with $\alpha_{s}=0.24$, $p_T=200$ GeV and R=0.4}
\end{center}
\end{table}

\begin{figure}[!htb]
\begin{center}
\includegraphics[width=0.48\linewidth]{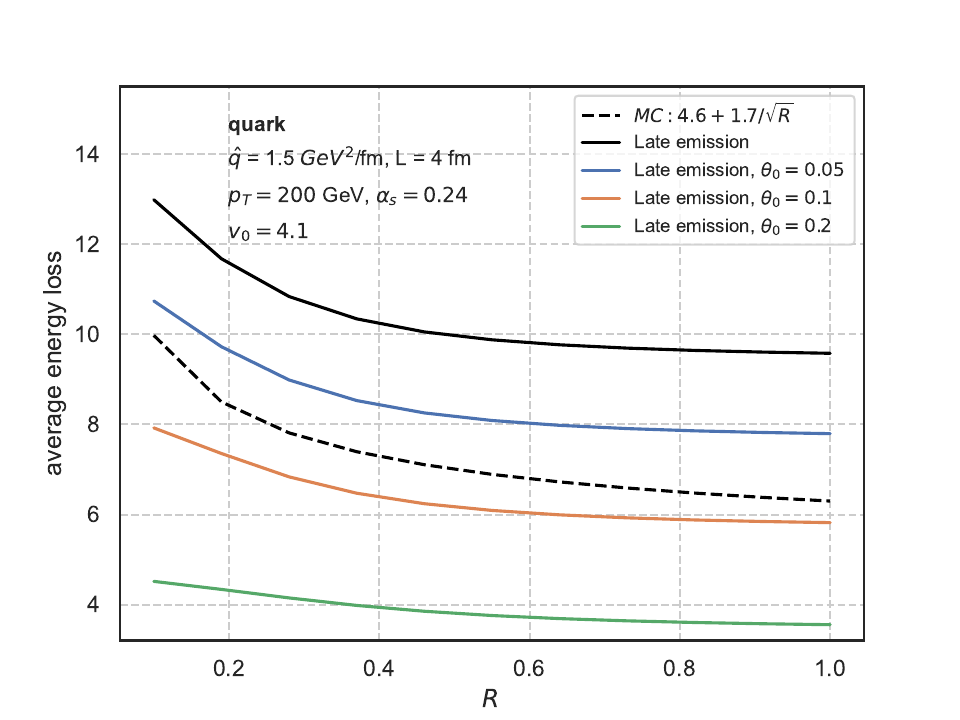}
\includegraphics[width=0.48\linewidth]{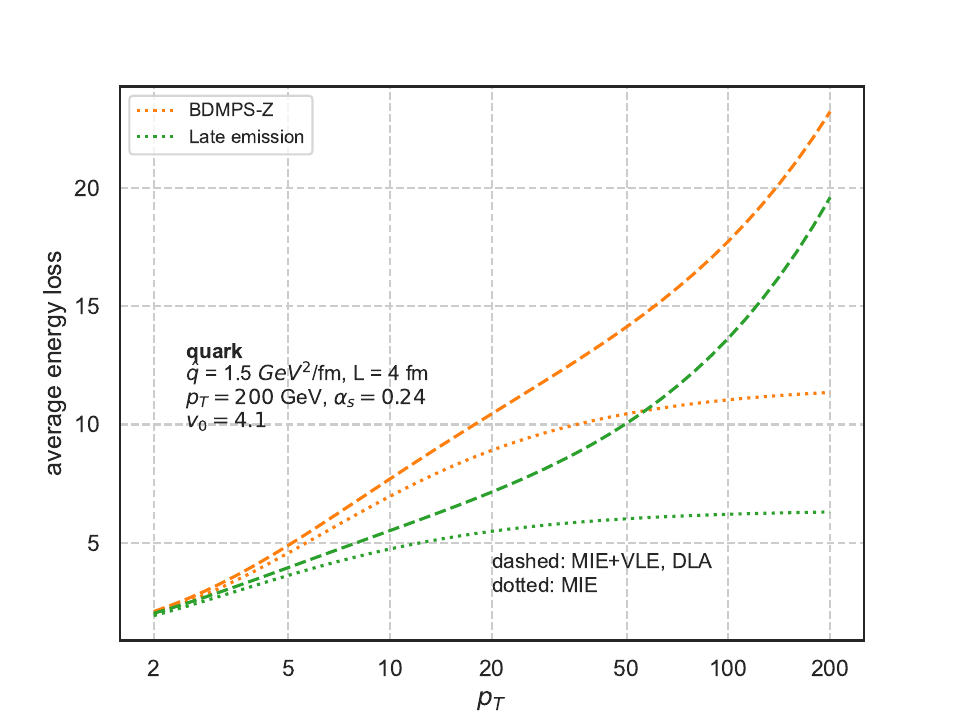}
\caption{The average energy loss as a function of the jet radius $R$, only MIE included (left) and the jet energy $p_T$ (right)}
\label{fig:eloss-full}
\end{center}
\end{figure}

  
The comparison for the energy loss $\epsilon_{MIE}$ and $\epsilon_{jet}$ as a function of $p_T$ for the late emission approaches and BDMPS-Z, with  DLA  multiplicity, is shown in Fig. \ref{fig:eloss-full}. %
and both $\omega_{br}$ and $\epsilon_q$ for late emission approximation with large dead-angle for $L=$ 4 and 6 fm is shown in Tab.\ref{tab:wbr-eloss}. %
For our numerical results, we see that the energy loss from the jet with DLA multiplicity is qualitatively agreed with the MC fit from Ref.~~\cite{Caucal:2019uvr}. For higher precision, one might need the full next-to-leading logarithms (NLL) calculation. We leave a more accurate study of the energy loss for the future. 

\section{The calculation of the $z_g$ distribution in dense medium}
\label{sec:z_g}
\par In this section, we explain our approach for the calculation of the $z_g$ distribution in the dense QCD  medium. In particular, we show that for the massless case, we obtain the $z_g$  distributions in agreement with \cite{Caucal:2019uvr} while using the late emission approximation and applying the Sudakov safety technique only for the VLE part of the shower. We shall also see that in massive case, there is no need to use the Sudakov safety technique, the distribution of $z_g$ remains the same.

\subsection{Definition of $z_g$ in the vacuum}

Let us review some basic ideas connected with $z_g$ in pp collisions (vacuum case). The $z_g$ is a known jet substructure observable that can test the $1\rightarrow2$ QCD splitting function. This observable is based on soft drop grooming~\cite{Larkoski:2014wba}, which means sequentially declustering the jet constituents with the Cambridge-Aachen (C/A) algorithm until the subjets satisfy the SD condition defined as
\begin{equation}
    z_{g}=\frac{min\left(p_{1},p_{2}\right)}{p_{1}+p_{2}}>z_{cut}\theta_{g}^{\beta},\,\theta_{g}=\frac{\Delta R_{12}}{R}
\end{equation}

This process is designed to remove the soft large-angle radiation from a jet until the algorithm finds the two-prong substructure, where $p_1$ and $p_2$ are the transverse momenta of the corresponding two subjets,  $\Delta R_{12}$ is their angular separation and R is the jet radius. For $\beta=0$ case, the soft drop grooming coincides with the modified mass drop tagger. 
\par For heavy ion collisions, to account for the detector resolution effects, it is common to set a cut-off, $\theta_{cut}=0.1$ from the minimal resolution angle $\Delta R_{12}>0.1$. Hence, naturally, we will adopt this cut-off in our calculation.

Originally, the $z_g$ distribution in vacuum was calculated with the Sudakov safe technique~\cite{Larkoski:2015lea}, because $z_g$ is not infrared collinear (IRC) safe observable for $\beta\geq0$ due to the collinear singularities. Hence, $z_g$ is calculated with its safe companion observable $\theta_g$ as
\begin{equation}\label{eq:Sudakov-safe}
f\left(z_{g}\right)=\int_{0}^{R}d\theta_{g}\Delta\left(R,\theta_{g}\right)P\left(z_{g,}\theta_{g}\right)\Theta\left(z-z_{cut}\theta_{g}^{\beta}\right),
\end{equation}
where $P\left(z_{g,}\theta_{g}\right)$ is the joint differential probability that is given by
\begin{equation}
d^{2}P=\frac{2\alpha_{s}C_{R}}{\pi}\bar{P}\left(z\right)dz\frac{d\theta}{\theta}\equiv P\left(z,\theta\right)dzd\theta.
\end{equation}
where $\bar{P}\left(z\right)$ is the symmetrised Altarelli-Parisi splitting function and $\Delta\left(R,\theta_{g}\right)$ is the Sudakov form factor. The latter is equal to
\begin{equation}\label{eq:Sudakov-safe2}
log\Delta\left(R,\theta_{g}\right)=-\int_{\theta_{g}}^{R}d\theta\int_{z_{cut}}^{1/2}dzP\left(z,\theta\right)\Theta\left(z-z_{cut}\theta_{g}^{\beta}\right),
\end{equation}
and is the probability of no emission between angle $\theta_g$ and R. 

When we include the cut-off, $\theta_g>\theta_{cut}$, Eq.\ref{eq:Sudakov-safe} can be re-written as 
\begin{equation}\label{eq:Sudakov-safe3}
f\left(z_{g}\right)=\frac{1}{1-\Delta\left(R,\theta_{cut}\right)}\int_{\theta_{cut}}^{R}d\theta_{g}\Delta\left(R,\theta_{g}\right)P\left(z_{g,}\theta_{g}\right).
\end{equation}
Then Eq.\ref{eq:Sudakov-safe3} coincides with Eq.\ref{eq:Sudakov-safe} when $\theta_{cut}\rightarrow0$, since in the collinear limit, the Sudakov form factor will reduce to 0. 

It is worth pointing out that for $\beta=0$ case at fixed coupling limit, the $z_g$ distribution can be calculated easily as
\begin{equation}
    f\left(z_{g}\right)=\frac{P\left(z_{g}\right)}{\int_{z_{cut}}^{1/2}dz_{g}P\left(z_{g}\right)},
    \label{mur}
\end{equation}
The Eq. \ref{mur} provides clear evidence that the $z_g$ distribution can be used to probe the splitting function. However, despite the fact that Eq.\ref{eq:Sudakov-safe} is finite, its perturbative expansion at any finite order is divergent due to $P\left(z,\theta\right)$ is collinear divergent, hence this observable is not collinear safe, and needs applying the all order resummation to absorb the collinear singularities with the Sudakov form factor, i.e. Sudakov safe.

\subsection{The in medium $z_g$ distribution.}
In the presence of a medium, as discussed in Sec.\ref{sec:intro}-\ref{sec:theory}, the full shower factorises into two parts: MIE and VLE which are created at different times.~\cite{Caucal:2018dla, Caucal:2019uvr}. The phase space boundary between MIE and VLE, as we briefly discussed in the introduction, is
\begin{equation}
t_{f}^{vac}\left(\omega,\theta\right)\equiv\frac{1}{\omega\left(\theta^{2}+\theta_{0}^{2}\right)}\ll t_f\left(\omega\right).
\end{equation}

\begin{figure}[htb!]
\begin{center}
\includegraphics[width=0.48\linewidth]{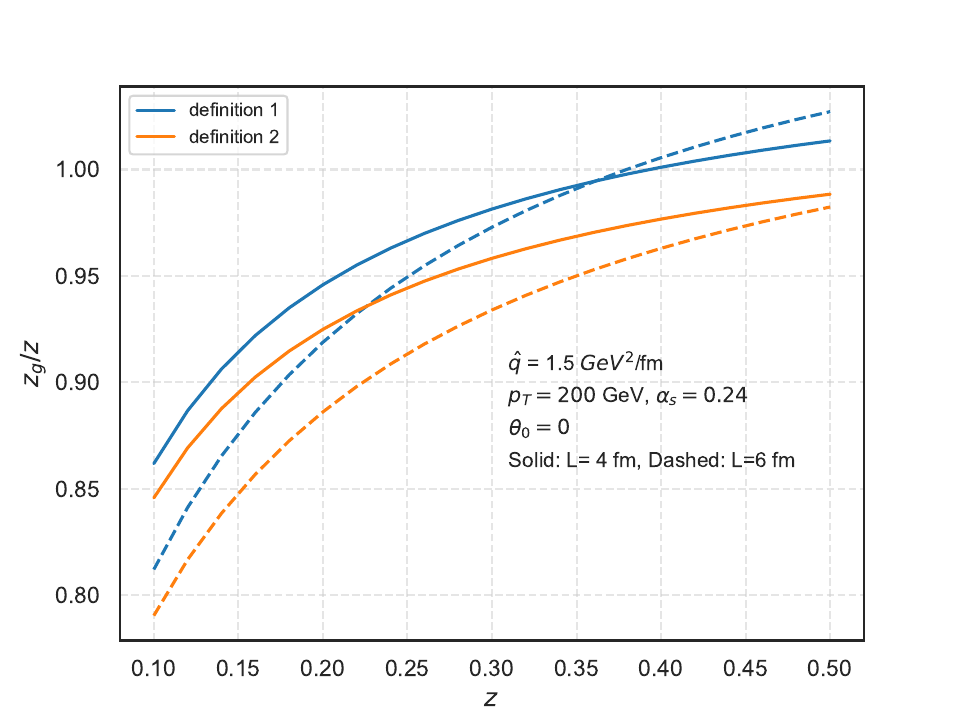}
\includegraphics[width=0.48\linewidth]{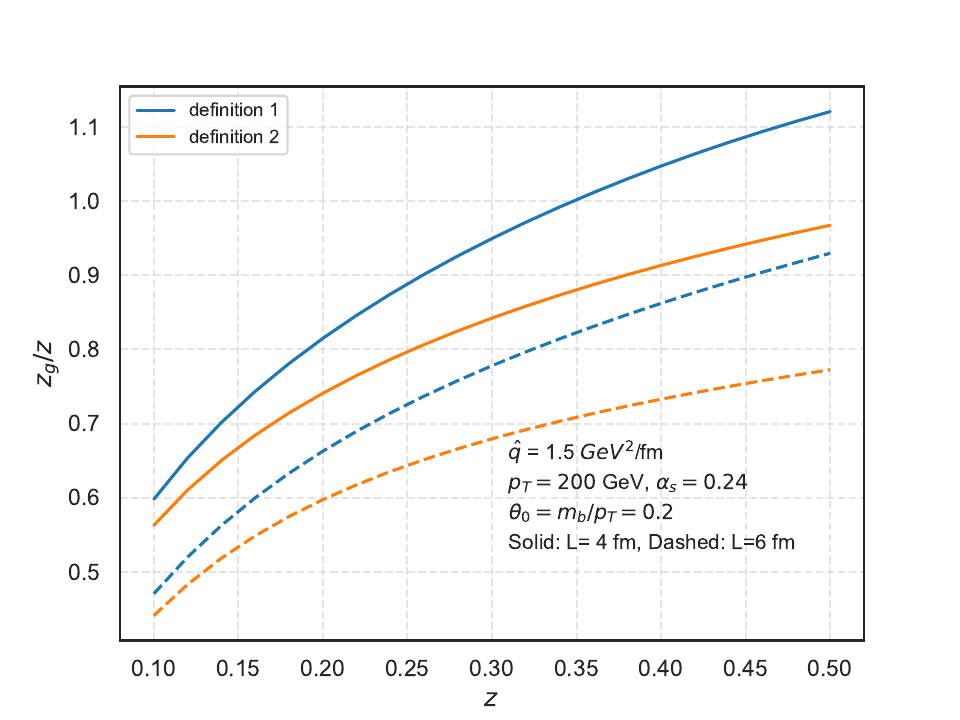}
\caption{The  $z_g/z$ value for two different definitions of $z_g$ shown in Eq.\ref{eqn:zg} (definition 1) and Eq.\ref{eqn:zg2} (definition 2) with L=4 and 6 fm
\label{fig:diff-zg}}
\end{center}
\end{figure}

For MIE part of $z_g$ distribution, it is straightforward to extend Eq.\ref{eq:Sudakov-safe} and Eq.\ref{eq:Sudakov-safe3}, to the medium case by replacing the joint differential probability for vacuum case by MIE splitting kernel
\begin{equation}
    P\left(z,\theta\right)=\frac{dI^{\rm MIE}}{dzd\theta},
\end{equation}
as discussed in Sec.\ref{sec:theory}. Details about different approaches for the MIE splitting kernel can be found in Section \ref{sec:joint-distribution}. 
\par Note that,  when $\theta_g > \theta_c$, where $\theta_c$ is the coherence angle, the two daughter partons act as independent sources of MIEs, i.e. both partons will lose energy, so we can write their observed energies as $p_{Ti}=\omega_i-\mathcal{E}_{i}\left(\omega,\theta_{g}\right). $

In the literature ~\cite{Caucal:2019uvr}, there are two different ways to define the measured in-medium $z_g$ from the physical splitting fraction $z$ with energy loss, as\footnote{In this article, we assume $p_{T1}<p_{T2}$}


\begin{equation}\label{eqn:zg}
z_{g}\equiv\frac{p_{T1}}{p_{T1}+p_{T2}}=\frac{zp_{T}-\mathcal{E}_{g}\left(zp_{T},\theta_{g}\right)}{p_{T}-\mathcal{E}_{i}\left(p_{T},\theta_{g}\right)}\equiv Z_{g}\left(z,\theta_{g}\right),
\end{equation} 
where we denoted the energy loss for $p_{T1}$ as $\mathcal{E}_{g}\left(zp_{T},\theta_{g}\right)$, and $\mathcal{E}_{i}\left(p_{T},R\right)$ in the denominator is the energy loss for the initial parton. On the other hand, in medium $z_g$ can also be defined as
\begin{equation}\label{eqn:zg2}
z_{g}\equiv\frac{zp_{T}-\mathcal{E}_{g}\left(zp_{T},\theta_{g}\right)}{p_{T}-\mathcal{E}_{g}\left(zp_{T},\theta_{g}\right)-\mathcal{E}_{i}\left(\left(1-z\right)p_{T},\theta_{g}\right)},
\end{equation} 

In practice, one can replace $\theta_g$ with R when evaluating energy loss, due to the SD process the soft emissions between $\theta_g$ and R can be neglected. 

The only difference between these two definitions is in the denominator, i.e. energy loss from the initial parton or the sum of the energy loss from the two daughter partons. However, the difference is small as shown in Fig.\ref{fig:diff-zg}, where we have used the energy loss $\epsilon_{MIE}\left(E,R\right)$, calculated in section 3 and shown in Fig.\ref{fig:energy-loss}-\ref{fig:eloss-full} , 
Since a lower energy jet corresponds to a larger dead-cone angle, which has smaller energy loss as discussed in Sec.\ref{ref:intro:energy-loss}, the difference between these two different definitions should be small. Note that the energy loss evaluated from Eq.\ref{eq:MIE-eloss} is not very accurate, especially for large dead-cone, i.e. low energy jet. Therefore, in this article, we will use Eq.\ref{eqn:zg} for the both MIE and VLE calculation.

\begin{figure}[htb!]
\begin{center}
\includegraphics[width=0.48\linewidth]{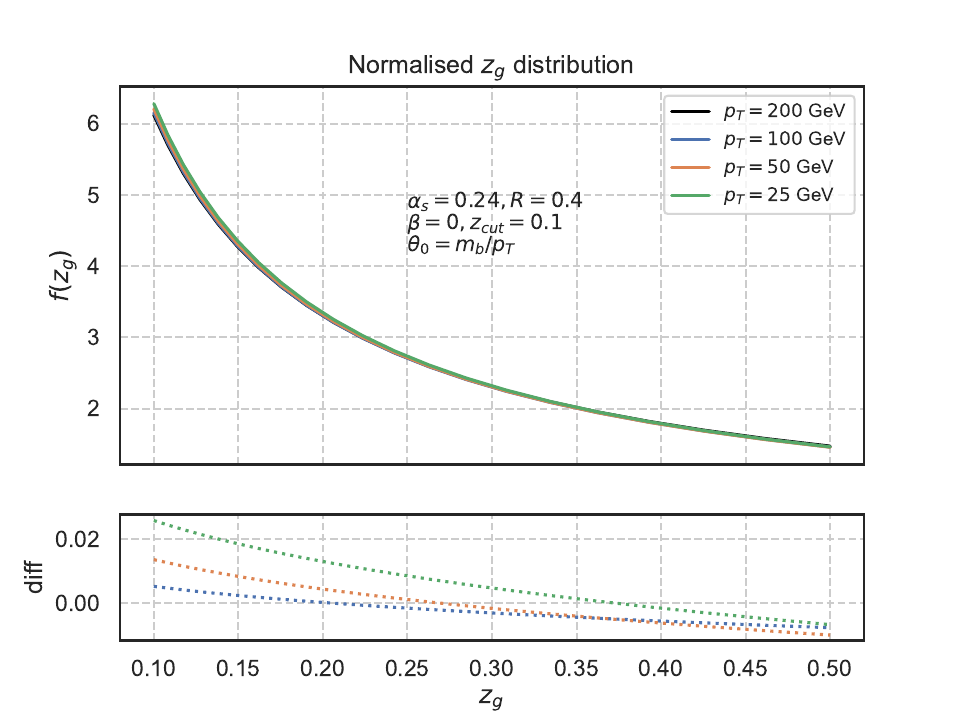}
\includegraphics[width=0.48\linewidth]{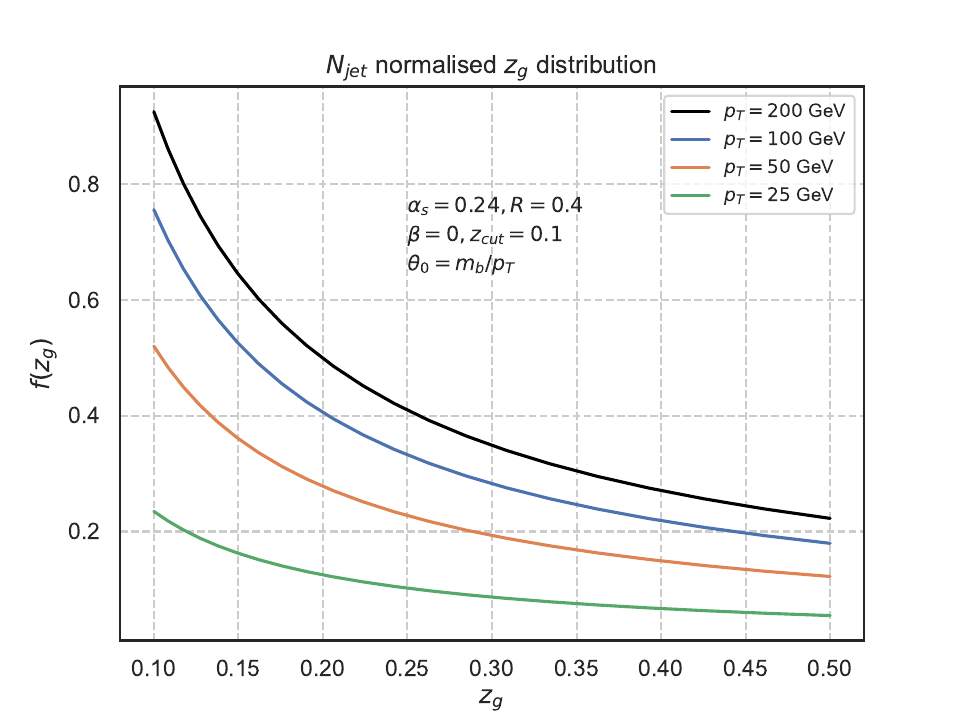}
\caption{The $z_g$ distributions in vacuum for quark jets for two different normalisation scheme}
\label{fig:zg-Vac}
\end{center}
\end{figure}
Note that in the later calculations we shall use the inverse of Eq.\ref{eqn:zg}:
\begin{equation}
Z(z_{g})=z_{g}-\frac{z_{g}\mathcal{E}_{i}-\mathcal{E}_{g}}{p_{T}}
\end{equation}
 It is straightforward to extend the Eq.\ref{eq:Sudakov-safe3} for vacuum  $z_g$ distribution to the QCD medium
 by redefining the joint differential probability with energy loss, based on Eq.\ref{eqn:zg}, as 
\begin{equation}\label{eqn:zg-redef}
f\left(z_{g}\right)=N\int_{\theta_{cut}}^{R}d\theta_{g}\Delta\left(R,\theta_{g}\right)\bar{P}\left(z_{g,}\theta_{g}\right)\Theta\left(z_{g}-z_{cut}\right),
\end{equation}
where
\begin{equation}\label{zig}
\bar{P}\left(z_{g,}\theta_{g}\right)=\int d\omega P\left(\omega,\theta_{g}\right)\delta\left(z_{g}-Z_{g}\left(\omega,\theta_{g}\right)\right),
\end{equation}
Here N is the normalisation factor $(1-\Delta\left(R,\theta_{cut}\right))^{-1}$. One can also obtain the $N_{jet}$ normalised distribution, by setting 
$N=1$.
\par The delta function in Eq. \ref{zig}
can be easily taken leading to 
\begin{equation}
 \bar{P}(z_{g},\theta_{g})=J(z_{g})P(Z(z_{g}),\theta_{g}) 
\end{equation}
where the Jacobian is equal to 
\begin{equation}\label{eq:jacobian}
    J=1-\mathcal{E}_i/p_T
\end{equation}
\subsection{Calculation for the $z_g$ distribution: vacuum and VLE-only}
Let's start with the SD condition triggered by the vacuum. For heavy flavour jets, we shall use the dead cone approximation: 
\begin{equation}
P^{vac}\left(z,\theta^2\right)\rightarrow\frac{P^{vac}\left(z,\theta^2\right)}{\left(1+\frac{\theta_{0}^{2}}{\theta^{2}}\right)^{2}}.
\end{equation}

The numerical results for normalised $z_g$ distribution, i.e. normalised to 1, and $N_{jet}$ normalised $z_g$ distribution are shown in Fig.\ref{fig:zg-Vac} Interestingly, we see that the normalised $z_g$ distribution is not sensitive to $\theta_0$ and $p_T$, the difference between massless $p_T=200$ GeV and b-jets with $p_T=25$ GeV is less than $2\%$ (shown in the lower panel of the left plot of Fig.\ref{fig:zg-Vac}), this behaviour can be explained by a simple fixed order calculation with fixed coupling constant at LL accuracy, we have
\begin{align}
f\left(z_{g}\right)&=\int_{0}^{R}d\theta_{g}p^{vac}\left(z_{g},\theta_{g}\right)\Theta\left(z_{g}-z_{cut}\right)\\\nonumber
&=\frac{\alpha_{s}C_{R}}{\pi}\frac{1}{z_{g}}log\frac{R^{2}}{\theta_{0}^{2}},
\end{align}
for the normalised $z_g$ distribution, the dead cone dependence cancels itself, which suggests Eq.\ref{mur}, is still valid even for heavy flavour jets. More details about the resummed distributions for heavy flavour $z_g$ can be found in Ref.~\cite{Caletti:2023spr}. 

\begin{figure}[htb!]
\begin{center}
\includegraphics[width=0.48\linewidth]{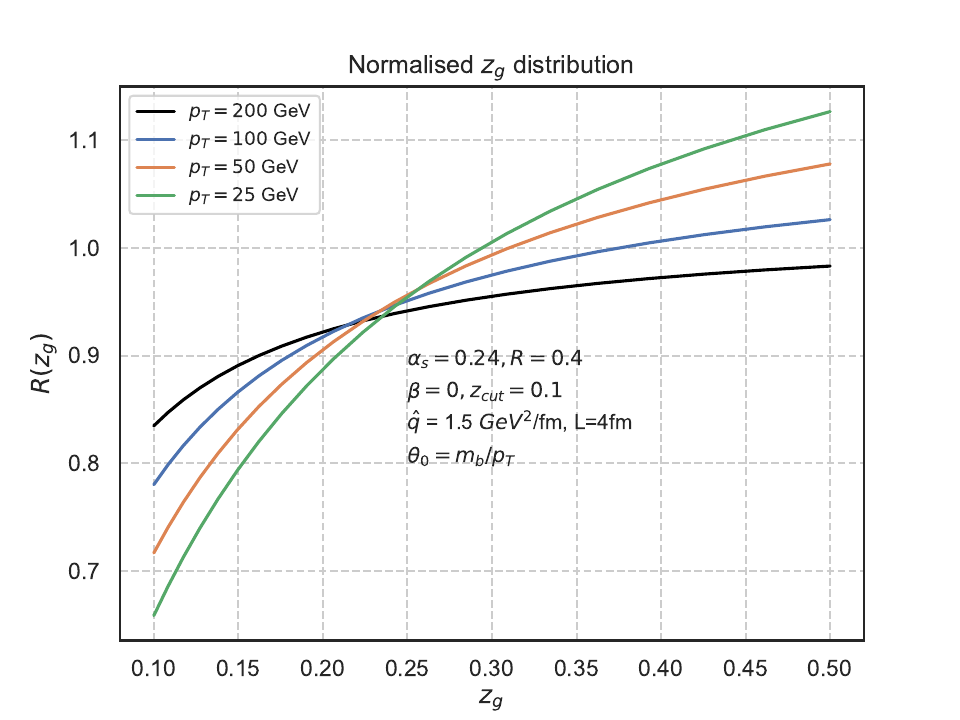}
\includegraphics[width=0.48\linewidth]{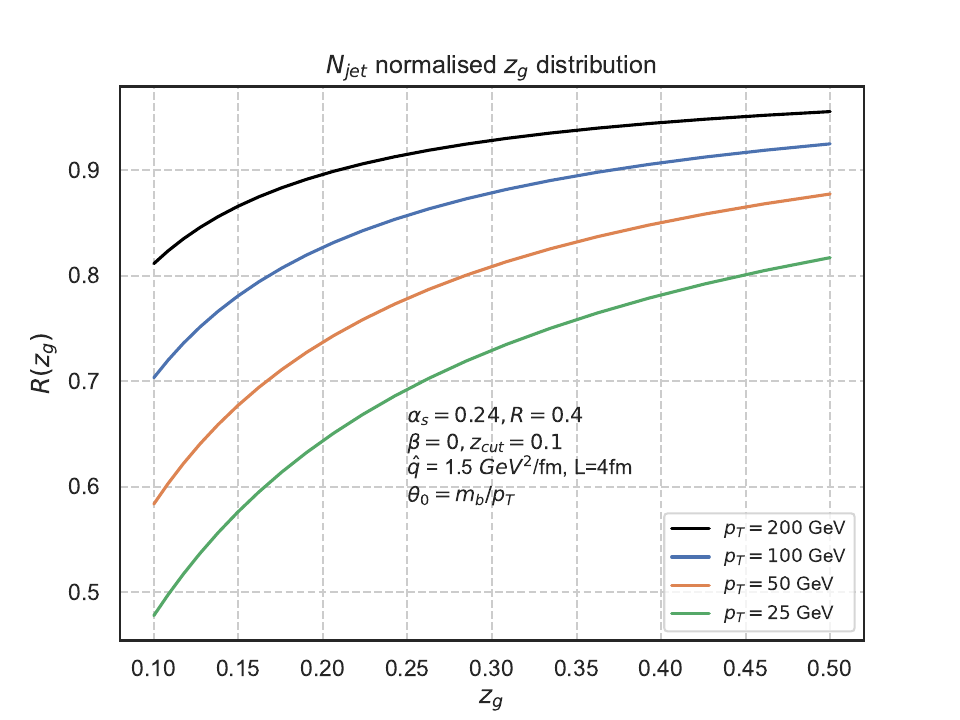}
\caption{R ratio of VLE for quark jets with two different normalisation schemes}
\label{fig:Rzg-VLE}
\end{center}
\end{figure}

To study the QCD medium effects in the $z_g$ distribution, i.e. the medium/vacuum ratio similar to the nuclear modification ratio called the R ratio, in this section we start by studying the VLE part of the shower. The results for VLE R ratio are shown in Fig.\ref{fig:Rzg-VLE}. We see the R ratio increases with the increase of $z_g$, and the $N_{jet}$ normalised distributions (right) are always smaller than one, as expected. Since at the fixed coupling constant limit, $z_g$ distribution is proportional to the splitting function, this behaviour can be analysed from the equation:
\begin{equation}
    R\left(z_{g}\right)\sim\left(1-\frac{\mathcal{E}_{i}}{p_{T}}\right)\frac{P\left(Z\left(z_{g}\right)\right)}{P\left(z_{g}\right)}
    \thickapprox\left(1-\frac{\mathcal{E}_{i}}{p_{T}}\right)\frac{z_{g}}{Z\left(z_{g}\right)},
\end{equation}
where the overall factor $1-\frac{\mathcal{E}_{i}}{p_{T}}$ comes from the Jacobian of the delta function, and $z_{g}/Z\left(z_{g}\right)$ is smaller than one as shown in Fig.\ref{fig:diff-zg}. %
Moreover, we see there is suppression for larger dead cone angle at $z\rightarrow z_{cut}$, which indicates $z_g$ distribution could be used to probe the dead cone angle. More details will be discussed later in this section for the full shower.

\subsection{Collinear finite: calculations with and without the Sudakov safe technique.}
As discussed in Sec.\ref{sec:theory}, the MIE part of the splitting kernel is collinear finite, so no Sudakov form factor is needed to be associated with the calculation. For the VLE part, there are no collinear singularities for the 
massive case also, since they are regulated by the dead cone angle,
Consequently, the convolution with the Sudakov form factor for VLE part 
is needed only if we are concerned about the massless case or want to get the smooth 
massless quark limit from heavy flavour jets.

\par In this section, we double-check our understanding.  We will perform some numerical studies and compare the results with the Sudakov safety technique calculation. 

\begin{figure}[htb!]
\begin{center}
\includegraphics[width=0.48\linewidth]{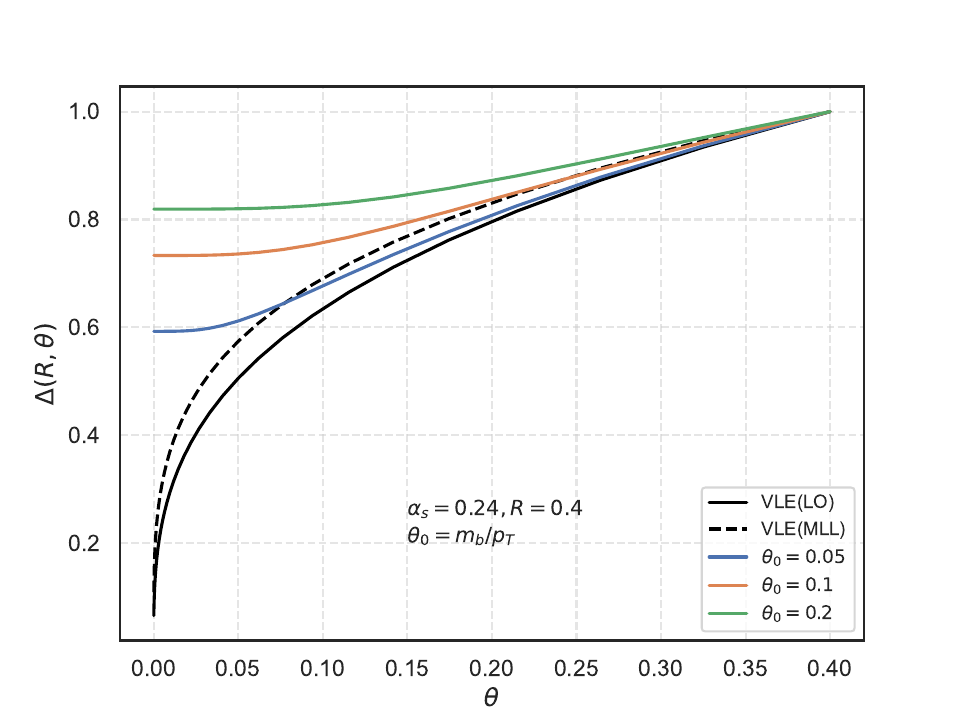}
\includegraphics[width=0.48\linewidth]{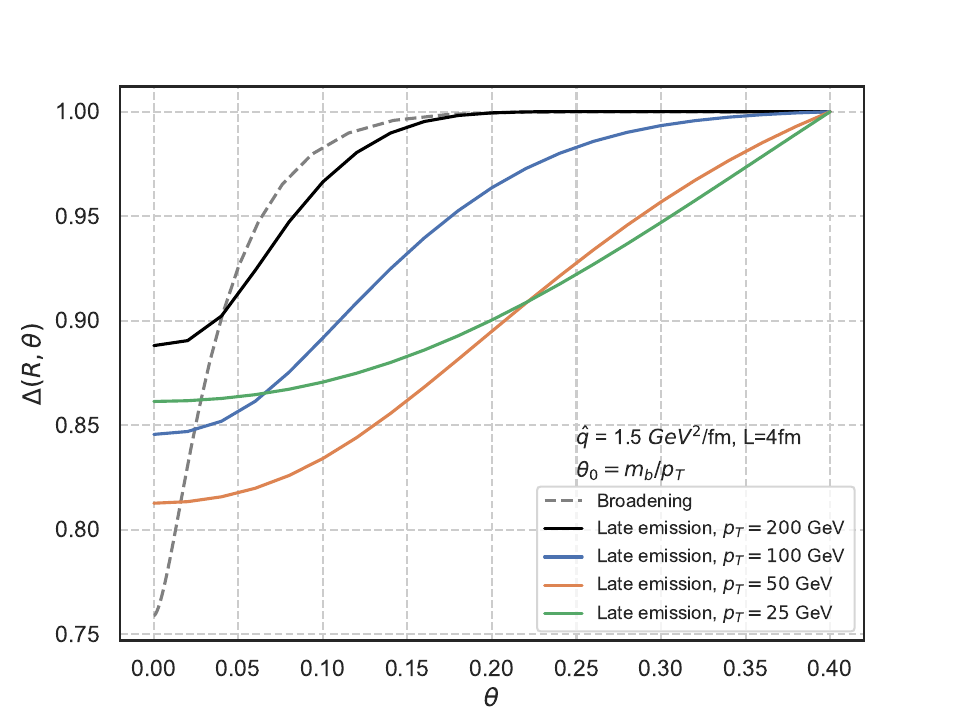}
\caption{The  Sudakov form factor for VLE with $p_T=200$ GeV (left) and MIE with $\theta_0=m_b/p_T$ (right), R=0.4}
\label{fig:sudakov-factors}
\end{center}
\end{figure}

\par We start with the Sudakov form factor $\Delta\left(\theta,R\right)$,  
which describes the possibility that no gluon has emitted between angle $\theta$ and the jet radius $R$. %
We compared the Sudakov form factor for VLE case for different masses using dead cone approximation, and the Sudakov form factor for MIE. We calculated the latter using both broadening approach and late emission approximation for massless case and late emission approximation for nonzero mass. %
The results are depicted in Fig.\ref{fig:sudakov-factors}.  As can be expected, the Sudakov form factor for heavy flavour jets is close to one. %
The  MIE Sudakov form factor stays close to one even in the massless limit.
\par
\begin{figure}[!htb]
\begin{center}
\includegraphics[width=0.48\linewidth]{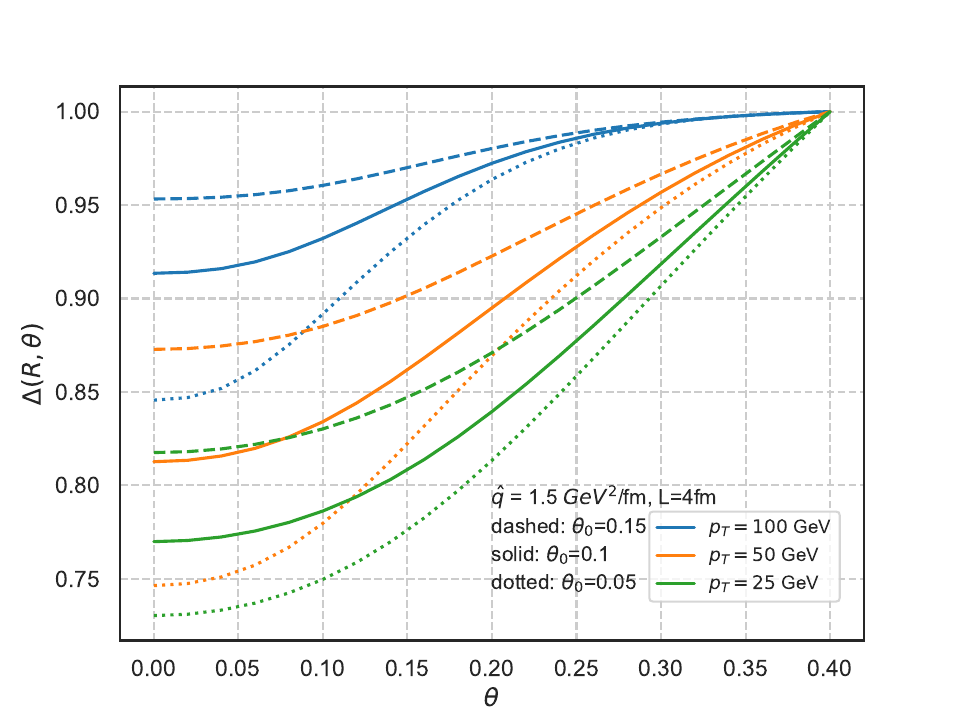}
\caption{Sudakov form factor for MIE, with fixed jet energy $p_T$, and varying $\theta_{0}$ by $\pm50\%$ }
\label{fig:sudakov-factors-test}
\end{center}
\end{figure}

  Note that the Sudakov form factor decreases when the jet energy decreases, and increases when dead cone angle increases, which is why on the right side of Fig.\ref{fig:sudakov-factors}, the $\theta_0=0.2$ curve is higher than $\theta_0=0.1$. In order to see this behaviour more clearly, we varied the dead-cone angle as $\theta_{0}=\left(0.05,0.1,0.15\right)$, with the jet energy $p_T=m_b/\theta_0$.
The result is shown in Fig.\ref{fig:sudakov-factors-test} for jet energies $p_T=$ 100, 50 and 25 GeV.

\par It is interesting to see the difference between the in-medium $z_g$ distributions calculated with and without the Sudakov safety technique.  Let us  calculate the $z_g$
distribution and its R ratio (i.e. ratio of medium and vacuum distributions) for both massless quark jets and heavy flavour jets. We shall see that in both cases there is a very small numerical difference between Sudakov safe calculation and calculation without Sudakov form factor.
Recall that for the broadening  approach 
the joint differential probability has the form
\begin{equation}\label{eq:p-br}
d^{2}P_{med}=\frac{\alpha_{s}C_{R}}{\pi}\sqrt{\frac{2\omega_{c}}{\omega^{3}}}P_{br}\left(\theta,\omega\right)\Theta\left(\omega_{c}-\omega\right)d\omega\theta\equiv P^{med}\left(\theta,\omega\right)d\omega d\theta,
\end{equation} 
with the broadening approach probability
\begin{equation}
P_{br}\left(\theta,\omega\right)=\frac{2\theta\omega^{2}}{qL}\Gamma\left(0,\frac{\omega^{2}\theta^{2}}{qL}\right)
\end{equation} 
where $\Gamma\left(0,x\right)$ is the incomplete Gamma function.

\begin{figure}[!htb]
\begin{center}
\includegraphics[width=0.48\linewidth]{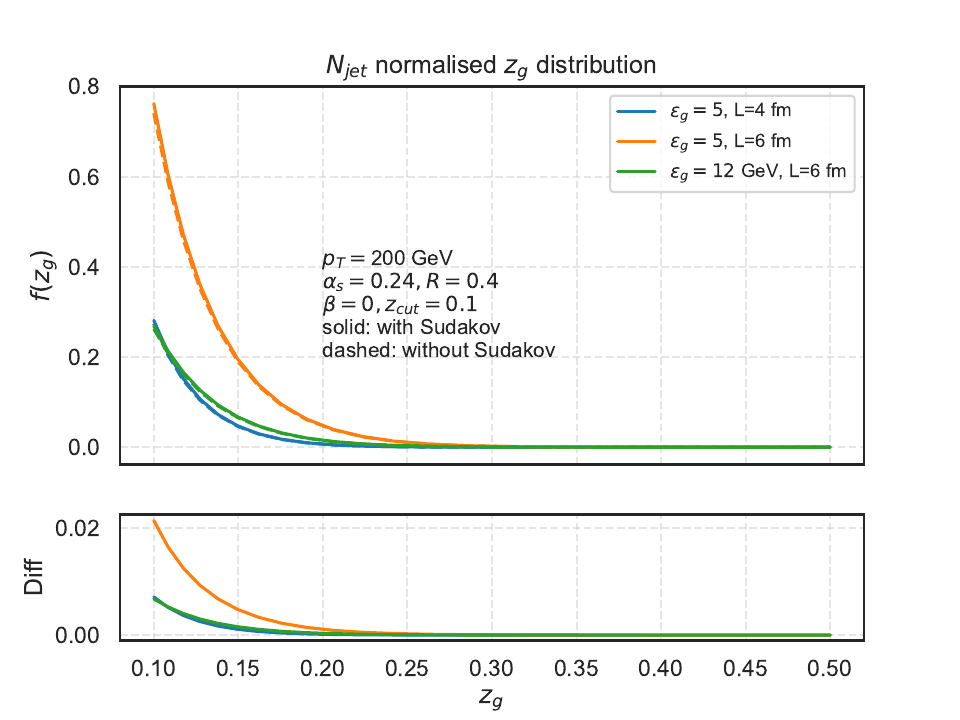}
\includegraphics[width=0.48\linewidth]{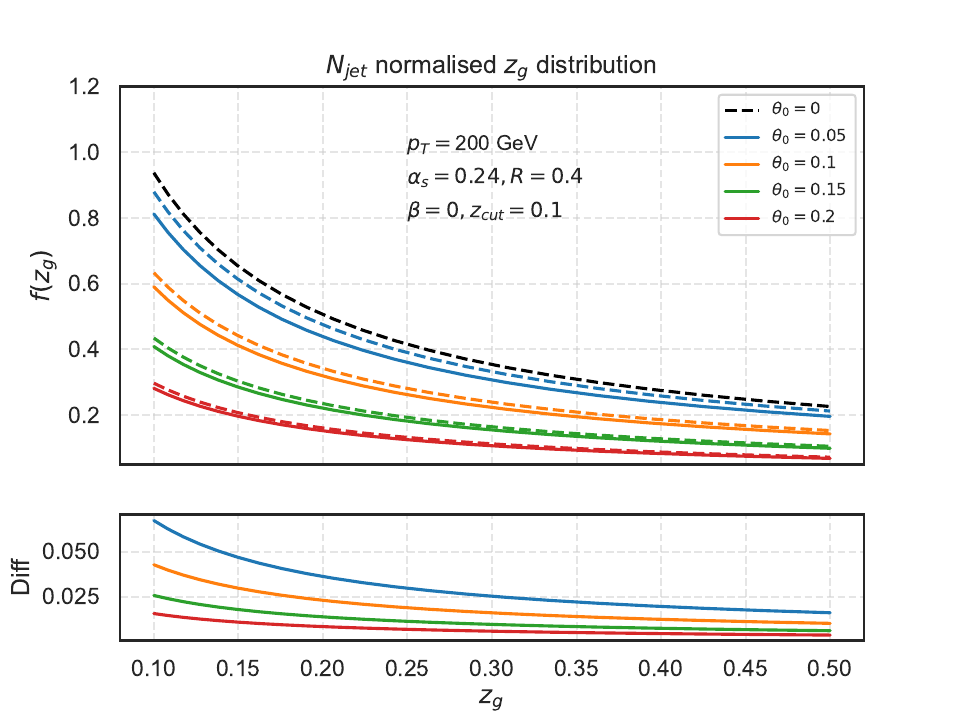}
\caption{$f(z_g)$ for MIE-only (left) with (dash line) and without (solid line) Sudakov form factor and VLE-only (right) for quark initialled jet}
\label{fig:soyez-zg}
\end{center}
\end{figure}

\begin{figure}[!htb]
\begin{center}
\includegraphics[width=0.48\linewidth]{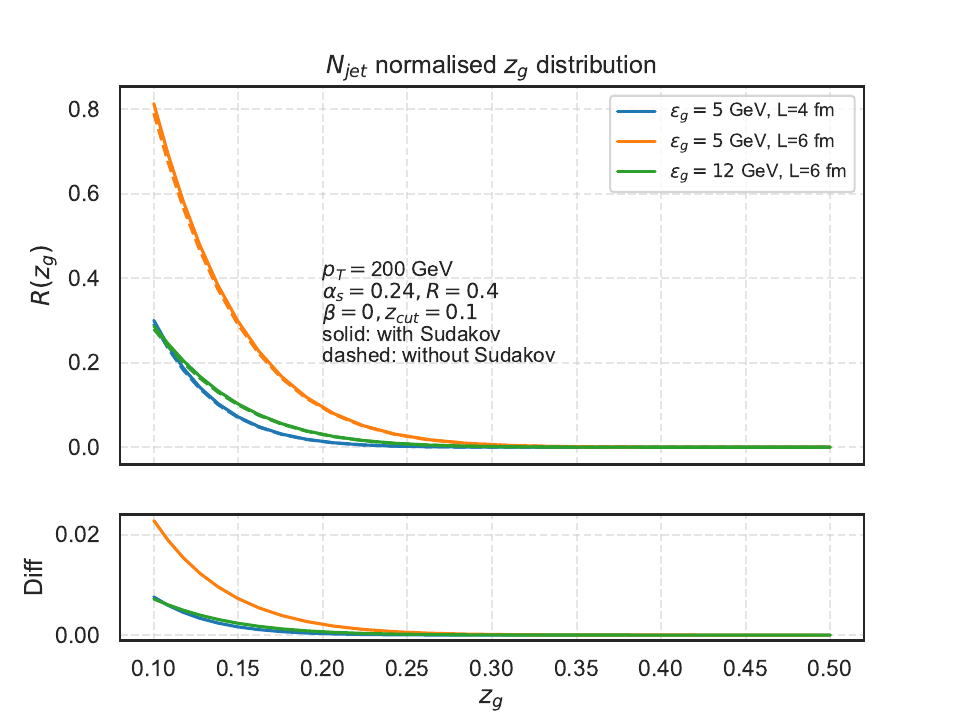}
\includegraphics[width=0.48\linewidth]{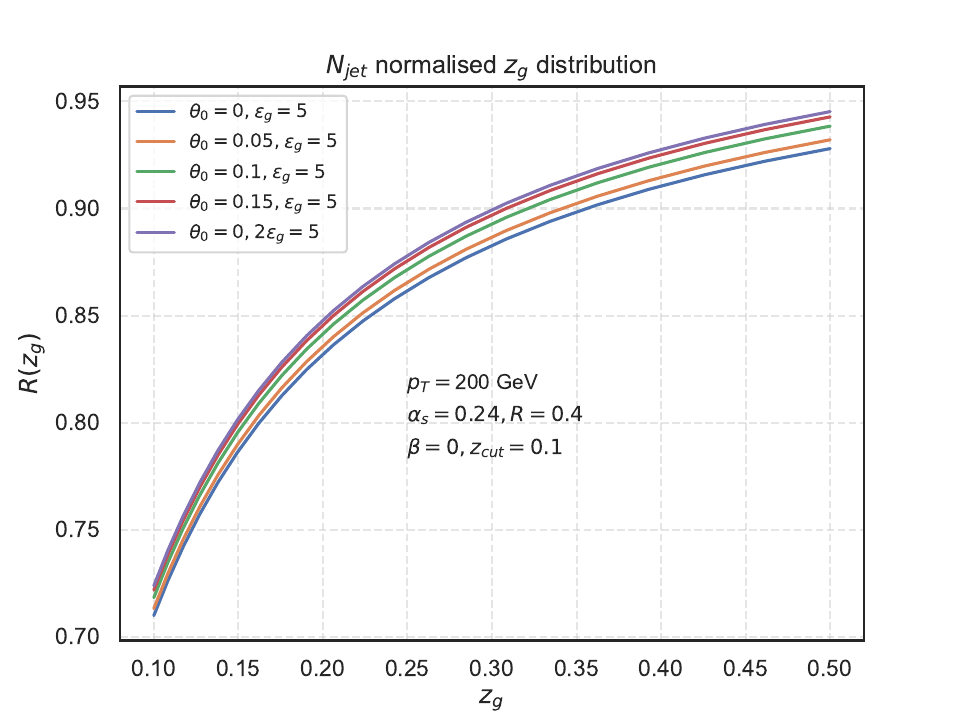}
\caption{R ratio for MIE-only (left) with (dash line) and without (solid line) Sudakov form factor and VLE-only (right) for quark initiated jet}
\label{fig:soyez-R}
\end{center}
\end{figure}

 The numerical result for the $f\left(z_{g}\right)$ and R ratio 
for the broadening approach are shown on the left-hand side of Fig.\ref{fig:soyez-zg} and Fig.\ref{fig:soyez-R}. As can be expected, the difference is negligible. Here we have calculated the energy loss $\epsilon_g=5$ GeV and $\epsilon_i=7.3$ GeV for $L= 4$ fm from Fig.\ref{fig:energy-loss}, which agrees with the MC fit in \cite{Caucal:2019uvr}.  For $L=6$ fm and using the same formula, we find the energy loss $\epsilon_g=12$ GeV and $\epsilon_i=16$ GeV. %
Since the  BDMPS-Z spectrum is proportional to the medium length, compared with L=4 fm case, we expect the L=6 fm MIE $z_g$ distribution will be more peaked around $z_g$ close to $z_{cut}$ region. %
However, due to jets losing more energy when propagating through a longer medium, we see that the numerical results for $L=4$ fm and $L=6$ fm are very close. Therefore, we included the calculation for $L = 6$ fm with $L = 4$ fm energy loss, shown as the orange line on the left side of Fig.\ref{fig:soyez-zg} and Fig.\ref{fig:soyez-R}. %
We see that for the massless case, the results are virtually the same for MIE $z_g$ distribution, due to medium-induced emissions having no collinear singularities, while for heavy flavour VLE $z_g$ the differences are very small and can be neglected, due to the dead cone effect.

\par  We see that  Sudakov safe technique is not needed for in-medium heavy flavour $z_g$. Hence, for the rest of this article, we will perform the calculation without the Sudakov safety technique for MIE-only part, and we shall include Sudakov form factor only in the   VLE-only and for the pure vacuum calculation, i.e. for the denominator of R-ratio.  For the full shower result, because of the factorisation picture, we sum up the VLE and MIE part, with the vacuum Sudakov form factor associated with the vacuum splitting kernel to absorb the collinear singularities at the massless limit
\begin{equation}
    P^{tot}\left(z,\theta_{g}\right)=\frac{dI^{vac}}{dzd\theta_{g}}\Delta\left(R,\theta_{g}\right)+\frac{dI^{mie}}{dzd\theta_{g}}
\end{equation}
A similar approach was used previously in \cite{Mehtar-Tani:2016aco}.
The $z_g$ distribution is now given by
\begin{equation}\label{eq:full-shower}
f\left(z_{g}\right)=N\int_{\theta_{cut}}^{R}d\theta_{g}P^{tot}\left(z_{g},\theta_{g}\right),
\end{equation} 
The normalised $z_g$ distribution is normalised by the joint probability distribution $P^{*}\left(R,\theta_{g}\right)$, and has the form
\begin{equation}\label{eq:norm-joint}
P^{*}\left(z_{g},\theta_{g}\right)=\frac{P^{tot}\left(z_{g},\theta_{g}\right)}{\int_{\theta_{cut}}^{R}d\theta\int_{z_{cut}}^{1/2}dzP^{tot}\left(z_{g},\theta_{g}\right)},
\end{equation}
while for the $N_{jet}$ normalised $z_g$, we only need to remove the denominator of Eq.\ref{eq:norm-joint}, this is equivalent to N=1 for Eq.\ref{eq:full-shower}. 

\par Note that the quark mass gives us a natural cut-off for the collinear singularities in vacuum. So if we are not interested in the massless limit, we can carry all calculations and define a joint probability 
without Sudakov safe technique at all.
\begin{equation}
P^{tot}\left(z_{g},\theta_{g}\right)=\ensuremath{\bar{P}^{vac}\left(z_{g},\theta_{g}\right)+\bar{P}^{mie}\left(z_{g},\theta_{g}\right)}, 
\end{equation}
except for the massless limit.
Numerically, we checked that there is no numerical difference between these two formulas for finite mass.
\par In the next section, we shall compare the $z_g$ distribution calculated for massless case using Eq. 
\ref{eq:full-shower}.
and the full Sudakov safety calculation.

\subsection{Calculation  of the  $z_g$ distribution: MIE-only and full shower}
\label{sec:zg-numerical}
In this section, we will calculate the $z_g$ distribution using the late emission approximation and compare it with the broadening approach results that we discussed in the last section. We shall also compare the $z_g$ calculation for the full shower using the 
full Sudakov safe technique and Eq. \ref{eq:full-shower}.
\par Recall the definition of $z_g$ in Eq.\ref{eqn:zg}, where the difference between MIE-only and VLE-only comes from the numerator, i.e. $\mathcal{E}_{g}$. For the MIE-only part, we use the time-averaged picture, i.e. $\mathcal{E}_{g}=\epsilon_{g}\left(zp_{T},\theta_{cut}\right)/4$ for the subjets with $\theta_g=\theta_{cut}$ \cite{Caucal:2019uvr}. Since the energy loss is proportional to $L^2$, for $p_T=200$ GeV and $L=4$fm, the broadening approach calculation in the last section corresponds to $\mathcal{E}_{g}=5$ GeV and $\mathcal{E}_{q}=7.3$ GeV according to Fig.\ref{fig:energy-loss}. For VLE-only, the energy loss is calculated directly from Eq.\ref{eq:MIE-eloss}, i.e. $\ensuremath{\mathcal{E}_{i}=\epsilon_{i}\left(zp_{T},R\right)}$.

\begin{figure}[!ht]
\begin{center}
\includegraphics[width=0.48\linewidth]{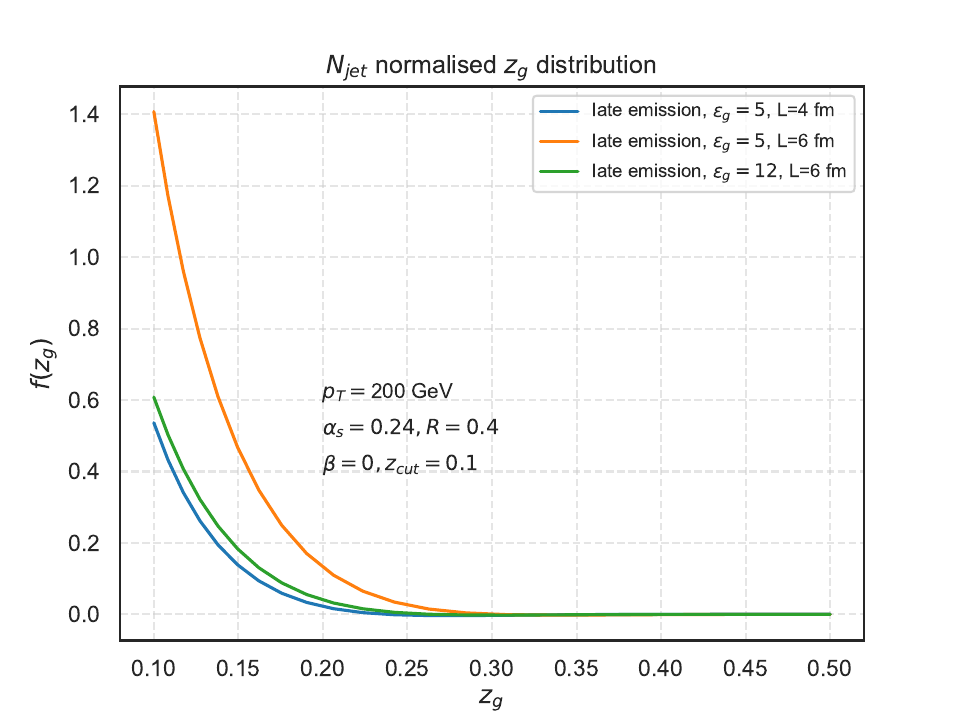}
\includegraphics[width=0.48\linewidth]{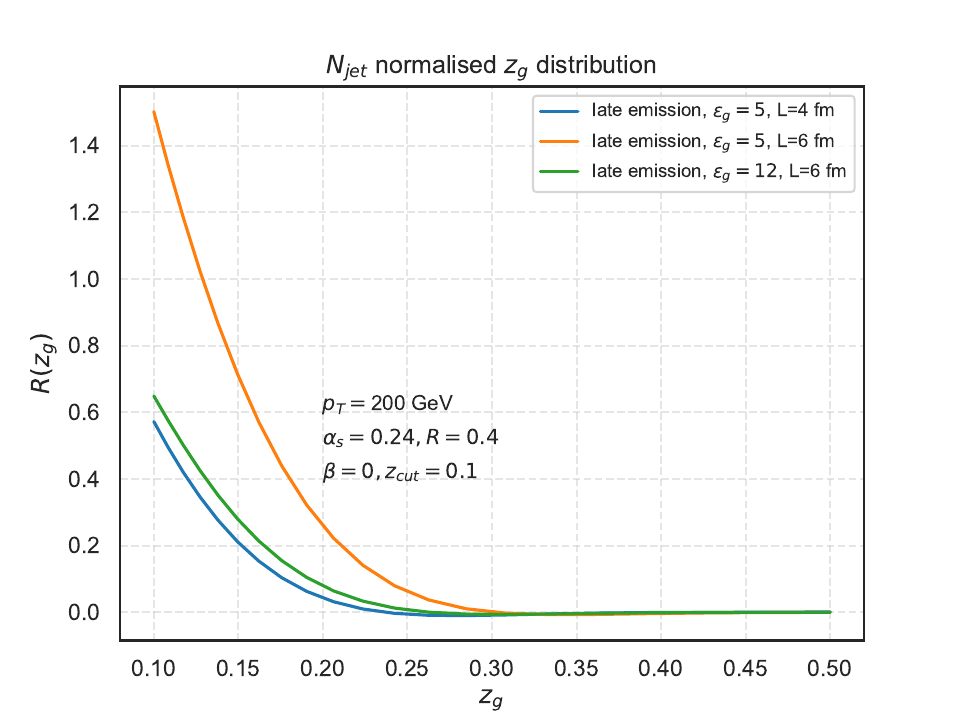}
\caption{$f(z_g)$ and R ratio for MIE-only calculated using late emission approximation}
\label{fig:soyez-zg2}
\end{center}
\end{figure}

For $L = 6$ fm, we  have the energy loss: $\mathcal{E}_{g}=12$  GeV and $\mathcal{E}_{q}=16$ GeV.
 To make the comparison, in our $z_g$ calculation, we use the same energy loss for both the broadening approach and the late emission approximation. The numerical results for the MIE-only part of $z_g$ distribution for the broadening approach and late emission for MIE approximation are depicted in Fig.\ref{fig:soyez-zg}-\ref{fig:soyez-zg2}. We see that late emission approximation results are slightly more peaked but otherwise agree with the results from the broadening approach.

\par Consider now the full shower for the massless case. 

We calculated the $z_g$ distribution using both approaches of Eq.\ref{eq:full-shower}, with only VLE part multiplied by the Sudakov form factor and Eq.\ref{eq:self-norm} which corresponds to the application of Sudakov safe technique for both VLE and MIE. 
We have compared these two formulas for the broadening possibility approach in Fig.\ref{fig:zg-full_massless}, and indeed the difference is negligible, as expected.


The full shower result for both broadening approach and late emission approximation is also depicted in Fig.\ref{fig:zg-full_massless}. Note that for the full shower result, we can't simply sum the MIE-only and VLE part that we have just calculated earlier in this section. Indeed, in the VLE inside the medium acting as a source, we need to include the multiplicity $1+\nu(z_g,\theta_g)$ for $P^{MIE}\left(z_{g},\theta_{g}\right)$. More details about the multiplicity have been discussed in Sec.\ref{sec:multi}.  Recall that "1" in the latter expression corresponds to the leading parton, and the second term is the total multiplicity of VLE shower inside the jet cone. We choose the lower bound for $\omega=z p_{T0}$, and the upper limit on $\theta_g$ integration is $\theta_g=R$, since we demand these gluons remain a part of the jet. In Sec.\ref{sec:energy-loss}, we see there that the difference for maximum multiplicity between heavy flavor and massless quark jets is small.


Note that we evaluated the total shower energy loss using late emission approximation with DLA multiplicity as shown in Fig.\ref{fig:eloss-full}. The comparison for the full shower result between the broadening possibility approach and late emission approximation is shown in Fig.\ref{fig:zg-full_massless}, where the difference coming from the MIE part, i.e. the peak region $z_{g}\rightarrow z_{cut}$. We see both approaches agree with each other well, the vacuum-like splitting inside the medium with incoherent energy loss controls the intermediate and large $z_g$ region i.e. the tail region $z_{g}$ close to 0.5, while at small $z_g$ region the medium induced radiation takes over, which leads to the enhancement at $z_{g}\rightarrow z_{cut}$. We also see the  curves become flatter in the limit of 
intermediate and large $z_g$ due to the VLEs with incoherent energy loss, and enhanced at small $z_{g}$ due to MIEs. 
\par The full Sudakov safety calculation was done for massless case in \cite{Caucal:2019uvr} using the general extension of the Sudakov safety technique.
\begin{align}\label{eq:self-norm}
f\left(z_{g}\right)	&=\mathcal{N}\int d\theta_{g}\Delta^{MIE}\left(R,\theta_{g}\right)\Delta^{VLE}\left(R,\theta_{g}\right)\ensuremath{}\\\nonumber
&\times	\int_{z_{cut}}^{1/2}dz\left[P^{vac}\left(z,\theta_{g}\right)\delta\left(z_{g}-Z_{g}^{vac}\left(z,\theta_{g}\right)\right)+P^{mie}\left(z,\theta_{g}\right)\delta\left(z_{g}-Z_{g}^{mie}\left(z,\theta_{g}\right)\right)\right],
\end{align}
where $\mathcal{N}$ is the normalisation factor, defined as $\left(1-\Delta^{MIE}\left(R,\theta_{g}\right)\Delta^{VLE}\left(R,\theta_{g}\right)\right)^{-1}$ as shown in Eq.\ref{eq:Sudakov-safe2}, and one for the $N_{jet}$ normalised scheme, 
\par We have compared the calculation of the $z_g$ distribution using both approaches of Eq.\ref{eq:full-shower}, with only VLE part multiplied by Sudakov form factor and Eq.\ref{eq:self-norm} which corresponds to the application of Sudakov safety technique for both VLE and MIE. 
We have compared these two formulas for the broadening possibility approach in Fig.\ref{fig:zg-full_massless}, and indeed the difference is negligible, as expected.

\hspace{-1.5em}\stackunder{
    \begin{minipage}[b]{0.49\textwidth}\vspace{0pt}
      \centering
      \includegraphics[width=\textwidth]{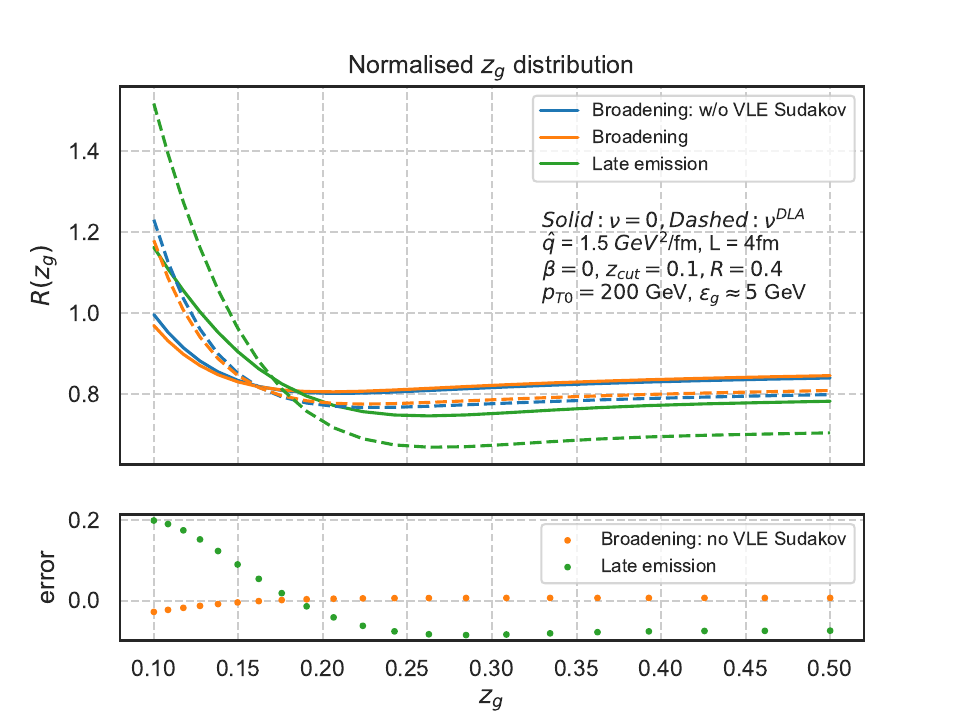}
    \end{minipage}
}{
}
\hspace{-1.em}
\stackon{
}{    \begin{minipage}[b]{0.48\textwidth}\vspace{0pt}
    \captionof{figure}{Full shower results with DLA multiplicity: a)  Eq.\ref{eq:full-shower} with both broadening approach (blue lines) and late emission approximation (green lines); b) Eq.\ref{eq:self-norm} from broadening approach (orange lines). The lower panel shows the difference between those two formulas for the broadening approach (green dots) and the difference between broadening approach and late emission approximation (orange dots)\label{fig:zg-full_massless}}
    \end{minipage}
}

\par
\begin{figure}[!htb]
\begin{center}
\includegraphics[width=0.48\linewidth]{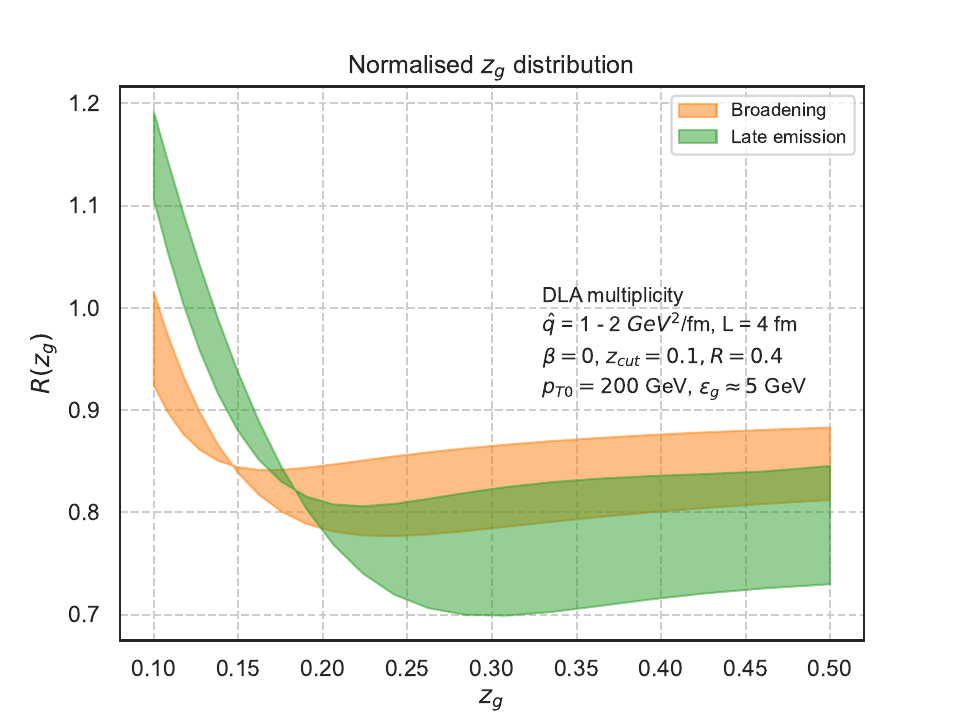}
\caption{R ratio for full shower normalised $z_g$ distribution with  DLA multiplicity for both  broadening  approach and late emission approximation with quenching parameter $\hat{q}$ between 1 and 2 $\text{GeV}^2$/fm}
\label{fig:zg-full_massless2}
\end{center}
\end{figure}

 In this article, we considered a static medium with length L = 4 and 6 fm, which is close to the average length of jets passing through the medium at LHC. To confirm our findings, here, we choose the quenching parameter $\hat{q}=1-2 \text{GeV}^2$/fm in order to estimate the theoretical uncertainty on the medium parameters. %
The results are shown in Fig.\ref{fig:zg-full_massless2}, to avoid the overlapping between those two aforementioned formulas for full shower, we only show the result for the broadening approach with Eq.\ref{eq:self-norm} and compare with our late emission approximation from Eq.\ref{eq:full-shower}. %
It would be interesting to extend our analysis to a non-static medium model as a future study.

Note that for a more realistic jet spectrum, jets passing SD conditions will lose more energy, hence, $z_g$ distribution will be more suppressed~\cite{Caucal:2019uvr},and the result will agree better with the experimental results. In this work, the realistic jet configuration is beyond our scope, therefore, we constrain ourselves for some qualitative understanding about how the heavy flavour jet substructure can help us study the mass effect in QGP. Details about the heavy flavour $z_g$ distribution will be discussed in the next section.

\section{Towards phenomenology: heavy flavour $z_g$ distribution.}
In the last section, we have calculated the $z_g$ distribution using the late emission approximation and compared it with the broadening approach for high energy massless quark jet, i.e. $p_T=200$ GeV. Now, we will extend to heavy flavour jets, i.e. b- and c- jets with $m_b=5$ GeV and $m_c=1.5$ GeV.  We chose the jet energy $p_{T}=\left(100,50,25\right)$ GeV.Hence the dead-cone configuration for b-jets is close to what we have discussed in the last two chapters, while for c-jets,  we take $m_c=1.5$ GeV with the maximum dead-cone angle $\theta_0=$0.06.
\par Note that for normalised jets and for $N_{\rm jet}$-normalised  with $\theta_{g}>0.1$, $z_g$ distribution can not probe the dead cone region except 
for $\theta=0.2$ case. The distributions that can be used to probe the gluons inside the dead cone correspond to $N_{jet}$ normalised distributions with $\theta_g<0.1$. (Recall that for $N_{jet}$ normalised distributions, we integrate over the angle variable $\theta_g$ from cut off to R if we speak about $\theta_g>0.1$ distribution, and
from $\theta_c$=0.04 and 0.02 for L=4 and 6 fm to $\theta_{cut}=0.1$ as an upper bound when we speak about $\theta_g<0.1 $.)

\begin{figure}[!htb]
\begin{center}
\includegraphics[width=0.48\linewidth]{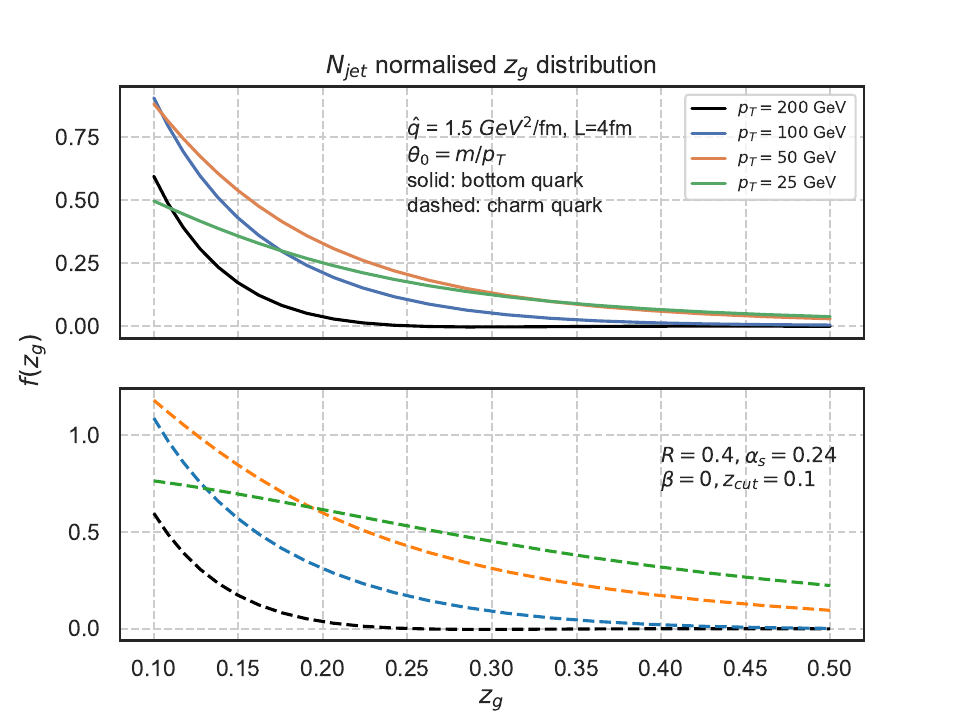}
\includegraphics[width=0.48\linewidth]{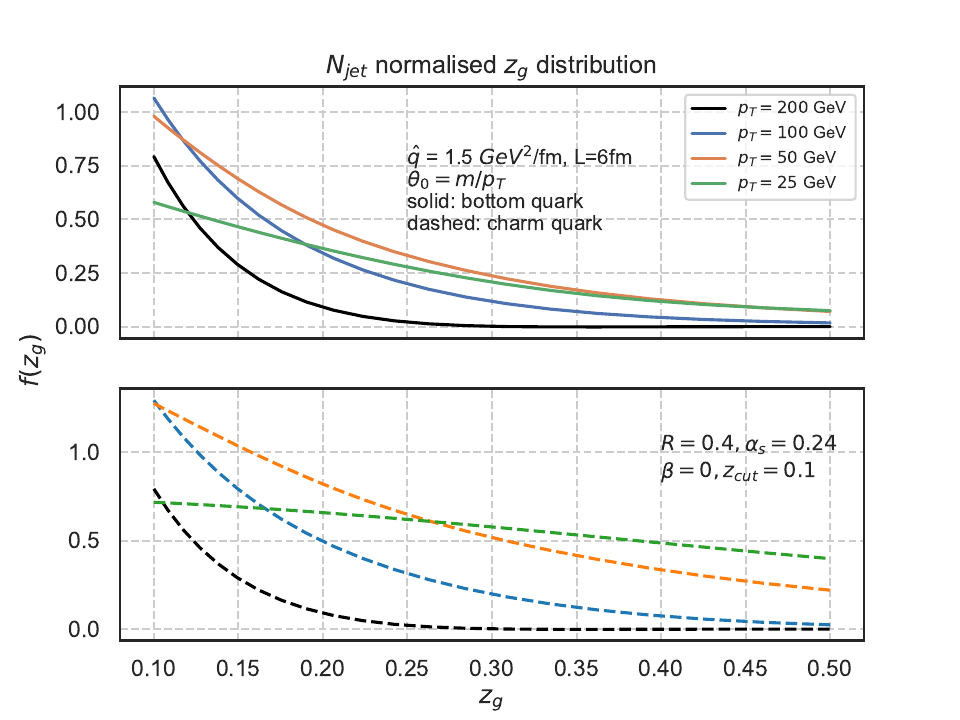}
\caption{MIE $z_g$ distribution from late emission approximation with L=4 and 6 fm for b- (upper panel) and c-jets (lower panel) \cw{MIE plots w/o mulitiplicity}}
\label{fig:zg-full}
\end{center}
\end{figure}

\begin{figure}[!htb]
\begin{center}
\includegraphics[width=0.48\linewidth]{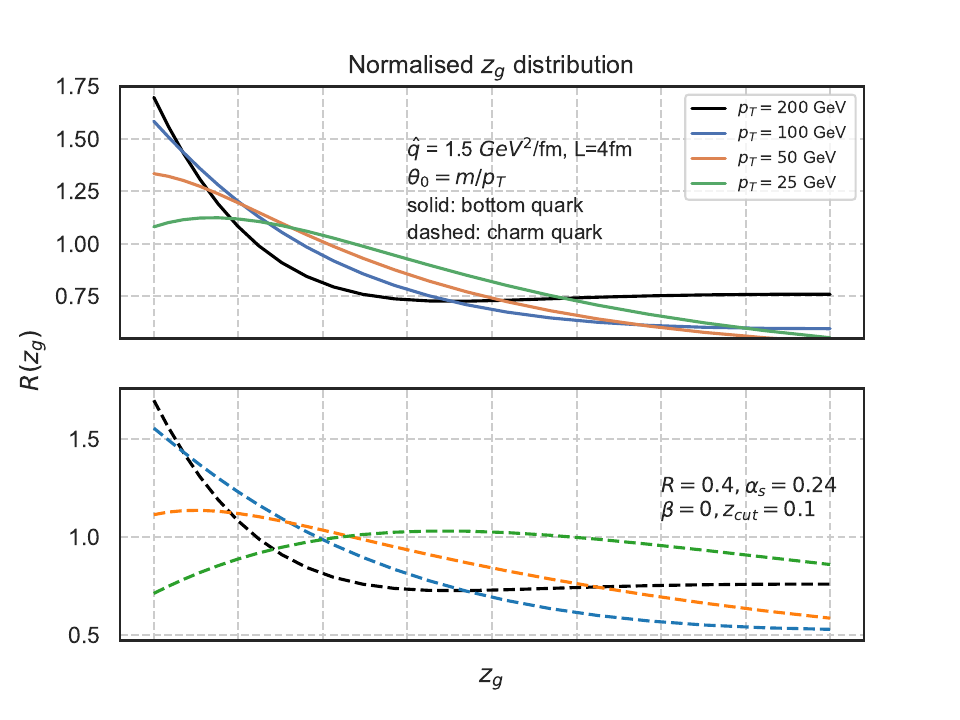}
\includegraphics[width=0.48\linewidth]{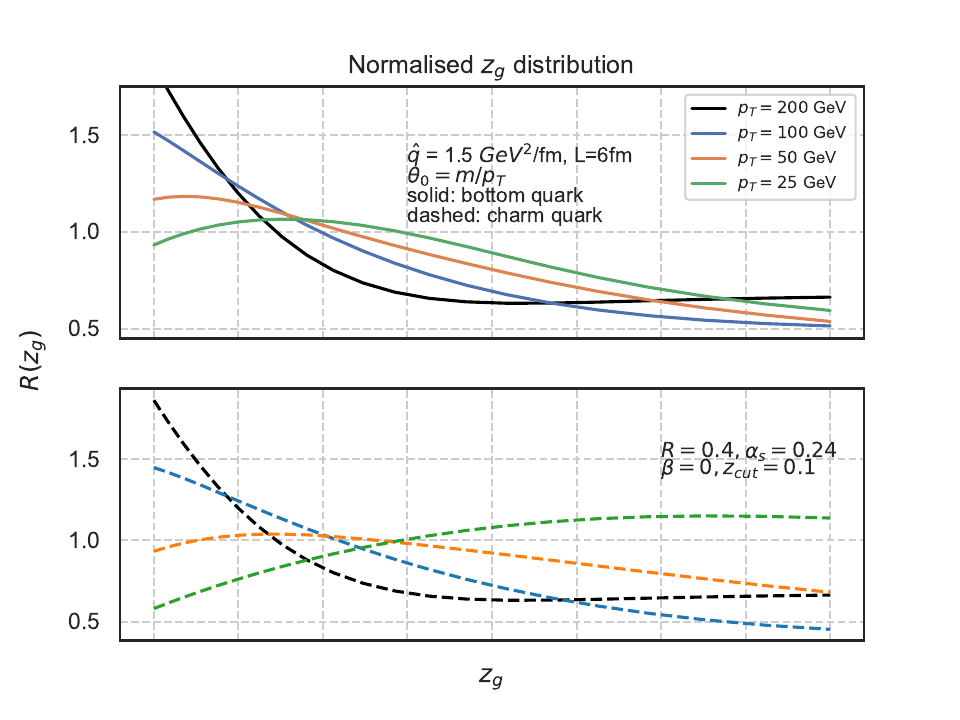}
\caption{Full shower R ratio of normalised $z_g$ distribution from late emission approximation with L=4 and 6 fm for b- (upper panel) and c-jets (lower panel) }
\label{fig:zg-full2}
\end{center}
\end{figure}

\subsection{Phenomenology with normalised $z_g$ distribution}
\label{sec:pheno}
As we discussed in the last chapter, the full shower result is a combination of MIE and VLE parts, due to the in-medium radiation factorised in time. In this section, we consider the normalised $z_g$ distribution for heavy flavour jets. In order to understand the full shower result, we start from the MIE part of $z_g$ distribution, the numerical results are shown in Fig.\ref{fig:zg-full}. We see that the MIE distribution increases at small $z_g$ region, but the increase rate depends on the dead cone angle. For $\theta\le 0.05$ the behaviour is similar to the massless jets, as it is expected from the energy distributions. 
For larger dead cone angles, i.e. lower energies jets, the MIE $z_g$ distribution increases less rapidly with the increase of the dead cone angle. 


\par For the full shower R ratio, we again start with the jet energy $p_T$=200 GeV. We see that our results agree with the massless quark jets, which we discussed in detail in section 4 since the dead cone is small for both b- and c-jets.
\par The features of the distributions for smaller energies and larger 
dead cone angles can be easily understood by considering the energy distributions depicted in Fig.\ref{fig:zg-Njet}. %
In the ASW/late emission approximation, the phase constraints lead to the maximum at some frequency that we shall denote as  $\omega_{min}$. This maximum occurs at the energies of order 5-8 GeV for L=4-6 fm.   The maximum position shifts slightly towards smaller $\omega$, when we increase the dead cone angle. Moreover,  for 25 and 50  GeV jet, we  find $\omega_{min}>z_{cut} p_T$.   In other words the $z_g$ distribution covers the whole maximum region, while for   100 GeV jets we find $\omega_{min}$ is still to the right of the maximum. 
Consequently, the picture for 200 and 100 GeV jets is similar to massless jets, while for 50 and 25 GeV jets where $\omega_{min}=$5 and 2.5 GeV respectively, we manage to catch the maximum, and 
even the region to the left of the maximum, leading to decreasing R ratio. Let us also comment on c quark jets, where for small $p_T=25$ GeV we again reach to the left of the energy distribution maximum, leading to maximum in the $z_g$ distribution curve, at $\theta_0=1.5/25=0.06$.


\begin{figure}[!htb]
\begin{center}
\includegraphics[width=0.48\linewidth]{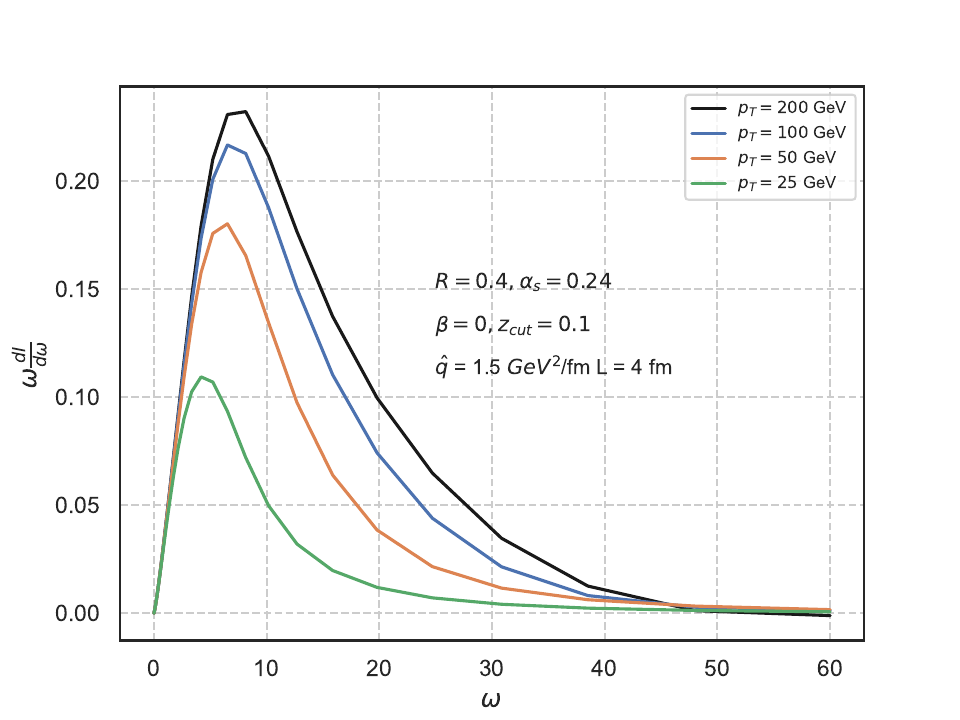}
\includegraphics[width=0.48\linewidth]{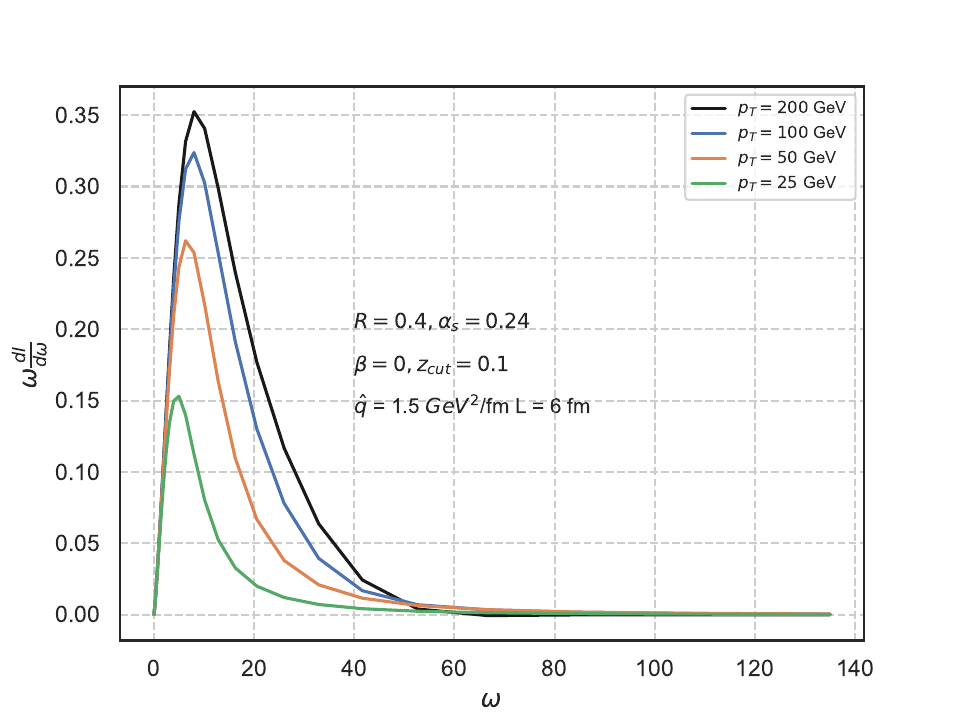}
\caption{The energy distribution $\omega\frac{dI}{d\omega}$ for L=4 and 6 fm, with the phase space for transverse momenta between $\omega\theta_{cut}$ and $\omega R$}
\label{fig:zg-Njet}
\end{center}
\end{figure}


Note that as shown in Fig.\ref{fig:zg-Vac}, the vacuum normalised $z_g$ distribution has no dead cone angle or jet energy dependence, so the relative changes for different jet energy are due to the ratio between MIE and VLE parts. Finally, we notice for $p_T$=25 GeV, charm jets decrease faster, this is because charm jets lose more energy compared with bottom jets, hence the R ratio of VLE decreases faster as shown in Fig.\ref{fig:Rzg-VLE} compared with b-jets, for the same jet energy, the dead cone is smaller.


In summary, the normalised  $z_g$ distribution is sensitive to the mass of the heavy
quark, and they are sensitive to the presence of MIE, which can be seen if we compare Figs. Fig.\ref{fig:Rzg-VLE} and the full shower
R ratios are depicted in Fig. \ref{fig:zg-full2}. However, these distributions do not probe the dead cone gluons
directly.

\subsection{Phenomenology with $\text{N}_{\text{jet}}$ normalised $z_g$ distribution}
\label{sec:pheno_njet}

As discussed in Ref.~\cite{Caucal:2019uvr}  for the massless case, the $\text{N}_{\text{jet}}$ normalised distribution contains more information due to the R ratio being the ratio between the jets passing the SD condition in medium and the total jets in the vacuum. 
 For the heavy flavour study, the situation is even more interesting, since, due to the filling of dead cone in the medium and the dead cone suppression in the vacuum, we expect $N_{jet}$ normalised distribution to be more sensitive to the mass effects in QGP.

\begin{figure}[!htb]
\begin{center}
\includegraphics[width=0.48\linewidth]{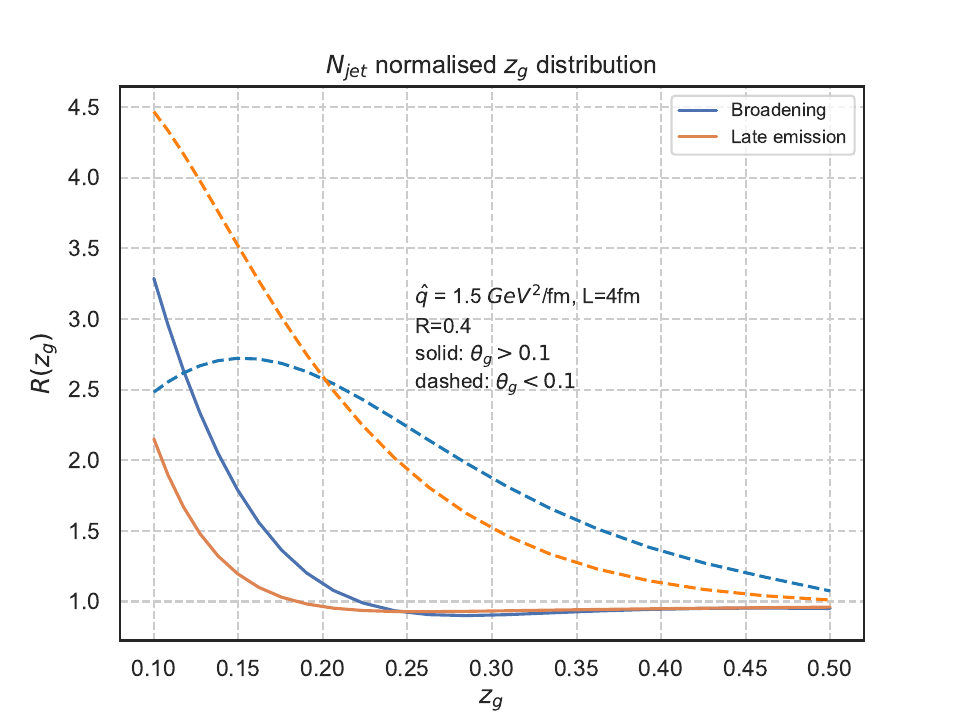}
\caption{The $\text{N}_\text{jets}$ normalised $z_g$ with $\theta_{g}>0.1$ (solid line) and $\theta_g<0.1$ (dashed line) calculated using late emission approximation and broadening approach for the massless case.}
\label{fig:zg-Njet}
\end{center}
\end{figure}

First, let us compare the broadening approach with our late emission approximation for massless quark jets, with both $\theta_{g}>0.1$ and $\theta_{g}<0.1$ as suggested by the literature~\cite{ALICE:2019ykw,Caucal:2019uvr}. The results are shown in Fig.\ref{fig:zg-Njet}, where we see the $\theta_{g}<0.1$ case is enhanced compared with $\theta_{g}>0.1$ case. The $\theta_g>0.1$ case captures more vacuum-like radiation with incoherent energy loss, hence the full shower result is suppressed compared with $\theta_g<0.1$ case, this result agrees with the literature well. 
Therefore, we expect the $\theta_g<0.1$ case is better suited for us to study the dead-cone effects.

\begin{figure}[!htb]
\begin{center}
\includegraphics[width=0.48\linewidth]{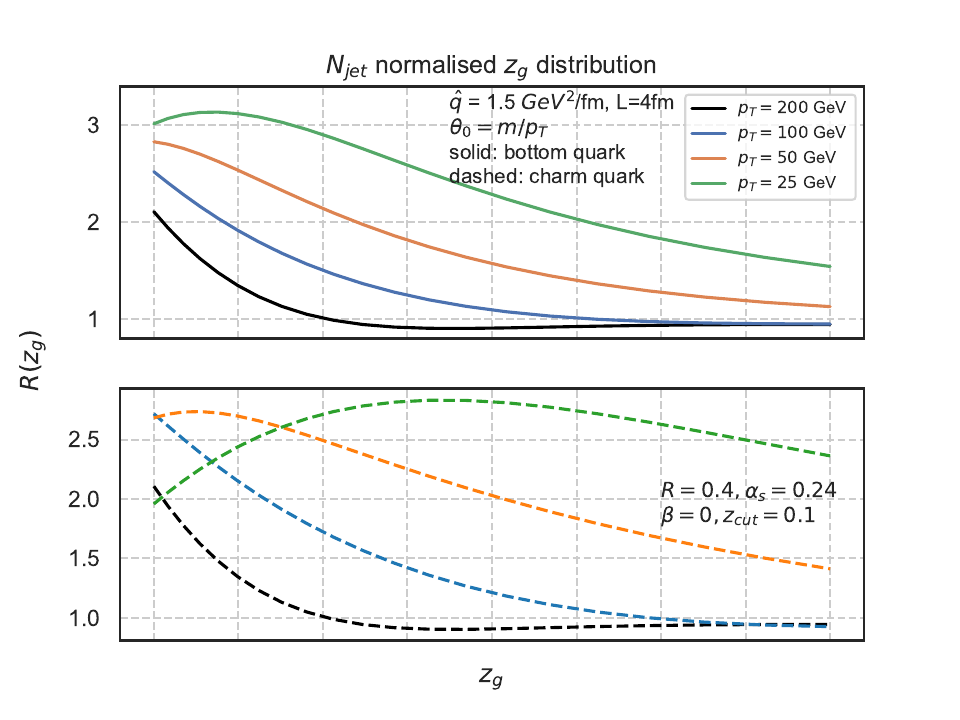}
\includegraphics[width=0.48\linewidth]{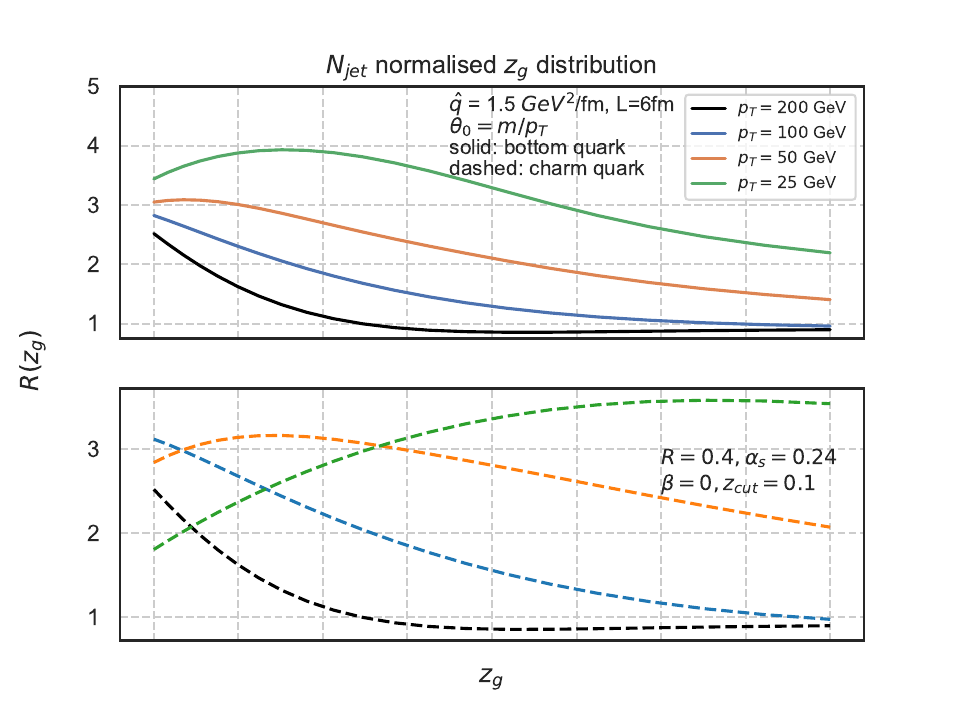}
\caption{Full shower R ratio with $N_{jet}$ normalised $\theta_g>0.1$ from late emission approximation with L=4 and 6 fm for b- and c-jets}
\label{fig:zg-Njet2}
\end{center}
\end{figure}

For $\theta_g>0.1$, our results for heavy flavour jets are presented in Fig.\ref{fig:zg-Njet2}, which by comparing the massless and massive case, we see the same behaviour as discussed in the last section. However, we see the curves are monotonically increasing with larger dead cone angle, due to vacuum $z_g$ distribution, i.e. the denominator, of the R ratio decreasing with a larger dead cone angle. Compared with the results shown in the last section, this means the overall shape of the numerical results is no longer controlled by the ratio between VLE and MIE, therefore, we see the result with a large dead cone is enhanced.  

For c-jets with $p_T$=25 GeV, we see the decreasing behaviour again around $z_g$ close to $z_{cut}$ region, and recall Fig.\ref{fig:zg-full}, the MIE $z_g$ distribution for c-jets increases around small $z_g$. However, due to c-jets having small dead cone angles, i.e. max $\theta_0$=0.06 with $p_T$=25 GeV, vacuum $z_g$ distributions are essentially massless without the dead cone suppression, which shows this decreasing behaviour.

In conclusion, we see that physically the behaviour is the same for normalised $z_g$ in the previous subsection, and changes between Fig.\ref{fig:zg-full} and Fig.\ref{fig:zg-Njet2} are due to different normalisation.

\begin{figure}[!htb]
\begin{center}
\includegraphics[width=0.48\linewidth]{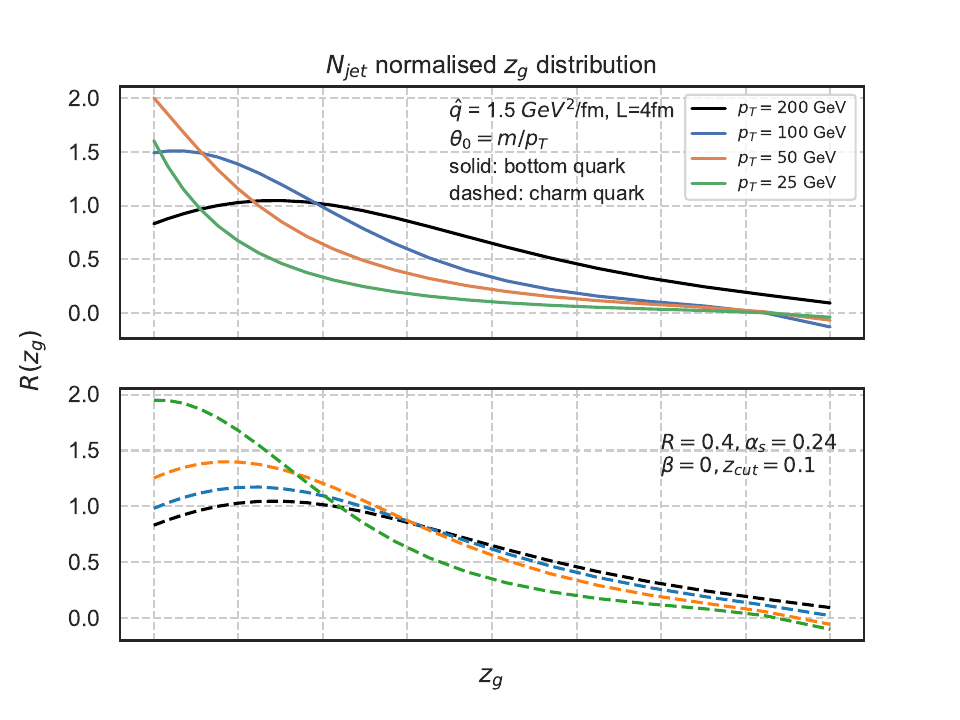}
\includegraphics[width=0.48\linewidth]{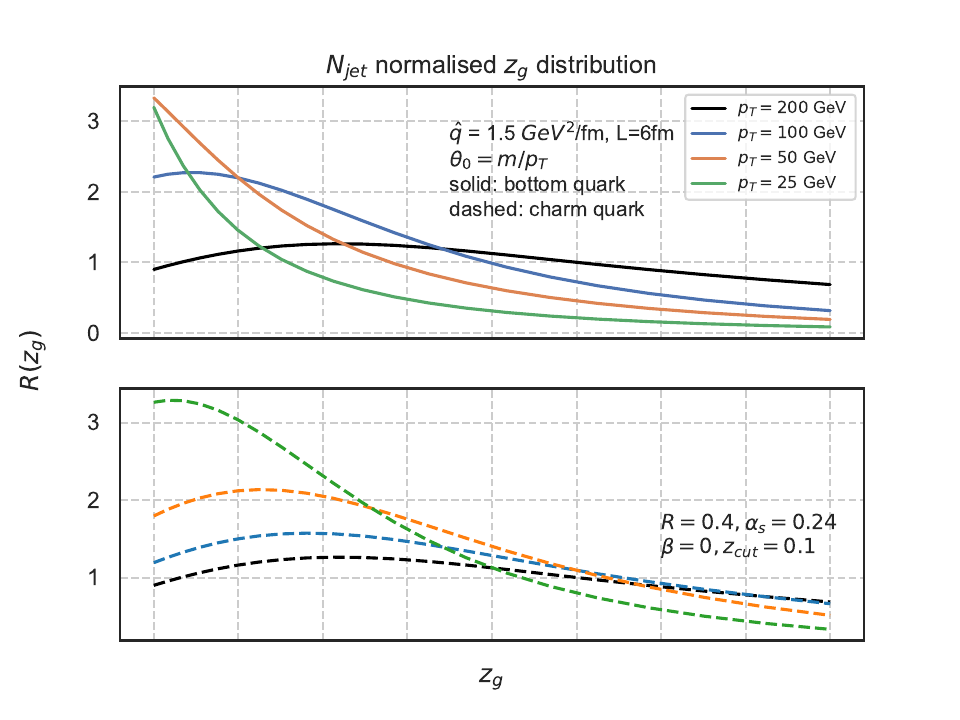}
\caption{MIE R ratio with $N_{jet}$ normalised $\theta_g<0.1$ from late emission approximation with L=4 and 6 fm for b- and c-jets}
\label{fig:zg-Njet3}
\end{center}
\end{figure}

Besides less vacuum-like radiation captured by SD condition for $\theta_g<0.1$, as we discussed for the massless quark jets earlier in this section. The situation is even more interesting for us since we look at the gluons inside the dead cone for b- and c-jets, i.e. $\theta_g<\theta_0$, except for b-jet with jet energy equal to 25 GeV case. Indeed, comparing MIE and full shower results in Fig.\ref{fig:zg-Njet3}-\ref{fig:zg-Njet4} shows that instead of VLE controlling most of the curve, we see MIE part takes up most of the contribution. Moreover, we see a huge increase for c-jets at $p_T$=25 GeV when compared with b-jets result, due to c-jets having a small dead cone angle $\theta_0<\theta_{cut}$ while b-jets' dead cone angle is larger than the cut-off, which cuts part of the phase space. Hence, we see the big difference with the other two configurations with $\theta_0<\theta_{cut}$, which slightly decreases at $z\rightarrow z_{cut}$. This indicates that we probed the dead cone by implying a minimal resolution distance close to the dead cone angle. 

We conclude that $\theta_g\le 0.1$ case gives us the most information about the gluon radiation inside the dead cone, which is controlled by MIE since VLE are suppressed by vacuum dead cone effect.
\par In other words, we see that the R ratios are monotonic with R ratio for $\theta_0=0.2$ being the steepest one when we approach the cut-off. This is due to the suppression of VLE i.e. the dead cone effect, that is weakened by the medium.

\begin{figure}[!htb]
\begin{center}
\includegraphics[width=0.48\linewidth]{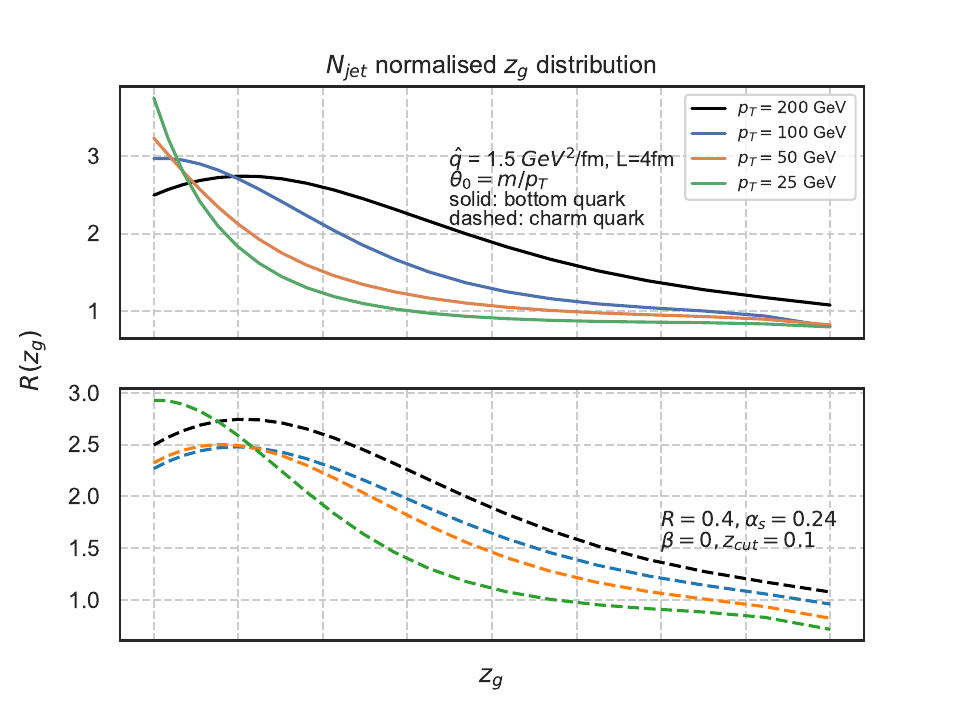}
\includegraphics[width=0.48\linewidth]{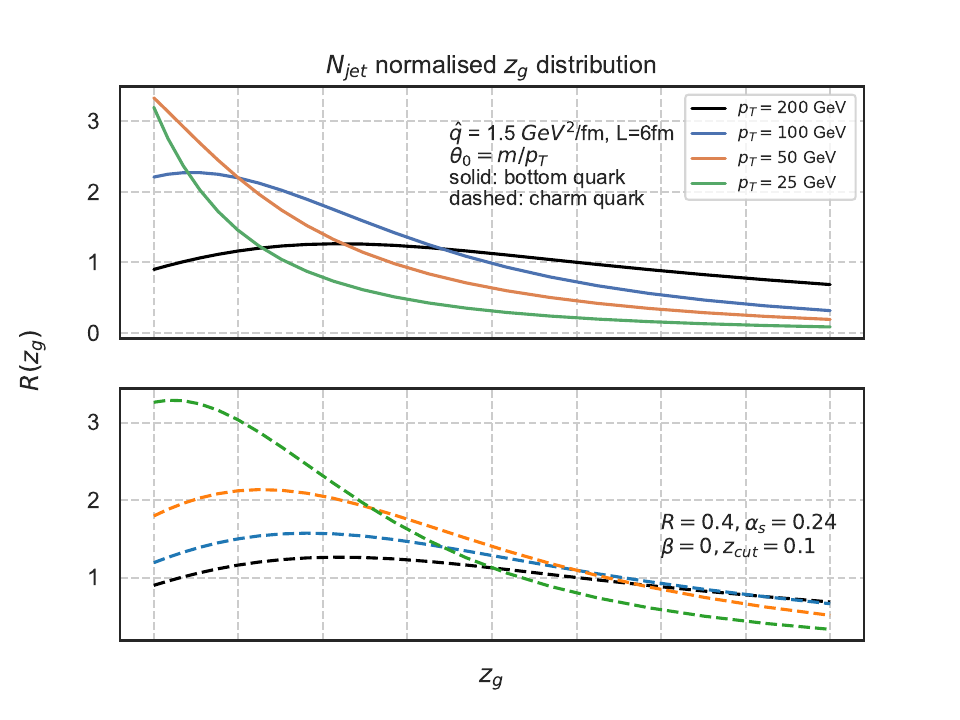}
\caption{Full shower R ratio with $N_{jet}$ normalised $\theta_g<0.1$ from late emission approximation with L=4 and 6 fm for b- and c-jets}
\label{fig:zg-Njet4}
\end{center}
\end{figure}

\begin{figure}[!htb]
\begin{center}
\includegraphics[width=0.48\linewidth]{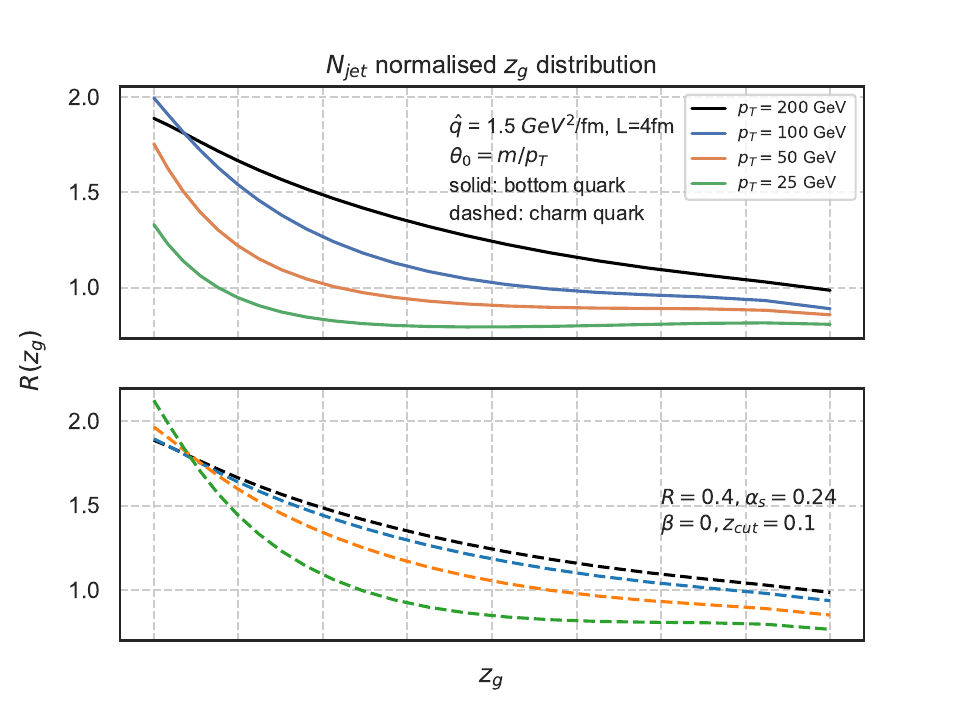}
\includegraphics[width=0.48\linewidth]{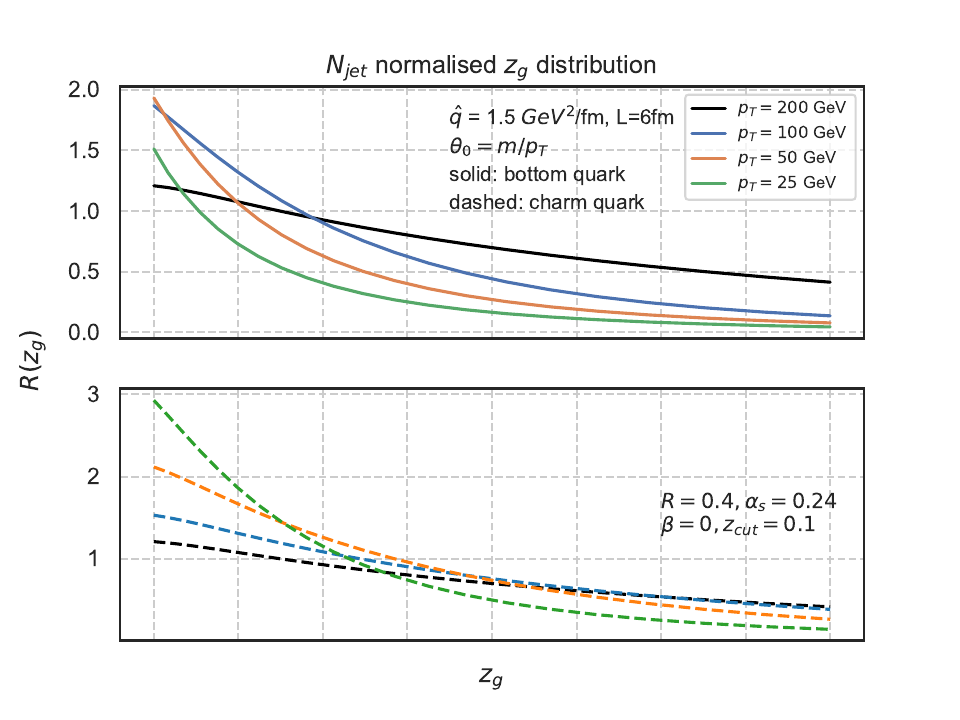}
\caption{Full shower R ratio with $N_{jet}$ normalised $\theta_g<0.2$ from late emission approximation with L=4 and 6 fm for b- and c-jets}
\label{fig:zg-Njet4}
\end{center}
\end{figure}
\par Finally, we also consider the R ratio for $\theta_g\le 0.2$, so that b-jets with $p_T=$25 GeV can be captured with our set up. %
As discussed above, we see that the result for $\theta_g<0.2$ is suppressed when compared with $\theta_g<0.1$ due to more vacuum-like radiation being captured due to the increasing phase space. Nevertheless, in Fig.\ref{fig:zg-Njet4}, we can still see the significant enhancement behaviour that R ratio becomes larger than one at small $z_g$ region. %
In addition, we also found that the relative order of the curves starts to change,  but $\theta=0.2$ is still the steepest one. 
\section{Summary and outlook}
\label{sec::conclusion}
The study of heavy flavour jet substructures is still in its early stages, with only a few observables calculated with all-order resummation techniques~\cite{Caletti:2023spr,Li:2017wwc,Craft:2022kdo,Cunqueiro:2022svx}. 
In this article, we have performed the calculation of the heavy flavour jet  $z_g$ distribution for dense QCD medium in the mean-field harmonic approximation. 
We used the late emission approximation of the ASW formula as our theoretical framework, and we found it in agreement with the physical picture of the factorisation between vacuum-like emissions and medium-induced emissions inside the medium well. Due to the radiation being factorised in time, the full shower result is simply a combination of MIE and VLE.

We demonstrated that in the massless limit, the broadening approach 
is in agreement with late emission approximation.
The extension of late emission approximation to nonzero dead cone angle is in agreement with `ASW formula and provides a natural theoretical framework for studying the heavy flavour jet substructure in the dense medium. %

\par Our calculations for $z_g$ distribution are semi-analytical, 
a higher precision study of  $z_g$ distribution and alternative novel observables that can help probe the heavy flavour jet substructure, along with the validation of MC simulation will be the next step of the research. 
In this article we studied both massless quark and heavy flavour jets, our findings can be summarised below:
\begin{itemize}
    \item Since there are no collinear singularities in the medium, we argued that the Sudakov safety technique is not needed for medium-induced emissions.
    \item We calculated both normalised and $N_{\rm jet}$ normalised  $z_g$ distributions, as well as R ratio (see section 5).
    We have seen that the form of curves for normalised jets 
    and for $N_{jet}$ normalised jets with $\theta_g>0.1$ are controlled by the VLE at large $z_g$ but for small 
    $z_g$ close to $z_{cut}$ one needs to include MIE to explain their behaviour.
    Note that normalised jets behaviour for small $z_g$ can be non-monotonic
      just because MIE contributions, see the discussion in section 5.
    \item We have seen that the most sensitive to dead cone dynamics 
    are the $z_g$ distributions with $\theta_g\le 0.2$ and especially $\theta_g\le 0.1$ which clearly show the weakening of the dead cone effect due to the presence of the medium.
\end{itemize}
\par It will be also interesting to include in our calculations the influence of $g\rightarrow \bar c c$ transitions considered
in \cite{Attems:2022ubu}. This effect corresponds to higher $\alpha_s$ corrections to our calculations, although numerically this effect may influence the comparison  with the experimental data.

Finally, note that although experimental measurements for massless $z_g$ distribution exist (see  ALICE and CMS for massless case \cite{Vertesi:2021brz,CMS:2017qlm}), as far as we know, no heavy flavour jet substructure observables have been studied experimentally, except the direct measurement of dead cone effect for splitting angle, which only studied in pp collisions \cite{ALICE:2021aqk}. It will be very interesting to carry the correspondent measurement for heavy flavour jet substructure for heavy iron collisions. 

\section*{Acknowledgments} We thank S. Marzani for the discussion on the $z_g$ distribution for heavy flavours in the vacuum case, and V. Khoze and Yu. Dokshitzer for the discussion on  jet  multiplicity. This work was supprted by BSF grant  2033344.
\appendix
\section{Comparison of late emission approximation to other approaches  for description MIE radiation.}
\subsection{Late emission approximation at massless limit versus BDMPS-Z and ASW}
\label{sec:joint-distribution}
\par It is interesting to compare the late emission approximation with the BDMPS-Z  approach for massless quark. Indeed, BDMPS-Z formalism leads to the following expression for the radiation spectrum:
\begin{equation}
\omega\frac{dI}{d\omega}=\frac{2\alpha_{s}C_{R}}{\pi}\log\vert\cosh\left(\Omega L\right)\vert
\end{equation}
Note that in the BDMPS-Z approach, the bulk and boundary contributions are given by
\begin{equation}
\omega\frac{dI}{d\omega}=\frac{2\alpha_{s}C_{R}}{\pi}\log\vert\sinh\left(\Omega L\right)\vert
\end{equation}
for the bulk contribution and 
\begin{equation}
\omega\frac{dI}{d\omega}=\frac{2\alpha_{s}C_{R}}{\pi}\log\vert\tanh\left(\Omega L\right)\vert
\end{equation}
for the boundary one.
Already for L=4 fm and $\alpha_s=0.24$ the boundary contribution in BDMPS-Z approach is negligible and decreases exponentially with L, as was mentioned in the previous section,
contrary to ASW formula, where the boundary contribution remains in the same order as the bulk contribution, although it does not scale with L, even for L=6 fm.
\par In order to restore the angular dependence we must take into  account  a Gaussian broadening proportional to $exp\left[-k_t^{2}/\bar{k}^{2}\right]/\bar{k}^{2}$, where $\bar{k}^{2}=\hat{q}(L-t)$ is the average transverse momentum accumulated through the medium over the distance $L-t$. We have~\cite{Blaizot:2014ula},\cite{Blaizot:2013hx},\cite{Blaizot:2013vha}
\begin{equation}
\tilde P(t,k_t,\hat{q},L)=\frac{1}{\pi\hat q(L-t)}e^{-k_t^2/\hat q (L-t)}
\end{equation}

\begin{figure}[b!]
\includegraphics[width=0.48\textwidth]{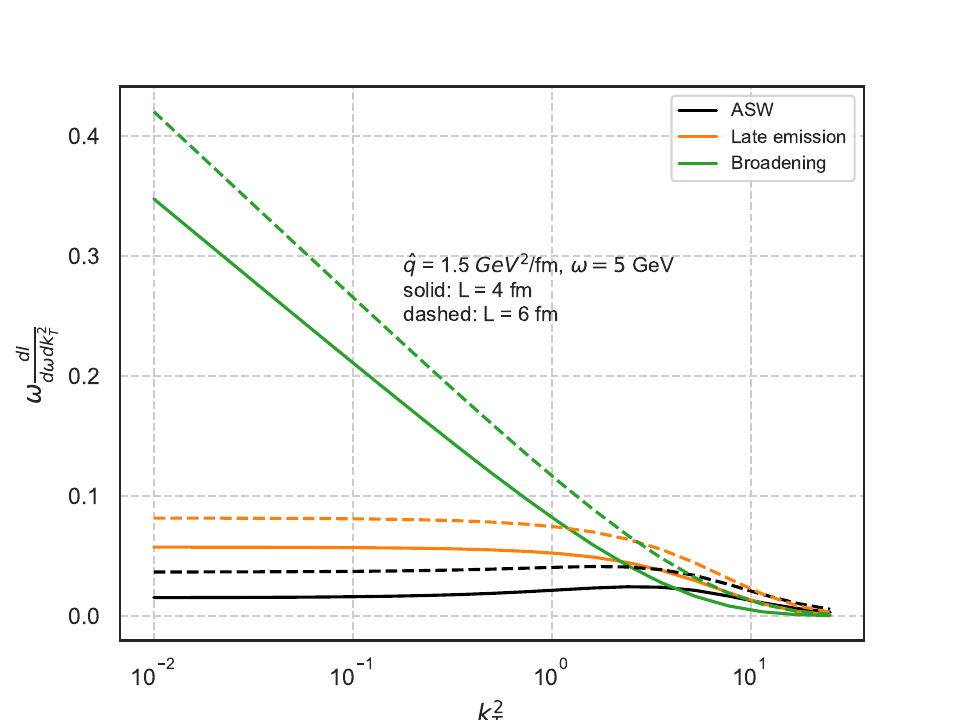}
\includegraphics[width=0.48\textwidth]{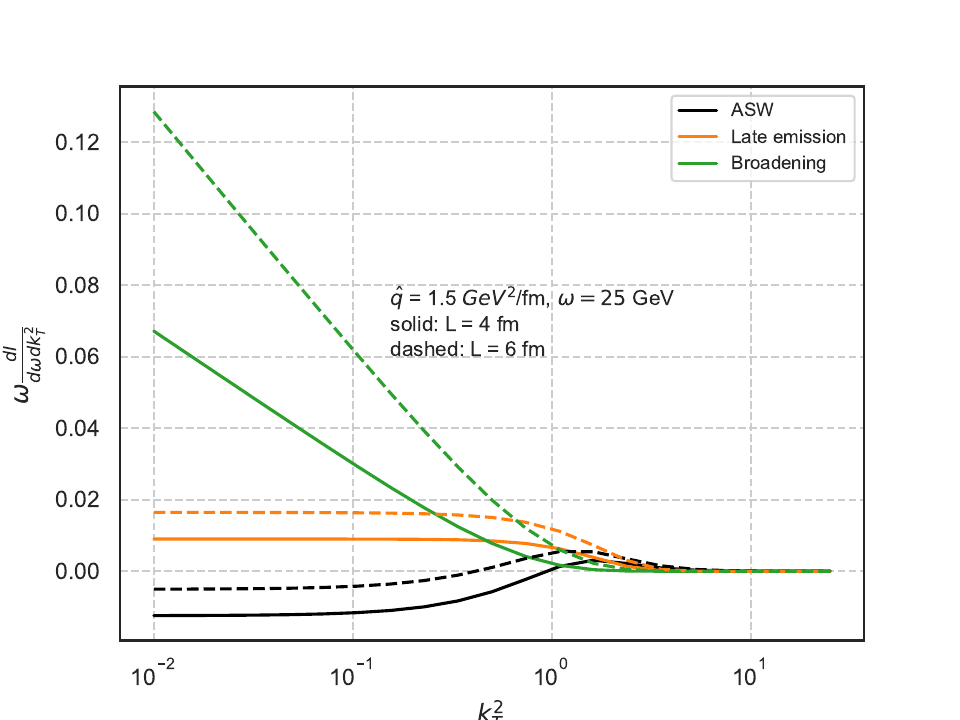}
\caption{The dependence on $k^2_t$ of $\omega\frac{dI}{d\omega dk^2_t}$ for  broadening approach , ASW and late emission approximation,
$\omega=(5, 25)$ GeV$\theta=0,L=(4, 6)$ fm and $\alpha_s$=0.24.
 \label{RA4}}
 \end{figure}
 
 The factor $\tilde{P}$ describes
 the diffusion of radiated BDMPS-Z gluons due to the scattering on medium centres. 
Integrating over emission times  $t$  from 0 to L we immediately obtain \cite{Mehtar-Tani:2016aco}
\begin{equation}
\bar{P}(k_t,\hat{q},L)=\int^L_0dt \tilde P(t,k_t,\hat q ,L)=\frac{1}{\pi\hat{q}L}\Gamma(0,\frac{k_t^{2}}{\hat{q}L}),
\label{br}
\end{equation}
where $\Gamma$ is the incomplete Gamma function,and
\begin{equation}
\omega\frac{dI}{d\omega d^{2}k_{t}}=\frac{2\alpha_{s}C_{R}}{\pi}\log\vert\cosh\left(\Omega L\right)\vert \bar {P}(k_t,\hat{q},L).
\end{equation}
We shall call this expression a broadening  approach.
Note that the broadening factor $\tilde{P}$ is normalised to one:
\begin{equation}
\int d^2k_t\tilde{P}(t,k_t,\hat q,L)=1.
\end{equation}
\par In other words, if we do not impose a phase constraint $k_t\le \omega$, the integration of the broadening factor over $k_t$
gives just BDMPS-Z energy distribution.
\begin{figure}[b!]
\centering
\includegraphics[width=0.48\textwidth]{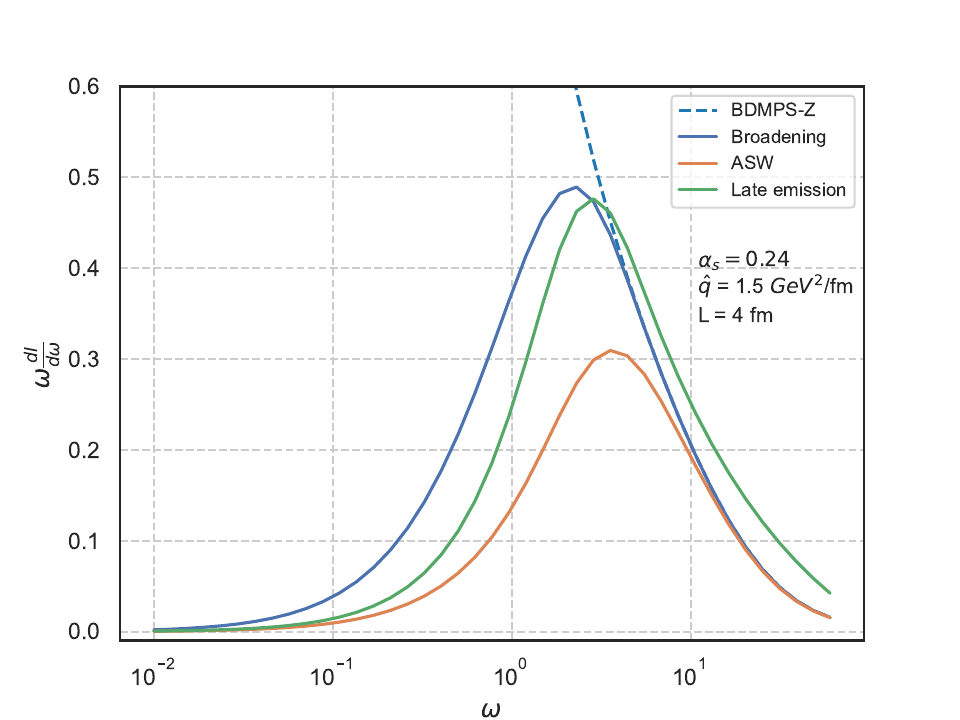}\\
\includegraphics[width=0.48\textwidth]{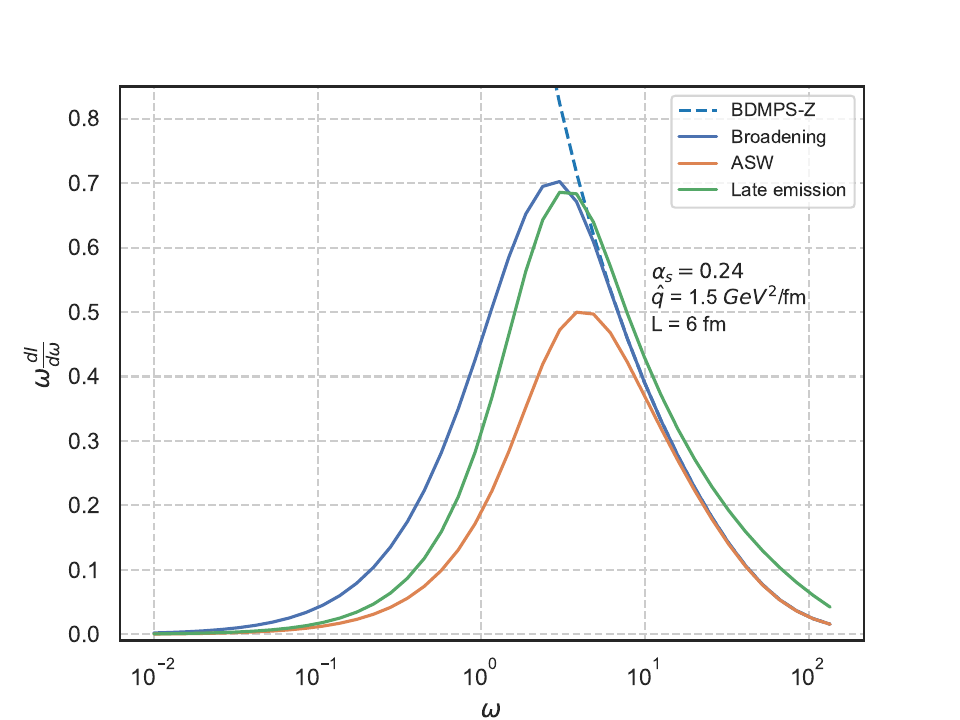}
\includegraphics[width=0.48\textwidth]{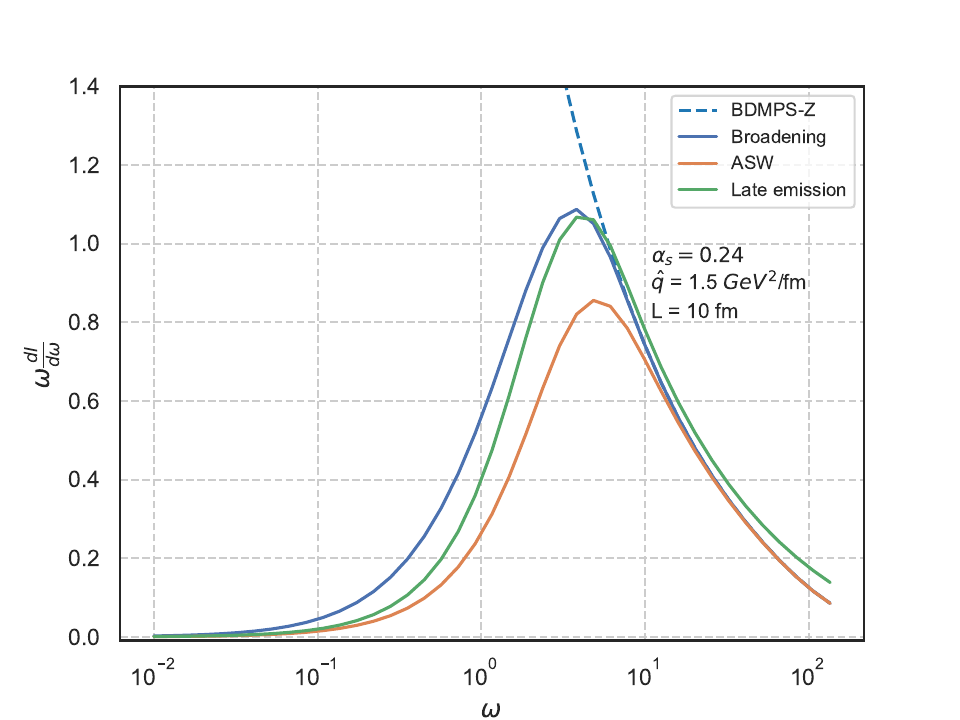}
\caption{The dependence on $\omega$ of $\omega dI/d\omega$ for BDMPS-Z, broadening approach, ASW, and the late emission approximation,
$\theta_0=0, L =$ 4, 6 and 10 fm, $\alpha_s$=0.24.
 \label{RA1}}
\end{figure}

For a numerical study, we compared the transverse momentum distributions for massless quark in different approaches, and the results were presented in Fig.\ref{RA4}. We depicted the  distribution $\omega\frac{dI}{d\omega d^2k_t} $ for   small $\omega =5$ GeV and large $\omega=25$ GeV frequencies for L=4 and 6 fm. We see that late emission approximation in the transverse plane behaves qualitatively similar to ASW formula, with the semblance increasing with the increase of L, while transverse distribution in the broadening approach ~\cite{Blaizot:2013hx} used in Ref.~\cite{Caucal:2019uvr} looks more singular. 

 In Fig.\ref{RA1}, we depicted the energy distributions. We see that indeed there is qualitatively good agreement between BDMPS-Z, ASW, and late emission approximation for zero dead cone angle. The agreement becomes quantitatively good when L increases so there are no uncertainties in the subtraction of nonscaling as L contributions. Note that in this comparison, we integrated over $k_t$ with the constraint $k_t<\omega$, hence the broadening approach result doesn't coincide with BDMPS-Z at small $\omega$ region.

It is worth noting that at the massless limit, we can obtain the broadening  approach formula from the late emission approximation, i.e. Eq.\ref{s1}.

In the massless limit, $\xi$ in Eq.\ref{s1} can be integrated analytically, since the integrand is a total derivative, 
\begin{equation}\label{eq:massless-ASW}
\omega\frac{dI}{d\omega d^{2}k_{t}}=\frac{\alpha_{s}C_{F}}{\pi^{2}\omega}Re\int_{0}^{L}dt\int\frac{d^{2}k'}{\left(2\pi\right)^{2}}P\left(\vec k_{t}-\vec k',t,L\right)e^{-\left(1+i\right)\frac{k'^{2}}{2k_{f}^{2}}},
\end{equation}
\par  This integral can be estimated using saddle point approximation. it is easy to see that the integral in $k'$ is saturated by 
\beq\vec k'\sim \vec k_t \frac{k_f^2}{\hat q (L-t)}
\eeq
Recall that for $\omega <<\omega_c$ we have $k_f^2<<\hat qL\equiv Q_s^2$, which means that if t is not very close to L, we can assume that $k_t>>k'$ in the integrand in \ref{eq:massless-ASW}. 
\begin{align}
\omega\frac{dI}{d\omega d^{2}k_{t}}	&=\frac{\alpha_{s}C_{F}}{4\pi^{3}}\sqrt{\frac{q}{\omega}}\int_{0}^{L}dtP\left(k_{t},t,L\right)\\
	&=\frac{\alpha_{s}C_{F}}{\pi^{2}}\sqrt{\frac{2\omega_{c}}{\omega}}\tilde{P}\left(k_{t},\hat{q},L\right)\nonumber
\end{align}
this is just the broadening approach. This explains why at transverse momenta larger than $\sim k_f$
these approaches give similar results, see Fig. \ref{RA4}.

Note that at the massless limit, our late emission formula, i.e. Eq.\ref{lea1}, will also be reduced to Eq.\ref{eq:massless-ASW}, with the integral over t changed to the interval $L>t>t_f$. Therefore, one can consider the late emission formula as a natural extension for the broadening possibility approach for heavy flavour.

\subsection{Late emission approximation versus dead cone factor approach.}
\par We shall also compare the late emission approximation with the dead-cone factor approach due to ~\cite{Dokshitzer:2001zm}.
Recall that in this approach, we assume  the transverse distribution in the form given by 
\beq
\omega \frac{dI}{d\omega d^2kt}=\omega \frac{dI_0}{d\omega d^2k_t}\frac{1}{(1+\theta_0^2/\theta^2)^2},
\eeq
where $\theta=k_t/\omega$. This form is similar to the form of distribution in the vacuum.
The authors of \cite{Dokshitzer:2001zm} argue that in order to estimate the energy loss for media induced radiation we just need to substitute $\theta$ by the effective angle 
$\theta_{BDMPSZ}=(\hat q/\omega^3)^{1/4}$,corresponding to the maximum of the BDMPSZ radiation.
Then  we obtain:
\beq
\omega\frac{dI}{d\omega}=\omega\frac{dI_{0}}{d\omega}\frac{1}{(1+\theta_0^{2}\sqrt{\omega^{3}/\hat q})^{2}}.
\eeq
We shall compare the energy distributions for different L and $\theta $. Here index 0 corresponds to the distribution with 
$\theta=0$.
First, consider the differential distributions in transverse momentum for a given frequency. We compare the difference between late emission approximation and use of
Dead-cone factor. We see the significant difference for small $k_t$, which is the dominant region of bremsstrahlung.
Note that this difference only weakly depends on L so we shall depict only the case of one L, say $L=4$ fm.
\begin{figure}
\centering
\includegraphics[width=0.48\textwidth]{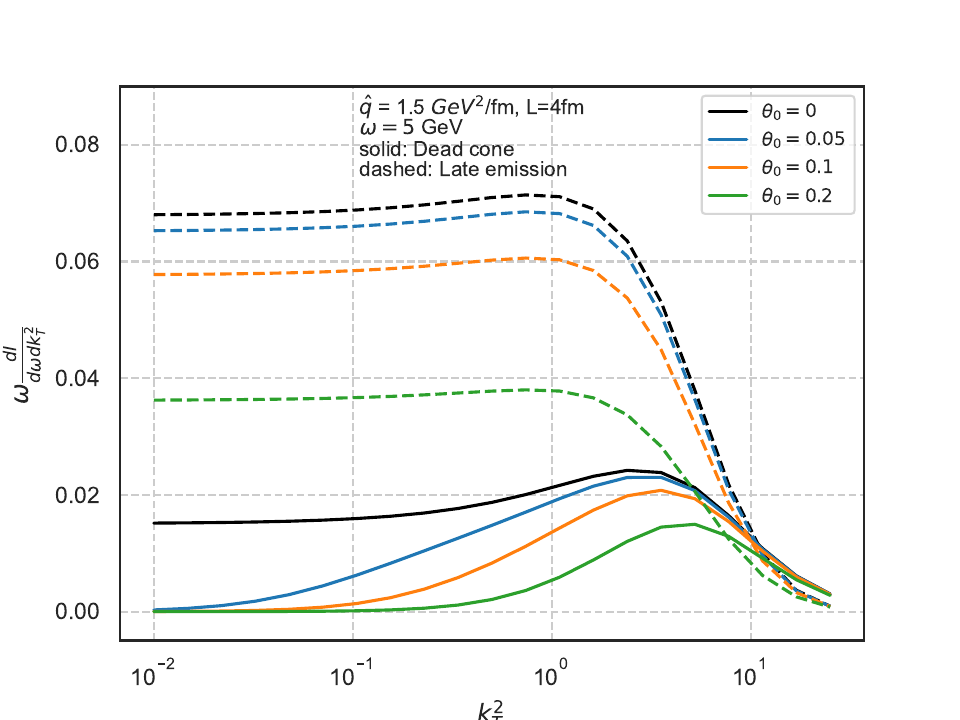}
\includegraphics[width=0.48\textwidth]{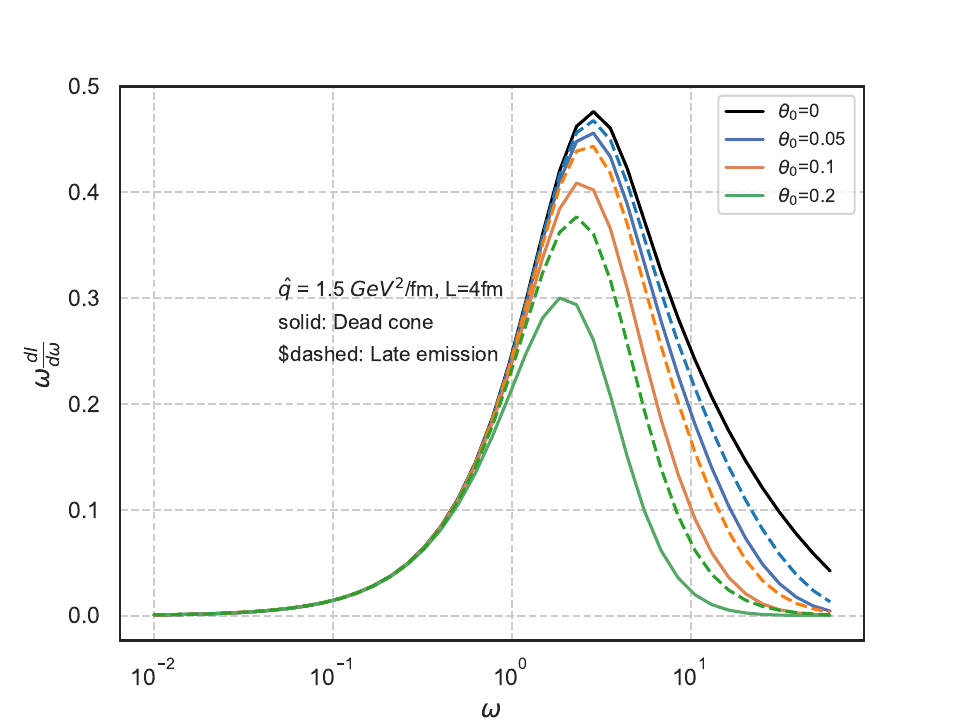}
\caption{The dependence on $k^2_t$ of the $\omega\frac{dI}{d\omega dk^2_t}$ and on $ \omega$  of the $\omega\frac{dI}{d\omega}$ for dead cone factor approach  versus  late emission approximation,
$ \theta=(0.05, 0.1, 0.2), L=4$ fm and $\alpha_s$=0.24.
 \label{R10} }
\end{figure}
We depict on the left side of Fig.\ref{R10} for $\omega\frac{dI}{d\omega d^2k_t}$ calculated in late emission approximation versus 
\beq
 \omega\frac{dI}{d\omega d^2k_t}= \omega\frac{dI_0}{d^2k_td\omega }\frac{1}{(1+\omega^2\theta^2/k_t^2)^2 }
 \eeq
 calculated using the Dokshitzer-Kharzeev dead cone factor. We see that for late emission, the gluons fill the dead cone as in the ASW case, while this effect is absent in the dead cone factor approach:
We see on the right side of Fig.\ref{R10} that the dead-cone approximation curve is always below, and the 
difference increases with the increase of the dead cone angle $\theta_0$. This result depends only very weakly on L.

\subsection{Late emission approximation and ASW formula}
\par In this section, we compare late emission approximation and ASW formula. We shall consider different $\theta=0,0.05,0.1,0.15,0.2$, and $L=20,30, 50 $GeV$^{-1}$, i.e. 4, 6 and 10 fm.

\par We first compare the energy distributions $\omega dI/d\omega$.
\begin{figure}
 \centering
\includegraphics[width=0.48\textwidth]{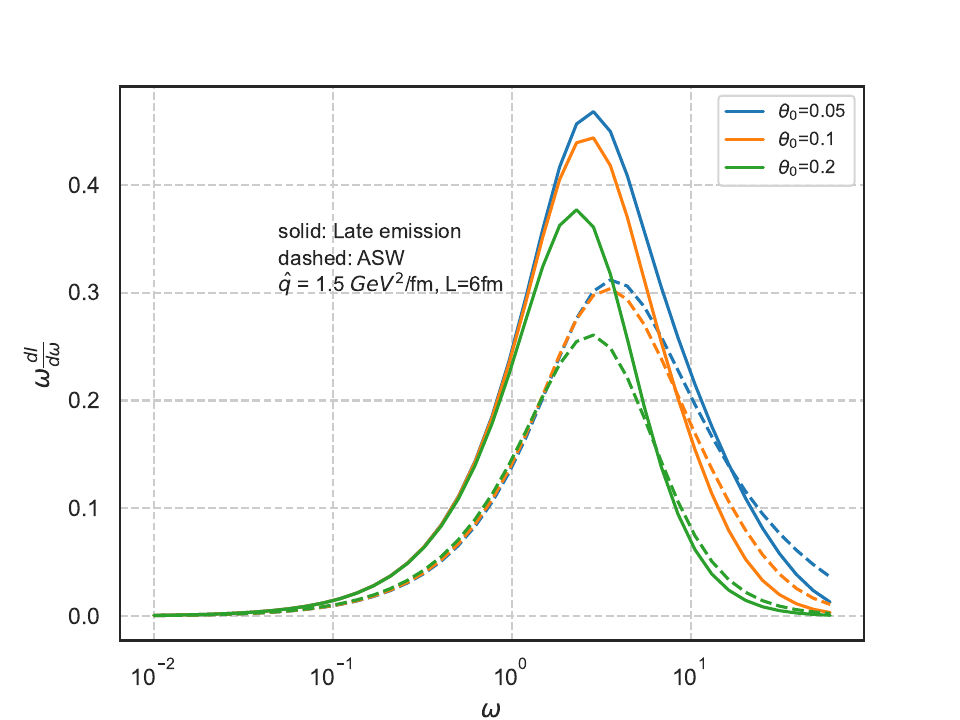}\\
\includegraphics[width=0.48\textwidth]{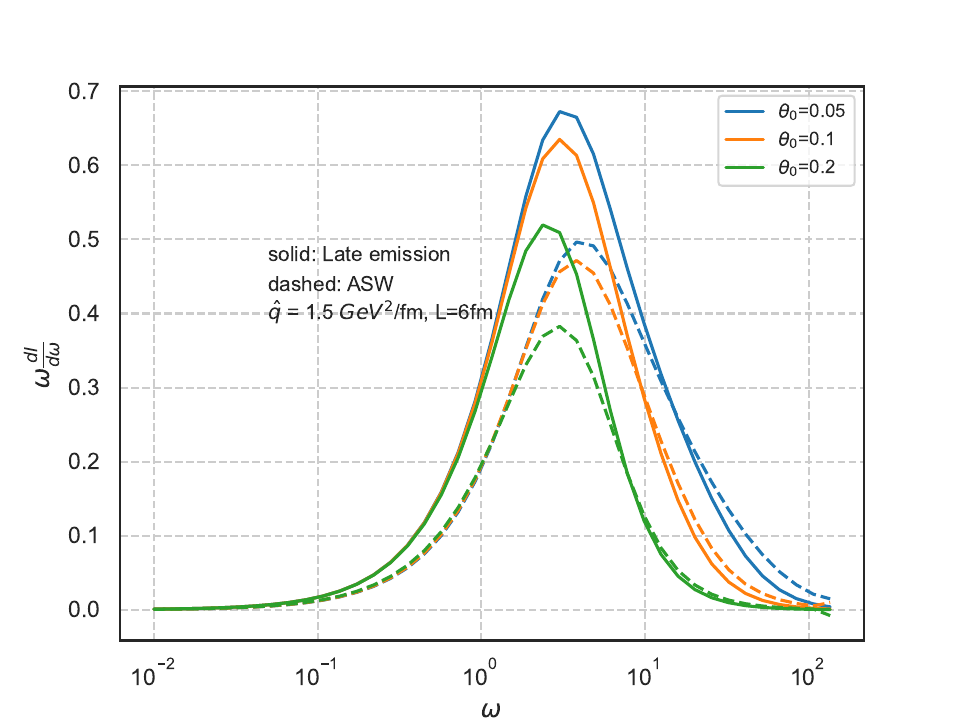}
\includegraphics[width=0.48\textwidth]{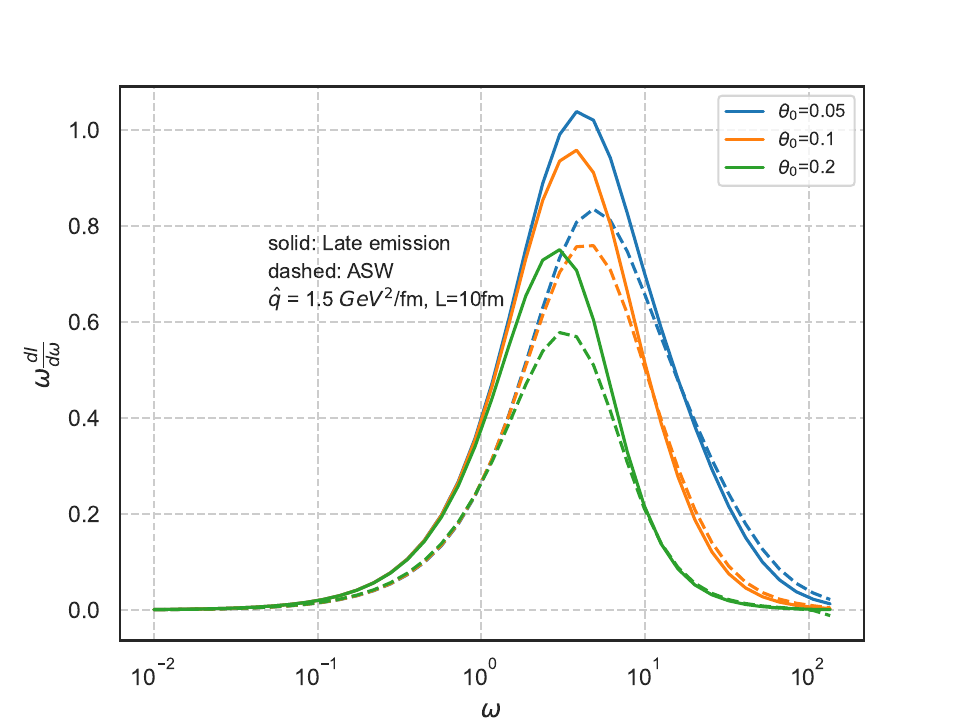}
\caption{ ASW and late emission 
for $\theta=(0.05,0.1,0.2),L=(4, 6, 10)$ fm, and $\alpha_s$=0.24.
 \label{R16} 
 }
\end{figure}
We see a rather good qualitative agreement between ASW and late emission approximation that 
improves quantitatively towards larger L as we expect. In Fig.\ref{R16}, they agree with each other pretty well quantitatively at L=6 fm and above. 
\par We see that late emission approximation for zero quark mass is in good agreement with ASW formula for transverse 
distributions. For finite masses, the results are in agreement with ASW up to not very large frequencies and are slightly smaller than ASW for large frequencies. Thus we are justified to use late emission approximation for the analysis of heavy quark jets.
\par Note that there are additional terms in ASW formula finite in L or even proportional to 1/L. These terms include boundary contribution and hard gluons that are emitted without re-scattering.  These terms are each separately large numerically, even for L=6 fm,
being of the same order as the bulk term, in contrast to BDMPS-Z approach, where the  boundary contribution 
$\sim \log{(\vert \tanh(\omega L)\vert )}$ is negligible already for L=3-4 fm.  This means that it is quite possible for a small L factorisation approach
based on the separation of medium-induced radiation and vacuum fails, or at least the validity of such factorisation remains to be proved for the ASW formula beyond late emission approximation. Indeed, the boundary contribution corresponds 
to radiation when gluon is emitted inside and absorbed outside medium or vice versa. We have seen these terms cancelled out with good accuracy in Sec.\ref{sec:early-emm}
 
\bibliographystyle{JHEP}
\bibliography{ref} 

\providecommand{\href}[2]{#2}\begingroup\raggedright\begin{thebibliography}{10}

\bibitem{Gyulassy:2000fs}
M.~Gyulassy, P.~Levai and I.~Vitev, \emph{{NonAbelian energy loss at finite
  opacity}}, \href{https://doi.org/10.1103/PhysRevLett.85.5535}{\emph{Phys.
  Rev. Lett.} {\bfseries 85} (2000) 5535}
  [\href{https://arxiv.org/abs/nucl-th/0005032}{{\ttfamily nucl-th/0005032}}].

\bibitem{Gyulassy:1999zd}
M.~Gyulassy, P.~Levai and I.~Vitev, \emph{{Jet quenching in thin quark gluon
  plasmas. 1. Formalism}},
  \href{https://doi.org/10.1016/S0550-3213(99)00713-0}{\emph{Nucl. Phys. B}
  {\bfseries 571} (2000) 197}
  [\href{https://arxiv.org/abs/hep-ph/9907461}{{\ttfamily hep-ph/9907461}}].

\bibitem{Gyulassy:2000er}
M.~Gyulassy, P.~Levai and I.~Vitev, \emph{{Reaction operator approach to
  nonAbelian energy loss}},
  \href{https://doi.org/10.1016/S0550-3213(00)00652-0}{\emph{Nucl. Phys. B}
  {\bfseries 594} (2001) 371}
  [\href{https://arxiv.org/abs/nucl-th/0006010}{{\ttfamily nucl-th/0006010}}].

\bibitem{Baier:1996kr}
R.~Baier, Y.L.~Dokshitzer, A.H.~Mueller, S.~Peigne and D.~Schiff,
  \emph{{Radiative energy loss of high-energy quarks and gluons in a finite
  volume quark - gluon plasma}},
  \href{https://doi.org/10.1016/S0550-3213(96)00553-6}{\emph{Nucl. Phys. B}
  {\bfseries 483} (1997) 291}
  [\href{https://arxiv.org/abs/hep-ph/9607355}{{\ttfamily hep-ph/9607355}}].

\bibitem{Baier:1996sk}
R.~Baier, Y.L.~Dokshitzer, A.H.~Mueller, S.~Peigne and D.~Schiff,
  \emph{{Radiative energy loss and p(T) broadening of high-energy partons in
  nuclei}}, \href{https://doi.org/10.1016/S0550-3213(96)00581-0}{\emph{Nucl.
  Phys. B} {\bfseries 484} (1997) 265}
  [\href{https://arxiv.org/abs/hep-ph/9608322}{{\ttfamily hep-ph/9608322}}].

\bibitem{Baier:1998yf}
R.~Baier, Y.L.~Dokshitzer, A.H.~Mueller and D.~Schiff, \emph{{Radiative energy
  loss of high-energy partons traversing an expanding QCD plasma}},
  \href{https://doi.org/10.1103/PhysRevC.58.1706}{\emph{Phys. Rev. C}
  {\bfseries 58} (1998) 1706}
  [\href{https://arxiv.org/abs/hep-ph/9803473}{{\ttfamily hep-ph/9803473}}].

\bibitem{Baier:1998kq}
R.~Baier, Y.L.~Dokshitzer, A.H.~Mueller and D.~Schiff, \emph{{Medium induced
  radiative energy loss: Equivalence between the BDMPS and Zakharov
  formalisms}},
  \href{https://doi.org/10.1016/S0550-3213(98)00546-X}{\emph{Nucl. Phys. B}
  {\bfseries 531} (1998) 403}
  [\href{https://arxiv.org/abs/hep-ph/9804212}{{\ttfamily hep-ph/9804212}}].

\bibitem{Baier:2001yt}
R.~Baier, Y.L.~Dokshitzer, A.H.~Mueller and D.~Schiff, \emph{{Quenching of
  hadron spectra in media}},
  \href{https://doi.org/10.1088/1126-6708/2001/09/033}{\emph{JHEP} {\bfseries
  09} (2001) 033} [\href{https://arxiv.org/abs/hep-ph/0106347}{{\ttfamily
  hep-ph/0106347}}].

\bibitem{Baier:2000mf}
R.~Baier, D.~Schiff and B.G.~Zakharov, \emph{{Energy loss in perturbative
  QCD}}, \href{https://doi.org/10.1146/annurev.nucl.50.1.37}{\emph{Ann. Rev.
  Nucl. Part. Sci.} {\bfseries 50} (2000) 37}
  [\href{https://arxiv.org/abs/hep-ph/0002198}{{\ttfamily hep-ph/0002198}}].

\bibitem{Zakharov:1996fv}
B.G.~Zakharov, \emph{{Fully quantum treatment of the Landau-Pomeranchuk-Migdal
  effect in QED and QCD}}, \href{https://doi.org/10.1134/1.567126}{\emph{JETP
  Lett.} {\bfseries 63} (1996) 952}
  [\href{https://arxiv.org/abs/hep-ph/9607440}{{\ttfamily hep-ph/9607440}}].

\bibitem{Zakharov:1997uu}
B.G.~Zakharov, \emph{{Radiative energy loss of high-energy quarks in finite
  size nuclear matter and quark - gluon plasma}},
  \href{https://doi.org/10.1134/1.567389}{\emph{JETP Lett.} {\bfseries 65}
  (1997) 615} [\href{https://arxiv.org/abs/hep-ph/9704255}{{\ttfamily
  hep-ph/9704255}}].

\bibitem{Dokshitzer:2001zm}
Y.L.~Dokshitzer and D.E.~Kharzeev, \emph{{Heavy quark colorimetry of QCD
  matter}}, \href{https://doi.org/10.1016/S0370-2693(01)01130-3}{\emph{Phys.
  Lett. B} {\bfseries 519} (2001) 199}
  [\href{https://arxiv.org/abs/hep-ph/0106202}{{\ttfamily hep-ph/0106202}}].

\bibitem{Armesto:2003jh}
N.~Armesto, C.A.~Salgado and U.A.~Wiedemann, \emph{{Medium induced gluon
  radiation off massive quarks fills the dead cone}},
  \href{https://doi.org/10.1103/PhysRevD.69.114003}{\emph{Phys. Rev. D}
  {\bfseries 69} (2004) 114003}
  [\href{https://arxiv.org/abs/hep-ph/0312106}{{\ttfamily hep-ph/0312106}}].

\bibitem{Aurenche:2009dj}
P.~Aurenche and B.G.~Zakharov, \emph{{Anomalous mass dependence of radiative
  quark energy loss in a finite-size quark-gluon plasma}},
  \href{https://doi.org/10.1134/S0021364009160048}{\emph{JETP Lett.} {\bfseries
  90} (2009) 237} [\href{https://arxiv.org/abs/0907.1918}{{\ttfamily
  0907.1918}}].

\bibitem{Mehtar-Tani:2012mfa}
Y.~Mehtar-Tani, C.A.~Salgado and K.~Tywoniuk, \emph{{The Radiation pattern of a
  QCD antenna in a dense medium}},
  \href{https://doi.org/10.1007/JHEP10(2012)197}{\emph{JHEP} {\bfseries 10}
  (2012) 197} [\href{https://arxiv.org/abs/1205.5739}{{\ttfamily 1205.5739}}].

\bibitem{Larkoski:2014wba}
A.J.~Larkoski, S.~Marzani, G.~Soyez and J.~Thaler, \emph{{Soft Drop}},
  \href{https://doi.org/10.1007/JHEP05(2014)146}{\emph{JHEP} {\bfseries 05}
  (2014) 146} [\href{https://arxiv.org/abs/1402.2657}{{\ttfamily 1402.2657}}].

\bibitem{Marzani:2019hun}
S.~Marzani, G.~Soyez and M.~Spannowsky, \emph{{Looking inside jets: an
  introduction to jet substructure and boosted-object phenomenology}},
  vol.~958, Springer (2019),
  \href{https://doi.org/10.1007/978-3-030-15709-8}{10.1007/978-3-030-15709-8},
  [\href{https://arxiv.org/abs/1901.10342}{{\ttfamily 1901.10342}}].

\bibitem{Mehtar-Tani:2016aco}
Y.~Mehtar-Tani and K.~Tywoniuk, \emph{{Groomed jets in heavy-ion collisions:
  sensitivity to medium-induced bremsstrahlung}},
  \href{https://doi.org/10.1007/JHEP04(2017)125}{\emph{JHEP} {\bfseries 04}
  (2017) 125} [\href{https://arxiv.org/abs/1610.08930}{{\ttfamily
  1610.08930}}].

\bibitem{Caucal:2019uvr}
P.~Caucal, E.~Iancu and G.~Soyez, \emph{{Deciphering the $z_g$ distribution in
  ultrarelativistic heavy ion collisions}},
  \href{https://doi.org/10.1007/JHEP10(2019)273}{\emph{JHEP} {\bfseries 10}
  (2019) 273} [\href{https://arxiv.org/abs/1907.04866}{{\ttfamily
  1907.04866}}].

\bibitem{Caucal:2018dla}
P.~Caucal, E.~Iancu, A.H.~Mueller and G.~Soyez, \emph{{Vacuum-like jet
  fragmentation in a dense QCD medium}},
  \href{https://doi.org/10.1103/PhysRevLett.120.232001}{\emph{Phys. Rev. Lett.}
  {\bfseries 120} (2018) 232001}
  [\href{https://arxiv.org/abs/1801.09703}{{\ttfamily 1801.09703}}].

\bibitem{Blaizot:2013hx}
J.-P.~Blaizot, E.~Iancu and Y.~Mehtar-Tani, \emph{{Medium-induced QCD cascade:
  democratic branching and wave turbulence}},
  \href{https://doi.org/10.1103/PhysRevLett.111.052001}{\emph{Phys. Rev. Lett.}
  {\bfseries 111} (2013) 052001}
  [\href{https://arxiv.org/abs/1301.6102}{{\ttfamily 1301.6102}}].

\bibitem{Casalderrey-Solana:2011ule}
J.~Casalderrey-Solana and E.~Iancu, \emph{{Interference effects in
  medium-induced gluon radiation}},
  \href{https://doi.org/10.1007/JHEP08(2011)015}{\emph{JHEP} {\bfseries 08}
  (2011) 015} [\href{https://arxiv.org/abs/1105.1760}{{\ttfamily 1105.1760}}].

\bibitem{Caletti:2023spr}
S.~Caletti, A.~Ghira and S.~Marzani, \emph{{On heavy-flavour jets with Soft
  Drop}},  \href{https://arxiv.org/abs/2312.11623}{{\ttfamily 2312.11623}}.

\bibitem{Gyulassy:1993hr}
M.~Gyulassy and X.-n.~Wang, \emph{{Multiple collisions and induced gluon
  Bremsstrahlung in QCD}},
  \href{https://doi.org/10.1016/0550-3213(94)90079-5}{\emph{Nucl. Phys. B}
  {\bfseries 420} (1994) 583}
  [\href{https://arxiv.org/abs/nucl-th/9306003}{{\ttfamily nucl-th/9306003}}].

\bibitem{Mehtar-Tani:2011lic}
Y.~Mehtar-Tani, C.A.~Salgado and K.~Tywoniuk, \emph{{The radiation pattern of a
  QCD antenna in a dilute medium}},
  \href{https://doi.org/10.1007/JHEP04(2012)064}{\emph{JHEP} {\bfseries 04}
  (2012) 064} [\href{https://arxiv.org/abs/1112.5031}{{\ttfamily 1112.5031}}].

\bibitem{Blok:2019uny}
B.~Blok and K.~Tywoniuk, \emph{{Higher-order corrections to heavy-quark jet
  quenching}}, \href{https://doi.org/10.1140/epjc/s10052-019-7061-4}{\emph{Eur.
  Phys. J. C} {\bfseries 79} (2019) 560}
  [\href{https://arxiv.org/abs/1901.07864}{{\ttfamily 1901.07864}}].

\bibitem{Dokshitzer:1991fd}
Y.L.~Dokshitzer, V.A.~Khoze and S.I.~Troian, \emph{{On specific QCD properties
  of heavy quark fragmentation ('dead cone')}},
  \href{https://doi.org/10.1088/0954-3899/17/10/023}{\emph{J. Phys. G}
  {\bfseries 17} (1991) 1602}.

\bibitem{Baier:2001qw}
R.~Baier, Y.L.~Dokshitzer, A.H.~Mueller and D.~Schiff, \emph{{On the angular
  dependence of the radiative gluon spectrum}},
  \href{https://doi.org/10.1103/PhysRevC.64.057902}{\emph{Phys. Rev. C}
  {\bfseries 64} (2001) 057902}
  [\href{https://arxiv.org/abs/hep-ph/0105062}{{\ttfamily hep-ph/0105062}}].

\bibitem{Dokshitzer:1988bq}
Y.L.~Dokshitzer, V.A.~Khoze and S.I.~Troian, \emph{{Coherence and Physics of
  QCD Jets}}, \href{https://doi.org/10.1142/9789814503266_0003}{\emph{Adv. Ser.
  Direct. High Energy Phys.} {\bfseries 5} (1988) 241}.

\bibitem{Khoze:2000iq}
V.A.~Khoze, W.~Ochs and J.~Wosiek, \emph{{Analytical QCD and multiparticle
  production}},  \href{https://arxiv.org/abs/hep-ph/0009298}{{\ttfamily
  hep-ph/0009298}}.

\bibitem{Khoze:1996dn}
V.A.~Khoze and W.~Ochs, \emph{{Perturbative QCD approach to multiparticle
  production}}, \href{https://doi.org/10.1142/S0217751X97001638}{\emph{Int. J.
  Mod. Phys. A} {\bfseries 12} (1997) 2949}
  [\href{https://arxiv.org/abs/hep-ph/9701421}{{\ttfamily hep-ph/9701421}}].

\bibitem{Cunqueiro:2022svx}
L.~Cunqueiro, D.~Napoletano and A.~Soto-Ontoso, \emph{{Dead-cone searches in
  heavy-ion collisions using the jet tree}},
  \href{https://doi.org/10.1103/PhysRevD.107.094008}{\emph{Phys. Rev. D}
  {\bfseries 107} (2023) 094008}
  [\href{https://arxiv.org/abs/2211.11789}{{\ttfamily 2211.11789}}].

\bibitem{Calvo:2014cba}
M.R.~Calvo, M.R.~Moldes and C.A.~Salgado, \emph{{Color coherence in a heavy
  quark antenna radiating gluons inside a QCD medium}},
  \href{https://doi.org/10.1016/j.physletb.2014.10.010}{\emph{Phys. Lett. B}
  {\bfseries 738} (2014) 448}
  [\href{https://arxiv.org/abs/1403.4892}{{\ttfamily 1403.4892}}].

\bibitem{Fister:2014zxa}
L.~Fister and E.~Iancu, \emph{{Medium-induced jet evolution: wave turbulence
  and energy loss}}, \href{https://doi.org/10.1007/JHEP03(2015)082}{\emph{JHEP}
  {\bfseries 03} (2015) 082} [\href{https://arxiv.org/abs/1409.2010}{{\ttfamily
  1409.2010}}].

\bibitem{Larkoski:2015lea}
A.J.~Larkoski, S.~Marzani and J.~Thaler, \emph{{Sudakov Safety in Perturbative
  QCD}}, \href{https://doi.org/10.1103/PhysRevD.91.111501}{\emph{Phys. Rev. D}
  {\bfseries 91} (2015) 111501}
  [\href{https://arxiv.org/abs/1502.01719}{{\ttfamily 1502.01719}}].

\bibitem{ALICE:2019ykw}
{\scshape ALICE} collaboration, \emph{{Exploration of jet substructure using
  iterative declustering in pp and Pb\textendash{}Pb collisions at LHC
  energies}}, \href{https://doi.org/10.1016/j.physletb.2020.135227}{\emph{Phys.
  Lett. B} {\bfseries 802} (2020) 135227}
  [\href{https://arxiv.org/abs/1905.02512}{{\ttfamily 1905.02512}}].

\bibitem{Li:2017wwc}
H.T.~Li and I.~Vitev, \emph{{Inverting the mass hierarchy of jet quenching
  effects with prompt $b$-jet substructure}},
  \href{https://doi.org/10.1016/j.physletb.2019.04.052}{\emph{Phys. Lett. B}
  {\bfseries 793} (2019) 259}
  [\href{https://arxiv.org/abs/1801.00008}{{\ttfamily 1801.00008}}].

\bibitem{Craft:2022kdo}
E.~Craft, K.~Lee, B.~Me\c{c}aj and I.~Moult, \emph{{Beautiful and Charming
  Energy Correlators}},  \href{https://arxiv.org/abs/2210.09311}{{\ttfamily
  2210.09311}}.

\bibitem{Attems:2022ubu}
M.~Attems, J.~Brewer, G.M.~Innocenti, A.~Mazeliauskas, S.~Park, W.~van~der
  Schee et~al., \emph{{The medium-modified $ g\to c\overline{c} $ splitting
  function in the BDMPS-Z formalism}},
  \href{https://doi.org/10.1007/JHEP01(2023)080}{\emph{JHEP} {\bfseries 01}
  (2023) 080} [\href{https://arxiv.org/abs/2203.11241}{{\ttfamily
  2203.11241}}].

\bibitem{Vertesi:2021brz}
{\scshape ALICE} collaboration, \emph{{Jet Substructure Measurements with
  ALICE}}, \href{https://doi.org/10.1134/S1063779623040354}{\emph{Phys. Part.
  Nucl.} {\bfseries 54} (2023) 670}
  [\href{https://arxiv.org/abs/2110.11606}{{\ttfamily 2110.11606}}].

\bibitem{CMS:2017qlm}
{\scshape CMS} collaboration, \emph{{Measurement of the Splitting Function in
  $pp$ and Pb-Pb Collisions at $\sqrt{s_{_{\mathrm{NN}}}} =$ 5.02 TeV}},
  \href{https://doi.org/10.1103/PhysRevLett.120.142302}{\emph{Phys. Rev. Lett.}
  {\bfseries 120} (2018) 142302}
  [\href{https://arxiv.org/abs/1708.09429}{{\ttfamily 1708.09429}}].

\bibitem{ALICE:2021aqk}
{\scshape ALICE} collaboration, \emph{{Direct observation of the dead-cone
  effect in quantum chromodynamics}},
  \href{https://doi.org/10.1038/s41586-022-04572-w}{\emph{Nature} {\bfseries
  605} (2022) 440} [\href{https://arxiv.org/abs/2106.05713}{{\ttfamily
  2106.05713}}].

\bibitem{Blaizot:2014ula}
J.-P.~Blaizot, Y.~Mehtar-Tani and M.A.C.~Torres, \emph{{Angular structure of
  the in-medium QCD cascade}},
  \href{https://doi.org/10.1103/PhysRevLett.114.222002}{\emph{Phys. Rev. Lett.}
  {\bfseries 114} (2015) 222002}
  [\href{https://arxiv.org/abs/1407.0326}{{\ttfamily 1407.0326}}].

\bibitem{Blaizot:2013vha}
J.-P.~Blaizot, F.~Dominguez, E.~Iancu and Y.~Mehtar-Tani, \emph{{Probabilistic
  picture for medium-induced jet evolution}},
  \href{https://doi.org/10.1007/JHEP06(2014)075}{\emph{JHEP} {\bfseries 06}
  (2014) 075} [\href{https://arxiv.org/abs/1311.5823}{{\ttfamily 1311.5823}}].

\end{thebibliography}\endgroup

\end{document}